\documentclass{aa}

\usepackage{graphicx}
\usepackage{xcolor}
\usepackage{txfonts}
\usepackage{natbib}

\usepackage[colorlinks=true, linkcolor=blue, citecolor=blue, urlcolor=blue]{hyperref}
\newcommand{\mSun}{\ensuremath{M_\odot}\xspace}
\newcommand{\mearth}{\ensuremath{M_\oplus}\xspace}

\newcommand\secref[1]{Sect.~\ref{#1}}

\newcommand\figref[1]{Fig. \ref{#1}}

\newcommand\figrefthree[3]{\ifnum\ifhmode\spacefactor\else2000\fi>1000 Figures~\ref{#1}, \ref{#2}, and \ref{#3}\else Figs.~\ref{#1}, \ref{#2}, and \ref{#3}\fi}

\newcommand\exto{Extended 1}
\newcommand\extn{Extended N}
\newcommand\como{Compact 1}
\newcommand\comn{Compact N}
\usepackage{orcidlink}

\graphicspath{ {./} }

\begin{document}

   \title{Implications for the formation of Oort cloud-like structures and interstellar comets in dense environments}

   \author{Santiago Torres 
          \inst{1}\,\orcidlink{0000-0002-3150-8988}
          }

   \institute{Institute of Science and Technology Austria (ISTA), Am Campus 1, 3400 Klosterneuburg, Austria\\
              \email{santiago.torres@ista.ac.at}
             }

   \date{}

%
 
  \abstract
  %
  {Most stars form in dense stellar environments, where frequent close encounters can strongly perturb and reshape the early architecture of planetary systems. The Solar System, with its rich population of distant comets, provides a natural laboratory to study these processes. We performed detailed numerical simulations using the \texttt{LonelyPlanets} framework that combines \texttt{NBODY6++GPU} and \texttt{REBOUND} to explore the evolution of debris disks around Solar System analogues embedded in stellar clusters. Two initial configurations are considered, the Extended model and the Compact model, each containing four giant planets and either an extended or compact debris disk. We find that compact disks primarily form Kuiper belt and scattered disk-like populations through planet–disk interactions, while extended disks are more strongly shaped by stellar encounters, producing Oort cloud-like structures and interstellar comets with ejection velocities of 1–3~km~s$^{-1}$. Stellar perturbations are most effective for encounter inclinations between $0^\circ$ and $30^\circ$, giving rise to distinct dynamical populations, like Sednoids, and inner Oort cloud analogues, and a characteristic tail in semimajor axis-eccentricity space. In coplanar encounters, the disk remains largely flattened, whereas polar flybys redistribute angular momentum vertically, producing nearly isotropic outer populations that resemble an emerging Oort cloud. Our results suggest that cometary reservoirs and interstellar objects are natural byproducts of planet–disk interactions and stellar flybys in dense clusters, linking the architecture of outer planetary systems to their birth environments.}


   \maketitle

\section{Introduction}
\label{intro}

\cite{J.H.Oort1950} proposed the existence of a cloud of comets that surrounds the Solar System, which was subsequently named after him. The Oort cloud is thought to contain about $10^{11}$--$10^{12} $ comets \citep{Brasser2013}, with a total mass of around 2--3\ \mearth \citep[e.g.,][]{Francis2005, Morbidelli2008}. Its shape is conjectured to be nearly spherical, with the cloud extending up to $0.5$\ pc from the Sun, limited by the Hill sphere of the Solar System \citep{J.H.Oort1950, Chebotarev1965}. Because these estimates of the Oort cloud properties are
highly uncertain (mainly because of the lack of observations of the long-period comets), the creation, evolution, and even the existence of the Oort cloud remains a puzzle today. One of the main open questions is related to the moment of its formation.

Several works have explored the origin and evolution of the Oort cloud, mainly by performing numerical simulations. These studies follow two main ideas of formation: primordial formation and late formation. The primordial model occurs in the early stages
of the Solar System when the Sun was still in its birth cluster \citep[see,
e.g.,][]{J.H.Oort1950,Hills1981,Heisler1986a, Duncan1987,Weissman1996,
Wiegert1999,Dones2004a,Levison2010}. In this scenario, the planetesimals were scattered due to the interaction with the growing giant planets and started populating the Oort cloud. The late model assumes that the Sun has already left its birth cluster. In this alternative scenario the Oort cloud was created in the later stages of the Solar System formation possibly due to gravitational instabilities when the giant plants experienced an orbital resonance that caused the minor bodies to be ejected into almost unbound orbits \citep[see, e.g.,][]{Levison2004,Brasser2006,Fouchard2006,Kaib2008,Morbidelli2008,Duncan2008,Brasser2013, Shannon2014,Dones2015,Nesvorny2018,Shannon2019,Vokrouhlicky2019,Wu2023,Wu2024}.

It is now widely accepted that the Solar System was born within a stellar cluster \citep{PortegiesZwart2009,Adams2010}, and that the Sun remained a member of this cluster during the formation of the planets \citep{Pichardo2012,PortegiesZwart2015a,Martinez-Barbosa2016}. Given that the typical lifetime of an open cluster is approximately 100 Myr and that stellar clusters are inherently dense environments, the likelihood of close stellar encounters during this period is significantly higher than it would be for a star evolving in isolation \citep{PortegiesZwart2015a}. Consequently, the Oort cloud probably experienced gravitational interactions with the Sun's stellar siblings during the early stages of its formation.

Several studies have investigated the evolution of the Solar System within dense stellar environments. For instance, \citet{Nordlander2017} examined the fate of a primordial Oort cloud across different models of the Sun's birth cluster (low-, intermediate-, and high-mass). They considered the long-term evolution of the cloud, beginning at 100 Myr and extending through the late heavy bombardment phase ($\sim$500 Myr). Their findings suggest that a primordial Oort cloud is unlikely to survive in any of the modeled clusters, with the low-mass cluster providing the most favorable conditions for Solar System's survival. This supports the hypothesis of a delayed Oort cloud formation \citep[e.g.,][]{Brasser2013}. Other authors \citep[e.g.,][]{Eggers1997,Levison2010,Jilkova2016a,Pfalzner2018a,Hands2019} have proposed that portions of the Oort cloud may have been captured through interactions with passing stars while the Sun remained in its birth cluster. Such stellar encounters have also been suggested as a possible origin for distant objects like Sedna. More recently, \citet{Wajer2024a,Wajer2024b} explored the dynamical evolution of planetesimals initially located in the Jupiter–Saturn region. Their work examined how stellar interactions in the birth cluster could lead to the formation of Sedna-like objects. Their simulations predict the existence of previously undetected classes of sednoids, including very small bodies and retrograde orbiters. The similarities and differences among these studies highlight a central challenge: the uncertainty in the initial conditions of the Sun's birth cluster and the poorly constrained history of stellar encounters experienced by the early Solar System.

\begin{table*}[]
  \centering
  \caption{Initial conditions for the Extended and Compact models. } 
  \label{table1}
  \begin{tabular}{lccccc}
    \hline
    \hline
    Model & Disk [au] & Planetary configuration [au] & Time & $N_{Enc}$ & $q_\star$ [au]\\
    \hline
    \exto   & 40-1000  &J:5.2, S:9.5, U:19.2, N:30.1  &  20 000 yr & Single  & 70 -- 400   \\
    \extn & 40-1000  & J:5.2, S:9.5, U:19.2, N:30.1   & 100 Myr  &  Multiple & Random \\
    \como   & 16--35  &J:5.5, S:8.1, N:11.5, U:14.2  & 20 000 yr & Single & 70 -- 400  \\
    \comn   & 16--35   & J:5.5, S:8.1, N:11.5, U:14.2 & 100 Myr  & Multiple & Random  \\
    \hline
    \hline
  \end{tabular}
  \tablefoot{The first column indicates the model type. The second column lists the initial disk size. The third column shows the initial orbital configuration of the giant planets: Jupiter (J), Saturn (S), Uranus (U), and Neptune (N). The following columns represent the integration time, the number of stellar encounters ($N_{Enc}$), and the perturber's distance at closest approach ($q_\star$).}
\end{table*}

In their detailed numerical study, \citet{PZ_ST_2021} provided a comprehensive picture of the chronology underlying the formation and evolution of the Oort cloud. Their simulations indicate that particles with semimajor axes ranging from $\sim 100$~au to several thousand au retain dynamical signatures indicative of the Sun’s origin within a dense stellar cluster of approximately $a \geq 1000$\ $M_{\odot}p^{-3}$. Furthermore, their results suggest that most of the outer Oort cloud formed after the Solar System’s departure from its natal environment. Nevertheless, the initial formation processes were initiated by planet–disk and stellar–disk interactions occurring while the Solar System still resided within its birth cluster. 

The formation and evolution of the Oort cloud offer a unique opportunity to better understand the formation of similar populations of comets in other planetary systems, which to date remain undetected. However, the detection of planetary systems hosting planets, debris disks, and exocomets \citep[see, e.g.,][]{BarradoNavascu1999,Lecavelier2022,Rebollido2024,JL_Torres2025} has opened the possibility of investigating the formation of Oort cloud-like structures and the dynamical processes that give rise to resonant bodies and interstellar comets or interstellar objects (ISOs). Motivated by these findings and by the hypothesis that the formation of the Oort cloud began during the early evolutionary stages of the Solar System while the Sun was still embedded in its birth cluster, we  used the Solar System as a laboratory to investigate the processes that led to the formation of the Oort cloud-like structures in a dense environment. Additionally, we explored the broader implications for the formation of substructures such as the Kuiper Belt, the scattered disk, and interstellar comets in other planetary systems. This approach will help improve our understanding and interpretation of future observations of Oort cloud-like structures and the formation of distinct dynamical populations, including Sednoids and ISOs. These populations are likely common in planetary systems, but remain largely undetected at present.

To address this, we investigated the dynamical evolution of a debris disk under the combined influence of both single and multiple random stellar encounters, as well as gravitational perturbations from the giant planets. As a reference framework, we adopted the Solar System and constructed two analogous configurations: an Extended model and a Compact model. The Extended model is based on the present-day orbits of the giant planets, Jupiter at 5.2 au, Saturn at 9.5 au, Uranus at 19.2 au, and Neptune at 30.1 au, and features an extended debris disk spanning $40 < a < 1000$ au. The Compact model follows a Nice-based configuration \citep{Gomes2005,Tsiganis2005,Morbidelli2005}, where the giant planets are initially placed on more compact orbits: Jupiter at 5.5 au, Saturn at 8.1 au, Neptune at 11.5 au, and Uranus at 14.2 au. This configuration includes a more confined debris disk extending from $16 < a < 35$ au.

The structure of this paper is as follows. In Sect. \ref{sec2} we describe our numerical implementation. In Sect. \ref{sec3} we present the results of our N-body simulations, which explore the evolution of Solar System analogues within a birth cluster environment under two modeled configurations. In Sect. \ref{sec4} we discuss the effects of a dense stellar environment on the formation of transitional interstellar comets \citep{Torres2019} and hyperbolic objects such as ’Oumuamua \citep{Chambers2016,Meech2017}. Finally, in Sect. \ref{summary} we summarize our results and present our conclusions.

\section{Numerical implementation}
\label{sec2}

To explore the formation of Oort cloud-like structures and interstellar comets, we examined the dynamical evolution of a debris disk subject to gravitational perturbations from planetary and stellar encounters. We developed two distinct models, the Compact and Extended configurations, each representing a Solar System analogue (SSA) composed of the four giant planets and a surrounding non-self-gravitating test particle disk. For each model, we conducted two simulation scenarios: one scenario featuring a single close stellar encounter between the SSA and another star (Sect. \ref{sec2.1}), represented by the \exto\ and \como\ models; and another scenario involving multiple stellar encounters (Sect. \ref{sec2.2}), represented by the \extn\ and \comn\ models. Detailed initial conditions for these four simulation models are presented in Table \ref{table1}.

\subsection{Single stellar encounter}
\label{sec2.1}

For the simulations and the generation of initial conditions for the \exto\ and \como\ models, we employed the N-body code \texttt{REBOUND} \citep{Rein2012} with the IAS15 integrator. IAS15 is a high-order, adaptive timestep integrator chosen specifically due to its ability to achieve extremely low relative energy errors ($\sim10^{-14}$). Despite being a nonsymplectic integrator, IAS15 demonstrates superior energy conservation compared to symplectic integrators, and its adaptive timestep capability facilitates accurate resolution of close stellar encounters and potential collisions. We carried out a total of 32 simulations, each modeling a stellar encounter with a Sun-like star. These encounters were characterized by impact parameters ranging from 70 au to 400 au, inclinations ($i$) of $0^\circ$, $30^\circ$, $60^\circ$, and $90^\circ$, and a relative velocity of approximately $1$~km/s, representative of typical conditions within open cluster environments \citep{Galactic_dynamics08}. The debris disk particles were represented as massless test particles ($1000$ per simulation), distributed randomly according to their initial semimajor axes. Each simulation ran for a total duration of $20,000$ years, with the closest approach occurring at $10,000$ years.

\subsection{Multiple stellar encounters}
\label{sec2.2}

To account for the effects of multiple stellar encounters, we simulated the \extn\ and \comn\ models using \texttt{LonelyPlanets} framework \citep{Cai2015,Cai2017a,Cai2018}, which integrates \texttt{NBODY6++GPU} \citep{Wang2015a} for stellar dynamics and \texttt{REBOUND} \citep{Rein2012} for planetary dynamics within the Astrophysical Multi-purpose Software Environment (AMUSE; \citealt{Portegies2009,Pelupessy2013a,amusebook2018}). The stellar environment was modeled as an open cluster using \texttt{NBODY6++GPU} with a Plummer sphere density distribution \citep{Plummer1911}. The initial mass function (IMF) for the stars ranged from $0.08$ to $20$~\mSun \citep{Kroupa2001}, comprising $2,000$ stars within a virial radius of $1$~pc. We placed the SSA into this stellar environment using the \texttt{LonelyPlanets} module and evolved the combined system over a period of $100$~Myr.

In the \texttt{LonelyPlanets} framework, SSA follow the global $N$-body dynamics of their host stars within the cluster. The resulting encounter statistics naturally reflect the underlying cluster demographics, which strongly favor low-mass perturbers. The Kroupa IMF yields an average stellar mass of only $\langle M\rangle\simeq0.5$--$0.6\,M_\odot$, such that in a $2000$-member cluster fewer than $\sim2\%$ of stars exceed $2\,M_\odot$ and only $\sim0.1\%$ exceed $8\,M_\odot$. SSAs are therefore more likely to reside in the cluster outskirts, where encounters are dominated by abundant low-mass stars and close approaches with massive perturbers remain rare \citep[e.g.,][]{Stock2020}. Moreover, massive stars evolve off the main sequence on timescales much shorter than the $\sim100$ Myr dissolution time of a typical young cluster, leaving only a brief window for interactions with planetary systems. Mass segregation further enhances this bias by driving the most massive stars into the cluster core on timescales of a few Myr \citep{Spurzem_Takahashi1995,Mouri2002}.

Once we set the initial conditions, we initially conducted a preliminary analysis using a grid of low-resolution simulations ($200$ simulations, each with $200$ test particles in the debris disk) to efficiently minimize computational resources. From these preliminary results, we selected $24$ systems for detailed, high-resolution simulations, each incorporating $2,000$ test particles in the debris disk. In every high-resolution run, the SSA was initialized at different positions within the cluster to capture the variety of local environments. The \texttt{LonelyPlanets} module employs a KD-tree algorithm to dynamically identify the five nearest neighbours to the SSA at each timestep. Following previous convergence tests in the \texttt{LonelyPlanets} framework, encounters with the five closest stars provide a good compromise between computational efficiency and accuracy in reproducing the cumulative perturbations experienced by planetary systems \citep[e.g.,][]{Cai2019,Dotti2019,Stock2020,Veras2020}.

The selected perturbers, together with the SSA, were then integrated using the IAS15 integrator \citep{Rein2014}, which ensures precise modeling of their mutual gravitational interactions, and low relative energy errors over simulation timescales of approximately $100$~Myr. The results for the Extended model are presented in Sect.~\ref{sec3.1}, while those for the Compact models are presented in Sect.~\ref {sec3.2}. To complement these cluster-based simulations, we also perform controlled single-encounter simulations with A- and B-type stars for the Compact and Extended models (Appendix~\ref{MS_ss}), to directly probe the impact of these rare but potentially encounters with massive stars.

\section{Dynamical evolution of a debris disk in dense environments}
\label{sec3}

Debris disks are the remnants of planet formation, such as the asteroid belt and Kuiper belt in the Solar System. Observational studies indicate that debris disk sizes typically span from approximately $10$\ au in their innermost regions to as much as $1000$\ au in their outer extents \citep[e.g.,][]{Stark2009,Hughes2018,Bertini2023}. The architecture and structure of these disks are shaped primarily by their dynamical histories during the early stages of their host stars' formation, especially within the birth clusters. Given that most stars, including our Sun, originate within such stellar environments \citep[e.g.,][]{Lada2003,PortegiesZwart2009,Adams2010,Pichardo2012}, these dense environments serve as ideal laboratories for investigating the formation of structures like Kuiper belt, scattered disk, the Oort cloud and the population of objects similar to Sedna. Using the numerical implementation described in Sect. \ref{sec2}, this section presents the results and detailed analysis of our simulations for both the Extended \ (Sect. \ref{sec3.1}) and Compact\ (Sect. \ref{sec3.2}) models.

\subsection{Extended models}
\label{sec3.1}

In this section we present the analysis and results of the \exto and \extn\ models (see Sect. \ref{sec2} and Table \ref{table1} for detailed initial conditions). First, we analyse the dynamical evolution of an extended debris disk (ranging from $40$ to $1000$ au) subject to a single close stellar encounter (Sect. \ref{sec3.1.1}). Subsequently, we investigate the disk evolution under multiple stellar encounters in an open cluster environment (Sect. \ref{Nenc_ss}).

\subsubsection{\exto} 
\label{sec3.1.1}

\begin{figure}
\centering 
\includegraphics[width=0.87\hsize]{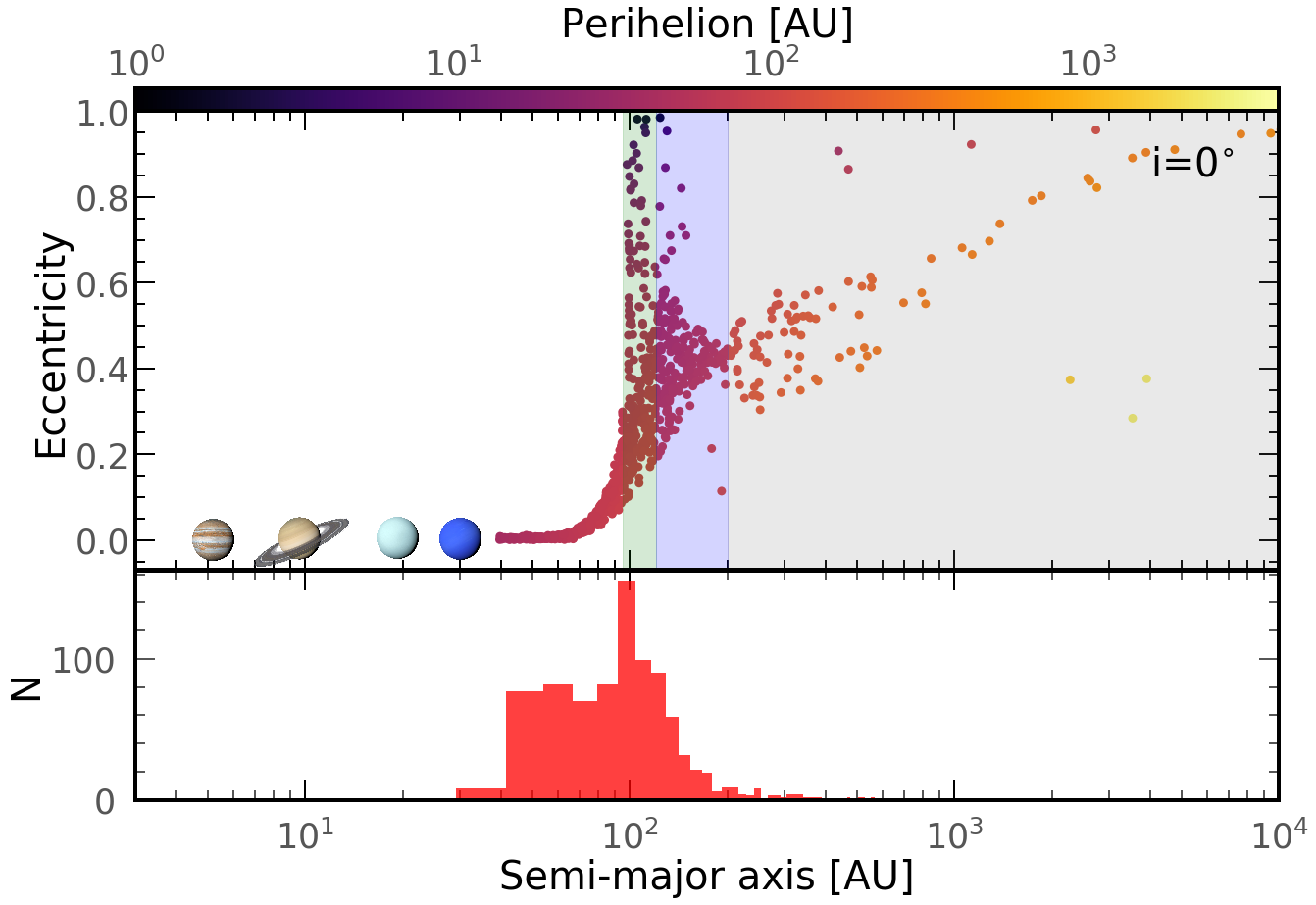}
  \includegraphics[width=0.87\hsize]{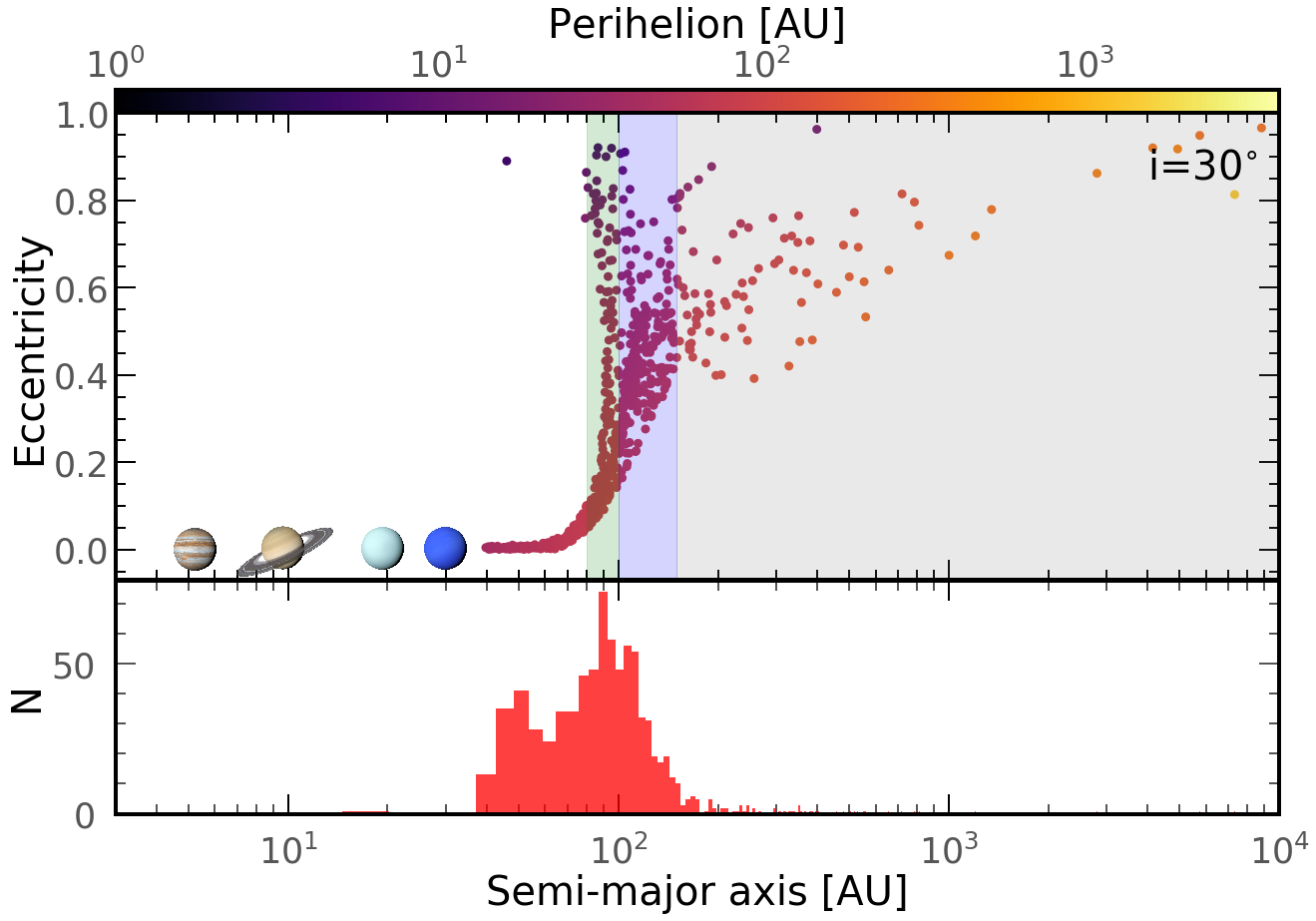}\\
  \includegraphics[width=0.87\hsize]{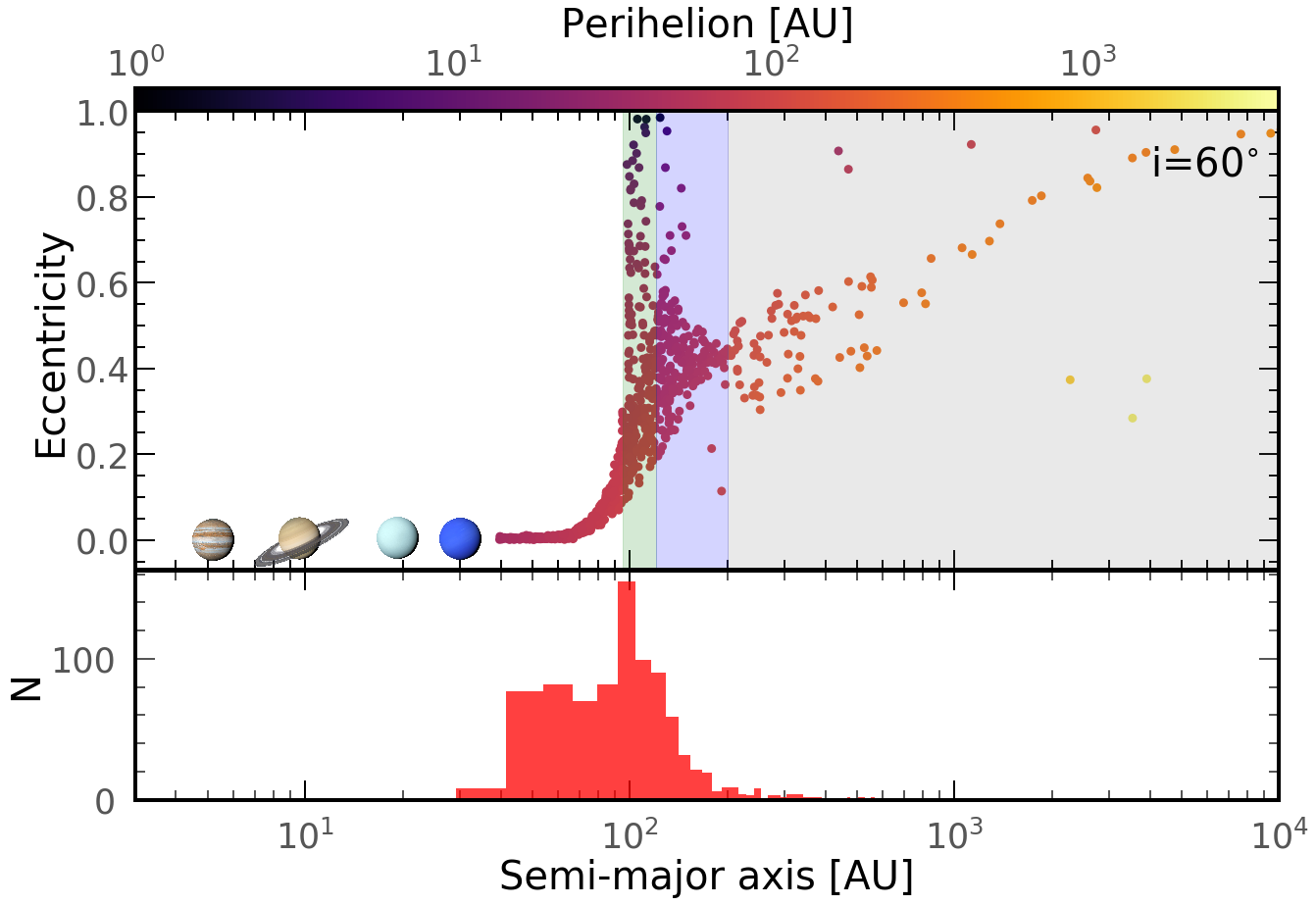}\\
  \includegraphics[width=0.87\hsize]{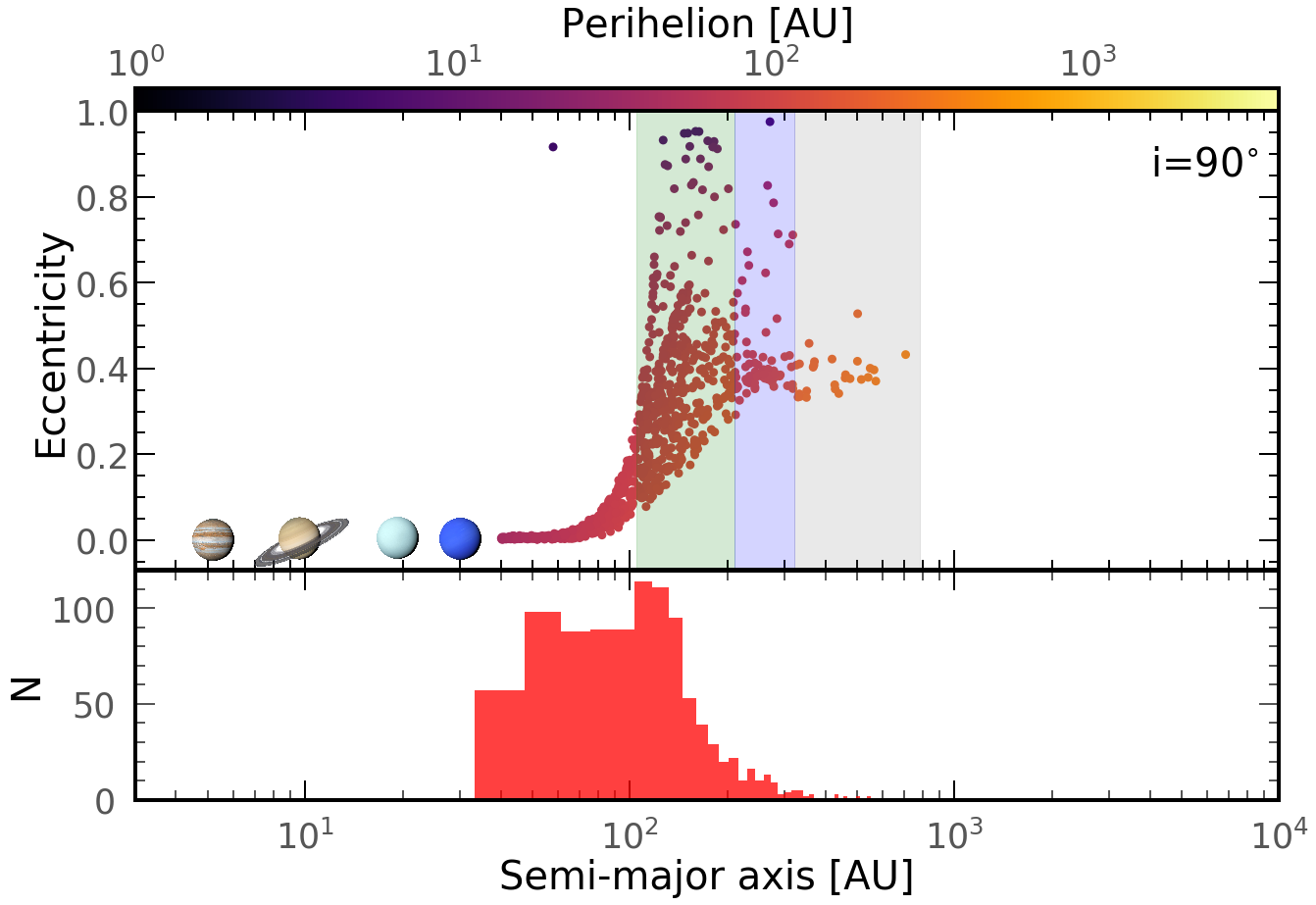} 
  \caption{\exto\ model. Semimajor axis as a function of eccentricity and particle count after an encounter with a $1$~\mSun star at 300 au. The color bar indicates the perihelion distance of the particles. Each panel corresponds to a different encounter inclination angle ($0^{\circ}$, $30^{\circ}$, $60^{\circ}$, and $90^{\circ}$).}
    \label{pss_ss}
\end{figure}  

\begin{figure}[ht!]
  \includegraphics[width=\columnwidth]{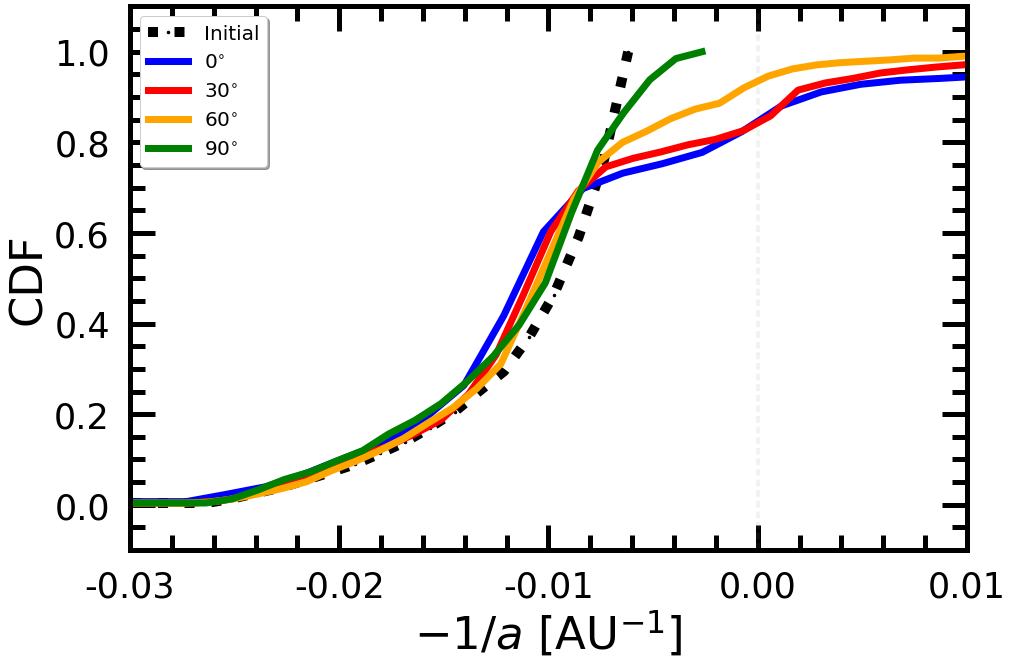}
  \caption{\exto\ model. Cumulative distribution of the final energy of particles in the disk. The colored lines correspond to different inclination angles of the encounter ($0^{\circ}$, $30^{\circ}$, $60^{\circ}$, and $90^{\circ}$), while the black dotted line represents the initial particle distribution as defined in Table \ref{table1}. Negative energy values indicate particles that remain bound to the system after the encounter, whereas positive values correspond to ISOs. The Kolmogorov-Smirnov (KS) probabilities relative to the initial distribution for the different angles are $0.90$\% for $0^{\circ}$, $0.19$\% for $30^{\circ}$, $0.07$\% for $60^{\circ}$, and $0.05$\% for $90^{\circ}$.} 
\label{e_ss}
\end{figure}

Following the methodology described in Sect. \ref{sec2} and using the initial conditions outlined in Table \ref{table1}, we conducted 32 simulations for the \exto\ model. In Figs. \ref{pss_ss} and \ref{e_ss} we present the most significant case, where the perturbing star passes within 300 au of the SSA.

Across all encounter angles, the majority of particles remain within the original disk, with retention rates of $73$\% for $0^{\circ}$, $76.4$\% for $30^{\circ}$, $81.2$\% for $60^{\circ}$, and $88$\% for $90^{\circ}$. These retained particles maintain near-zero eccentricities and semimajor axes between $40$ and $60$ au. Beyond this region, the stellar encounter induces the formation of three distinct dynamical structures, highlighted in green, blue, and gray in Fig. \ref{pss_ss}. In the green region, the particles exhibit eccentricities ranging from $0.1$ to $0.9$ with perihelion distances from $0$ to $50$ au, and are likely to become resonant bodies, analogous to trans-Neptunian objects in the Kuiper Belt. In the blue region, the particles have maximum eccentricities between $0.6$ and $0.7$ and perihelion distances from $50$ to $100$ au, with semimajor axes comparable to those of scattered disk objects in the Solar System. Finally, in the gray region, the particles acquire eccentricities from $\sim0.4$ to $1$ and perihelion distance of up to $1000$ au, similar to the comets in the inner Oort cloud, and in particular to Sedna-like objects. 

For a stellar encounter with angle at $0^{\circ}$, the inner region of the disk with particles with semimajor axes in the range $30 < a_c < 100$ au gains approximately 8.5\% more particles compared to the original distribution, while the inner Oort cloud region ($200 < a < 10,000$ au) acquires around 6.4\% of the initial particles. Additionally, 20.2\% of the particles become unbound (Fig. \ref{e_ss}). When the inclination angle is set to $30^{\circ}$, the perturbation remains qualitatively similar to the $0^{\circ}$. The trans-Neptunian region (green area in \figref{e_ss}) gains 1.7\% more particles compared to the $0^{\circ}$ encounter, while the inner Oort cloud region is less populated, with 5.2\% of the initial particles acquiring semimajor axes between $200$ and $10,000$ au.

The dynamical effects of the close encounter decreased further for inclination angles of $60^{\circ}$ and $90^{\circ}$. For a $60^{\circ}$, the population in the inner Oort cloud increases by 3\% and 4.2\% compared to the $0^{\circ}$ and $30^{\circ}$ cases, respectively. When the encounter angle is set to $90^{\circ}$, approximately 12\% of the original particles acquire semimajor axes within the Oort cloud region. The trans-Neptunian region experiences a comparable population increase for $60^{\circ}$ (6.8\%) and $90^{\circ}$ (5.2\%) inclination angles.

The geometry of an encounter plays a crucial role in determining the fate of the debris disk, as the impact of stellar encounters is closely correlated with the inclination angle, as shown in Fig. \ref{e_ss} \citep[see also][]{Punzo2014,Pfalzner2018a}. Even with a relatively small impact parameter of $300$ au, the inclination angle of the encounter significantly influences the resulting structure and dynamics of the particles within the system.

These differences are reflected in the distribution of particles across various regions of the system following the encounter. For an inclination of $i=0^{\circ}$, approximately $0.4$\% of the initial particles acquire semimajor axes within the planetary region ($0 < a < 30$ au), whereas for $i=30^{\circ}$, $60^{\circ}$, or $90^{\circ}$, no particles remain in this region. In the trans-Neptunian region, the number of particles increases for $i=30^{\circ}$, while the most effective inclination angle for populating the inner Oort cloud region is $90^{\circ}$. At this inclination, the stellar encounter raises the semimajor axes of $12$\% of the particles to the range between $200$ and $10,000$ au. The structure of the initial disk remains largely consistent across all inclination angles. Particles with semimajor axes between $40$ and $65$ au remain unperturbed, whereas those in the range of $65-200$ au can attain eccentricities approaching $1$.

As several studies have shown \citep[see, e.g.,][]{Jilkova2015, Cai2018, Pfalzner2018a,Veras2020,Pfalzner2024}, a very close encounter between a star and a planetary system can lead to the ejection and potential capture of a significant fraction of particles from a debris disk. Such encounters also form a population of highly elliptical objects, known as  transitional interstellar objects (TIOs) \citep{Torres2019}, which may eventually evolve into ISOs due to the influence of galactic tides and subsequent stellar interactions.

Our simulations indicate that the formation of TIOs and ISOs is strongly dependent on the inclination angle of the stellar encounter, as we can see in \figref{e_ss}. The production of ISOs changes with the encounter angle: for $0^{\circ}$, approximately 20.2\%, while for $30^{\circ}$ and $60^{\circ}$, the fractions decrease to $18.4$\% and $9.4$\%, respectively. Interestingly, for an encounter angle of $90^{\circ}$, none of the particles became unbound from the system despite the close encounter.

\subsubsection{\extn}
\label{Nenc_ss}

Following the methodology described in Sect. \ref{sec2.1} and using the initial conditions outlined in Table \ref{table1} for the \extn\ model, we conducted a set of 200 simulations. Each simulation accounts for the effect of the planets and the five closest stars in the cluster, as described in Sect. \ref{sec2.2}, on the SSA's debris disk at any given time over a total integration period of $100$\ Myr.

\begin{figure}[ht!]
  \includegraphics[width=1.\hsize]{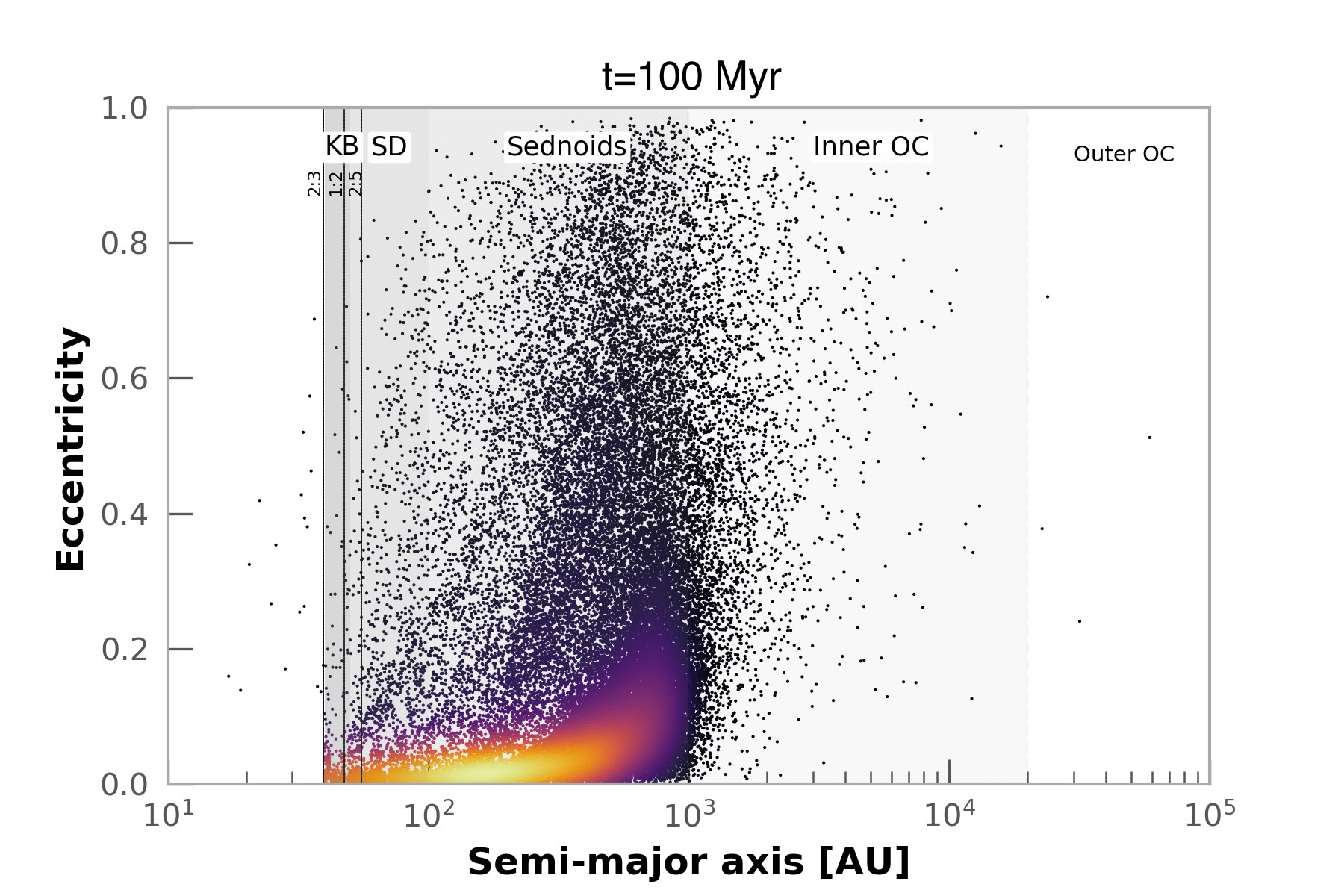}
  \includegraphics[width=1.\hsize]{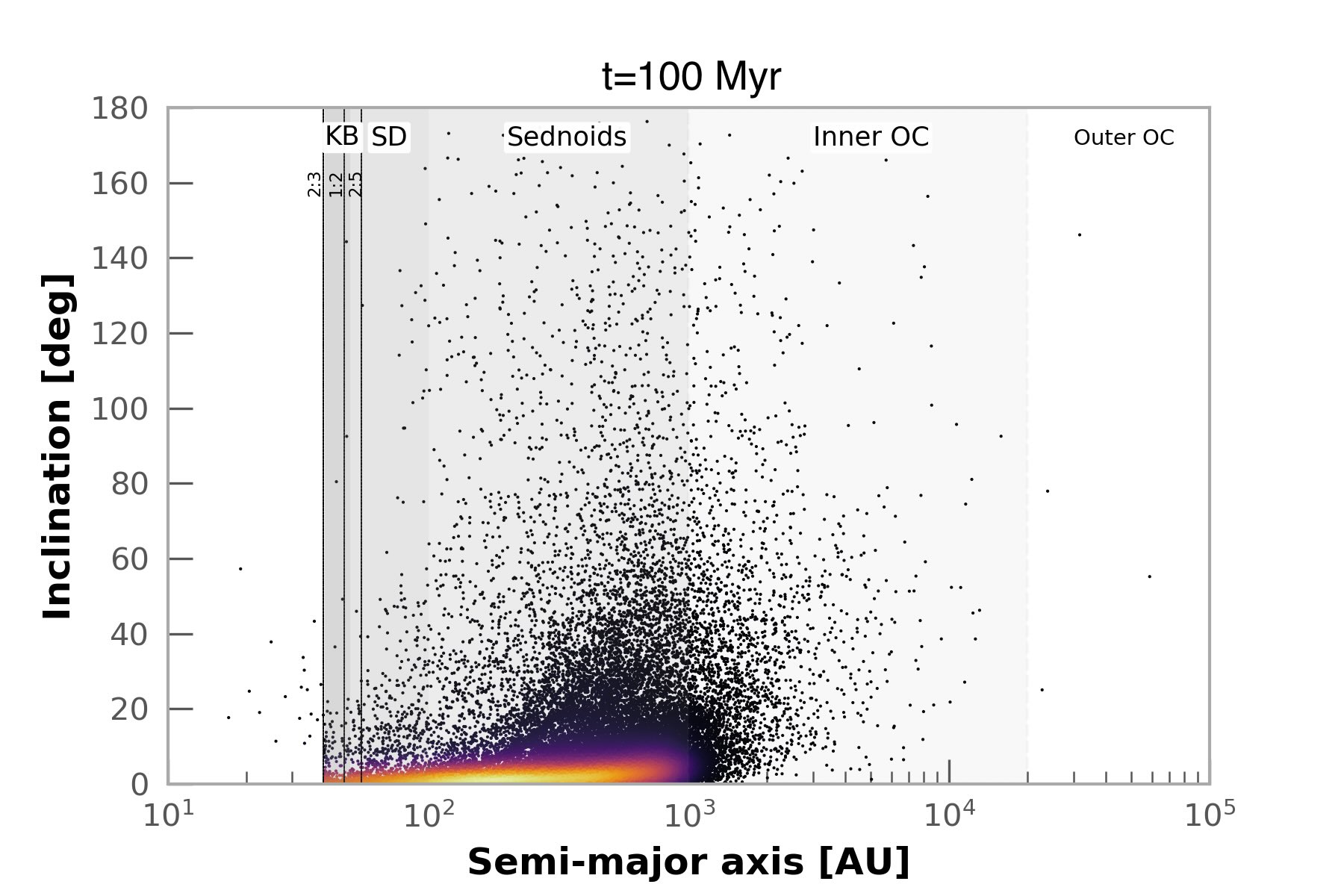}\\
  \caption{\extn\ model. Semimajor axis as a function of eccentricity (top panel) and orbital inclination (bottom panel) for all the particles in the 200 simulated systems. The gray areas represent the different regions of the Solar System: Kuiper belt (KB), scattered disk (SD), Sednoids, inner and outer Oort cloud (OC); the dashed lines show the 2:3, 1:2, and 2:5 resonances with Neptune. The integration time is set to 100~Myr. An animation is available on the \href{https://www.aanda.org/articles/aa/olm/2026/04/aa54978-25/aa54978-25.html}{A\&A website}.} \label{all_eai_p}
\end{figure}

Figure \ref{all_eai_p} shows the final distribution of the semimajor axis as a function of eccentricity and inclination for all particles across the $200$ simulated systems. The figure highlights the emergence of structures reminiscent of the outer Solar System. Particles within the range of $30$ to $40$ au become trapped in mean-motion resonances with Neptune (2:3, 3:5, 4:7, 1:2, and 2:5), while particles between approximately $40$ and $100$ au attain eccentricities of up to $0.4$ and inclinations reaching $20^{\circ}$. These distributions closely resemble the orbital characteristics of objects in the Kuiper belt and the scattered disk of the Solar System.

The inner region of the disk in all simulations experiences sufficient perturbations to develop substructures. In the outer disk, approximately $7.5$\% of the particles are ejected, significantly contributing to the population of the inner Oort cloud ($1000 < a < 20,000$~au). About $18$\% of the particles acquire highly eccentric orbits ($e > 0.8$); a fraction of these particles develop planet-crossing perihelia, potentially undergoing additional scattering events that place them on wider orbits with even higher eccentricities. Only a very small fraction ($\sim$0.02\%) reaches the outermost regions consistent with the current outer Oort cloud ($a > 20,000$~au). Furthermore, a substantial fraction of particles ($\sim$36\%) is ejected from their systems due to cumulative gravitational interactions with passing stars, thereby contributing significantly to the population of ISOs within the cluster environment.

\begin{figure*} 
  \includegraphics[width=0.5\textwidth]{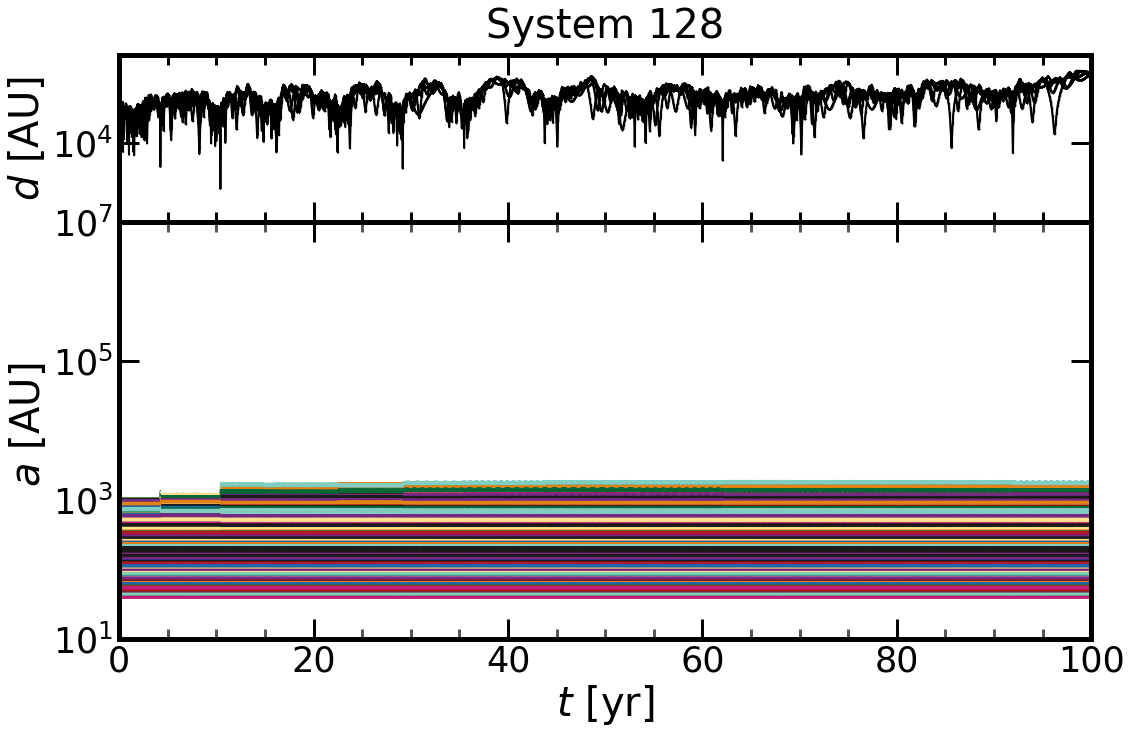}
  \includegraphics[width=0.5\textwidth]{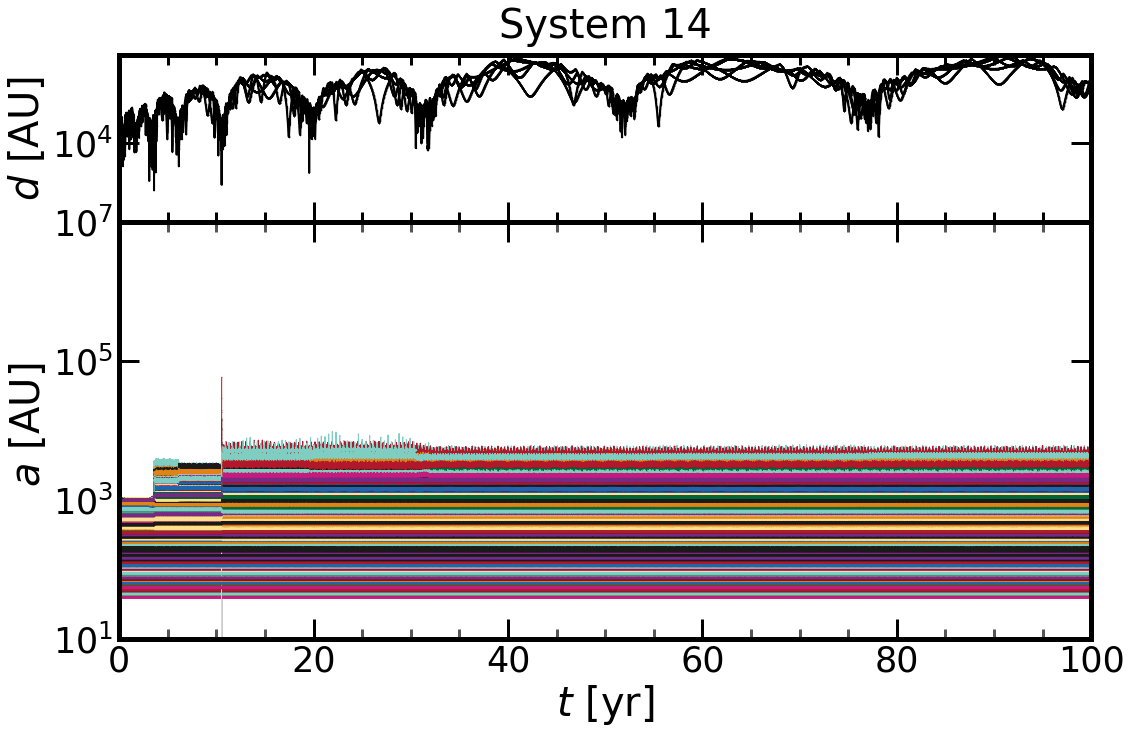}
  \includegraphics[width=0.5\textwidth]{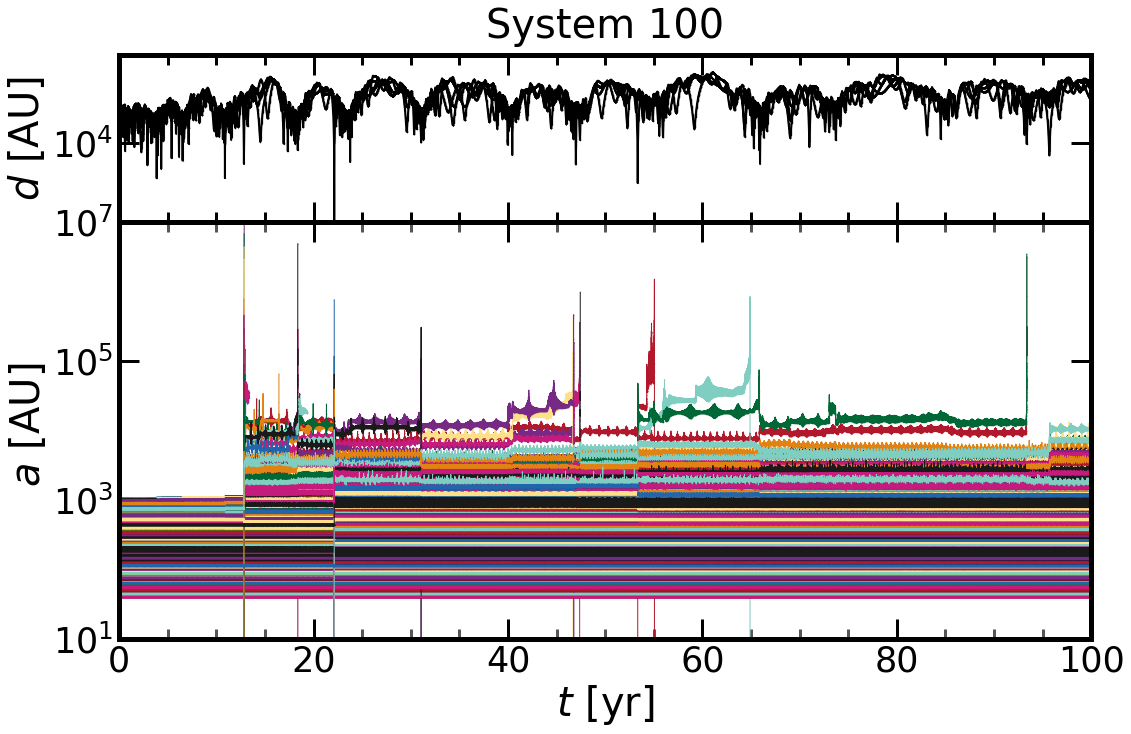}
  \includegraphics[width=0.5\textwidth]{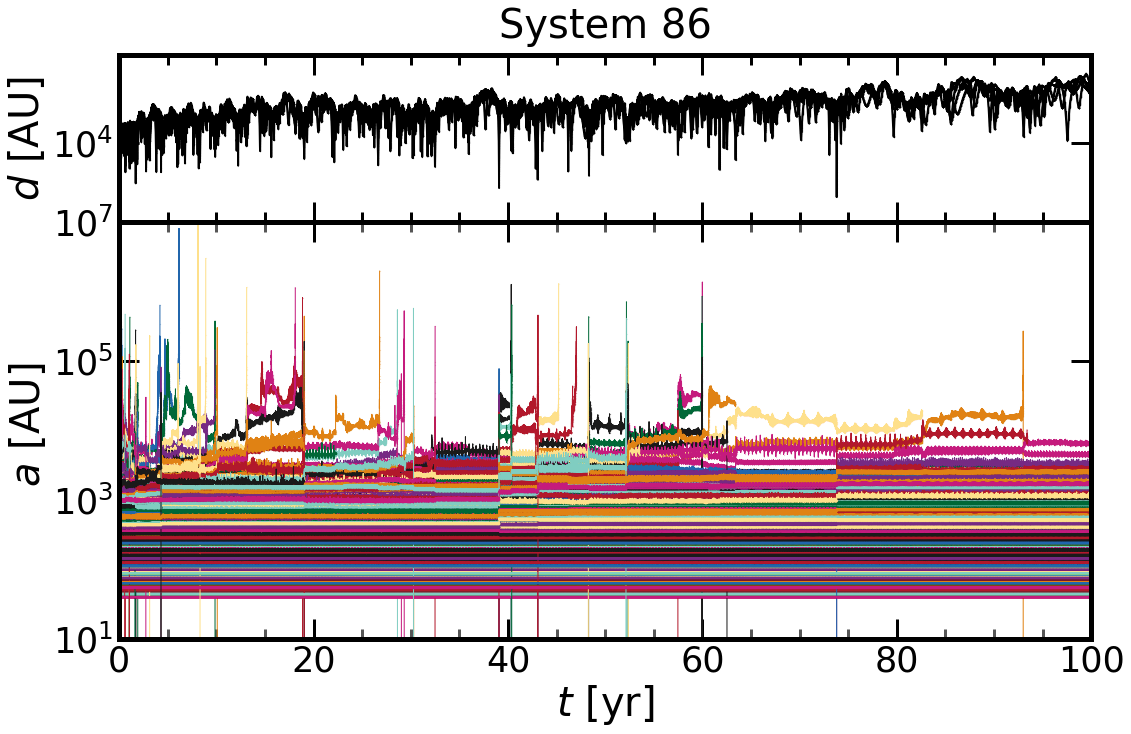}
  \includegraphics[width=0.5\textwidth]{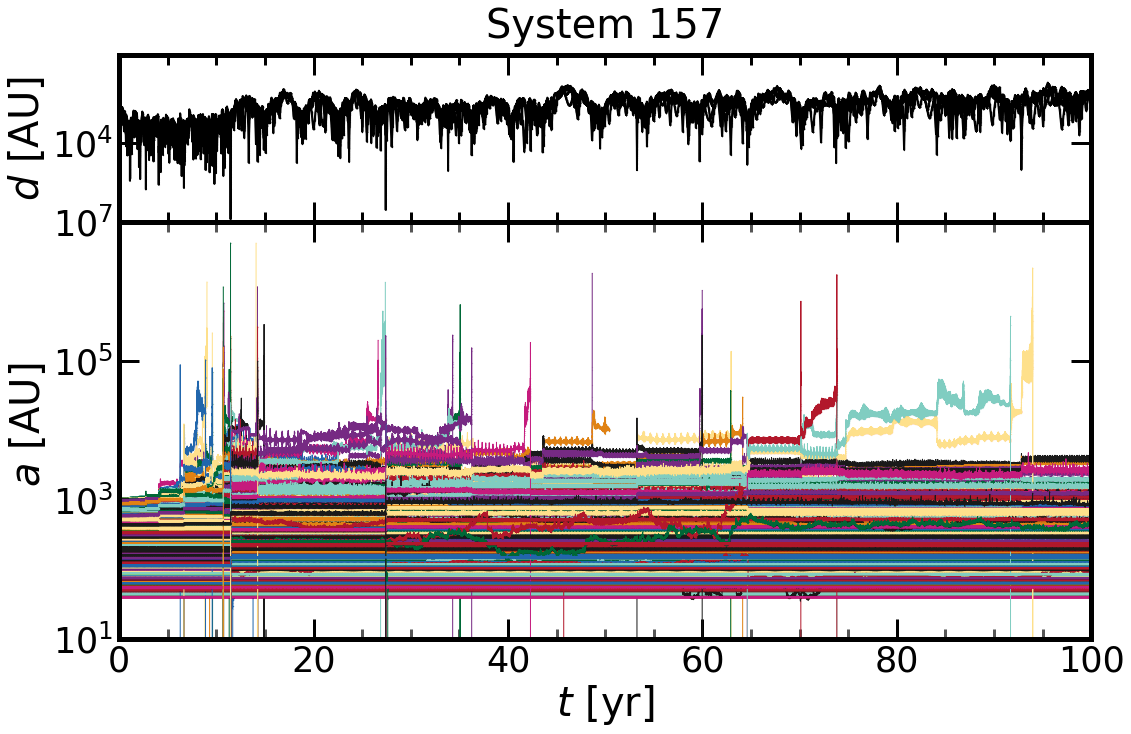}
  \caption{\extn\ model. Orbital evolution of particles in the disk over 100 Myr. The bottom panels in each plot depict the semimajor axis as a function of time, while the top panels show the distance of the perturber. The panels represent systems 128, 14, 100, 86, and 157 (as labeled). The colored lines correspond to individual particles within each system.}
  \label{tad_ss}
\end{figure*}

\begin{figure*}
  \includegraphics[width=0.34\textwidth]{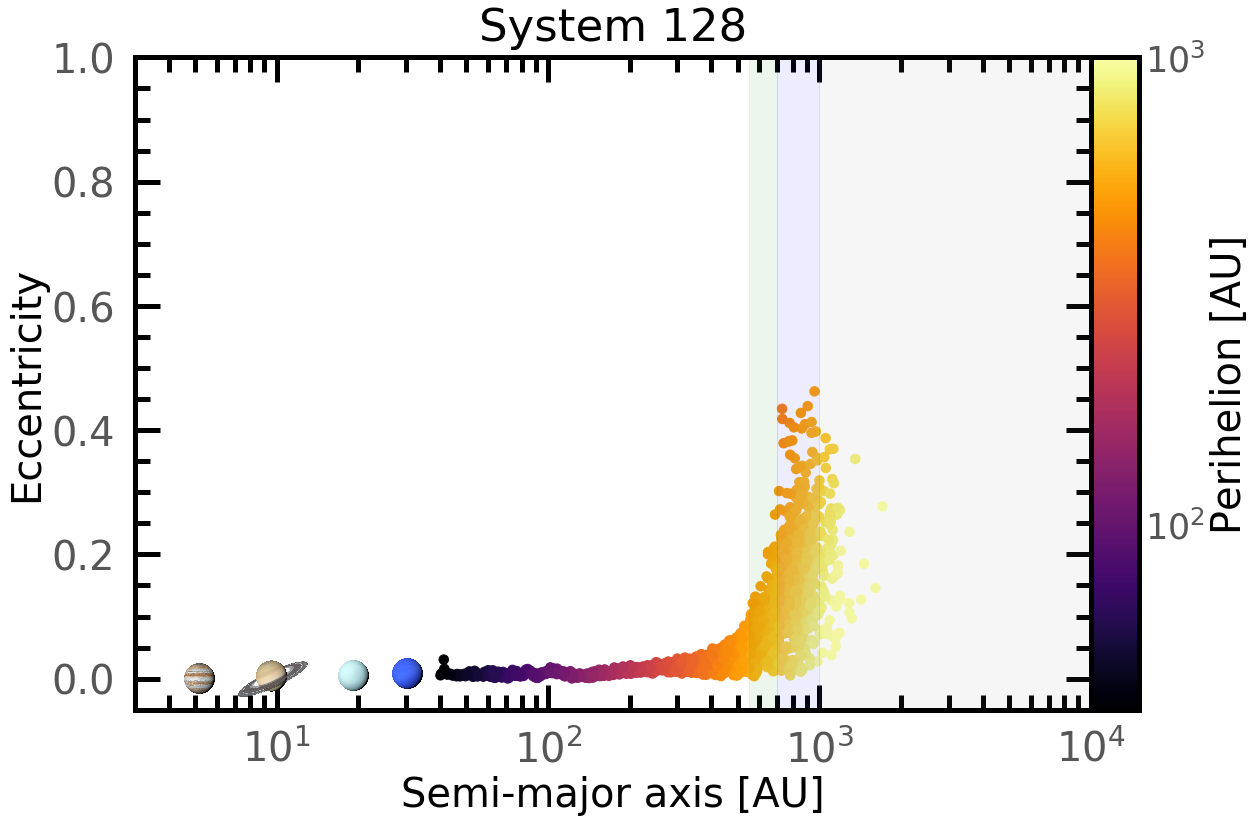}
  \includegraphics[width=0.34\textwidth]{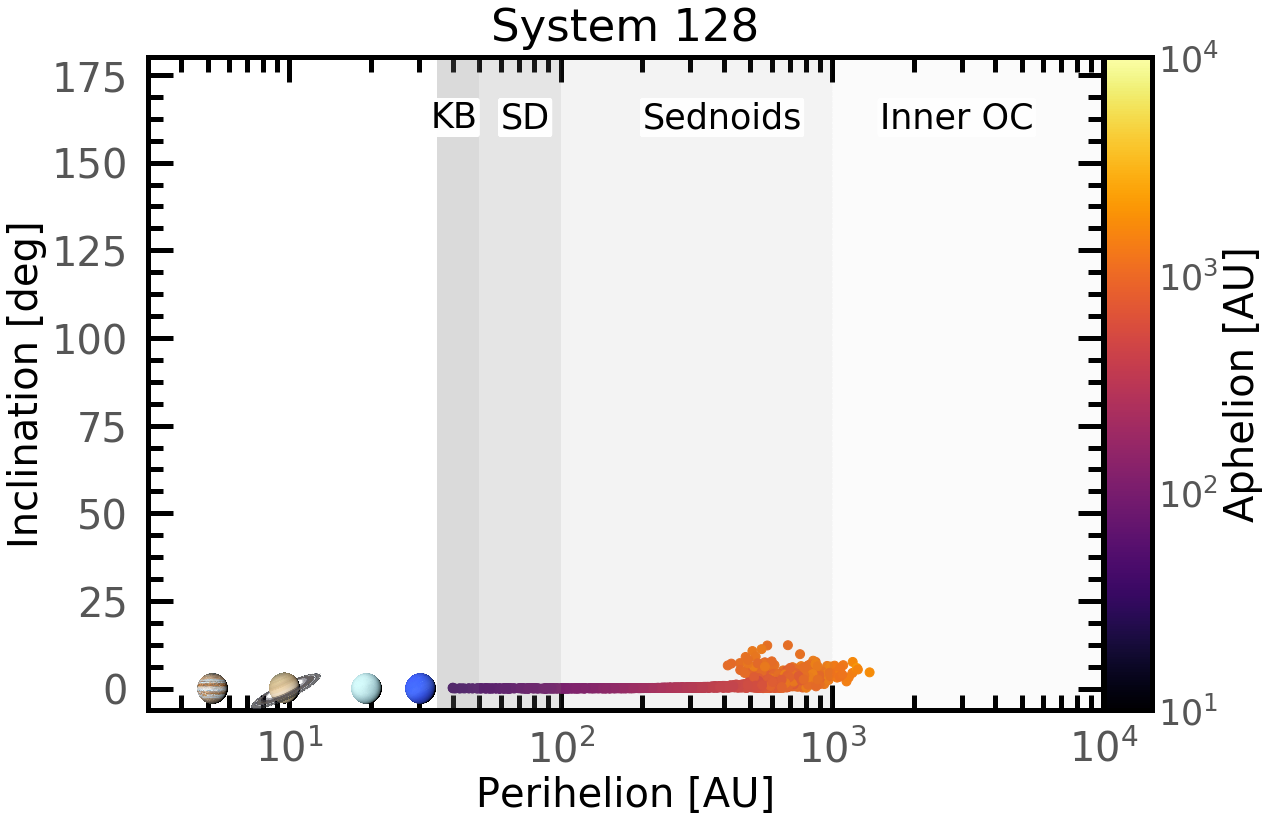}
  \includegraphics[width=0.34\textwidth]{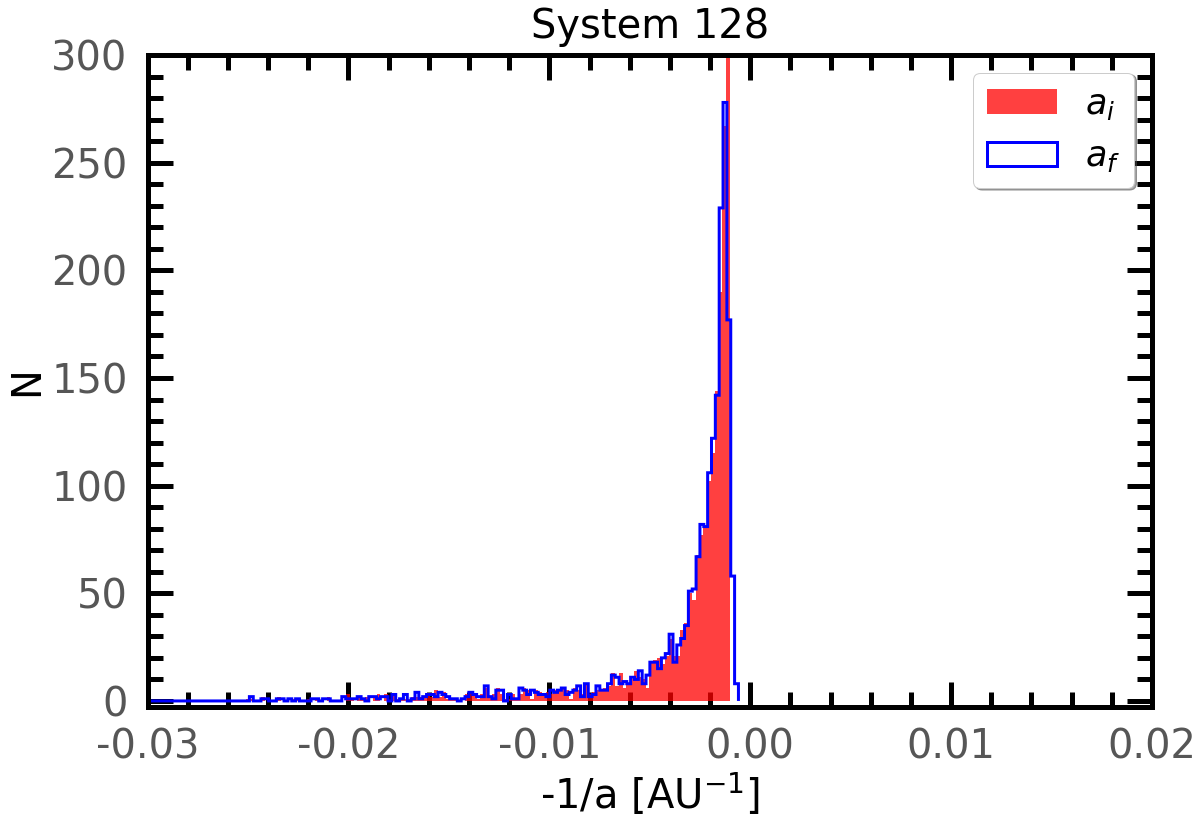}\\
  \includegraphics[width=0.34\textwidth,trim=0 0 0 0,clip]{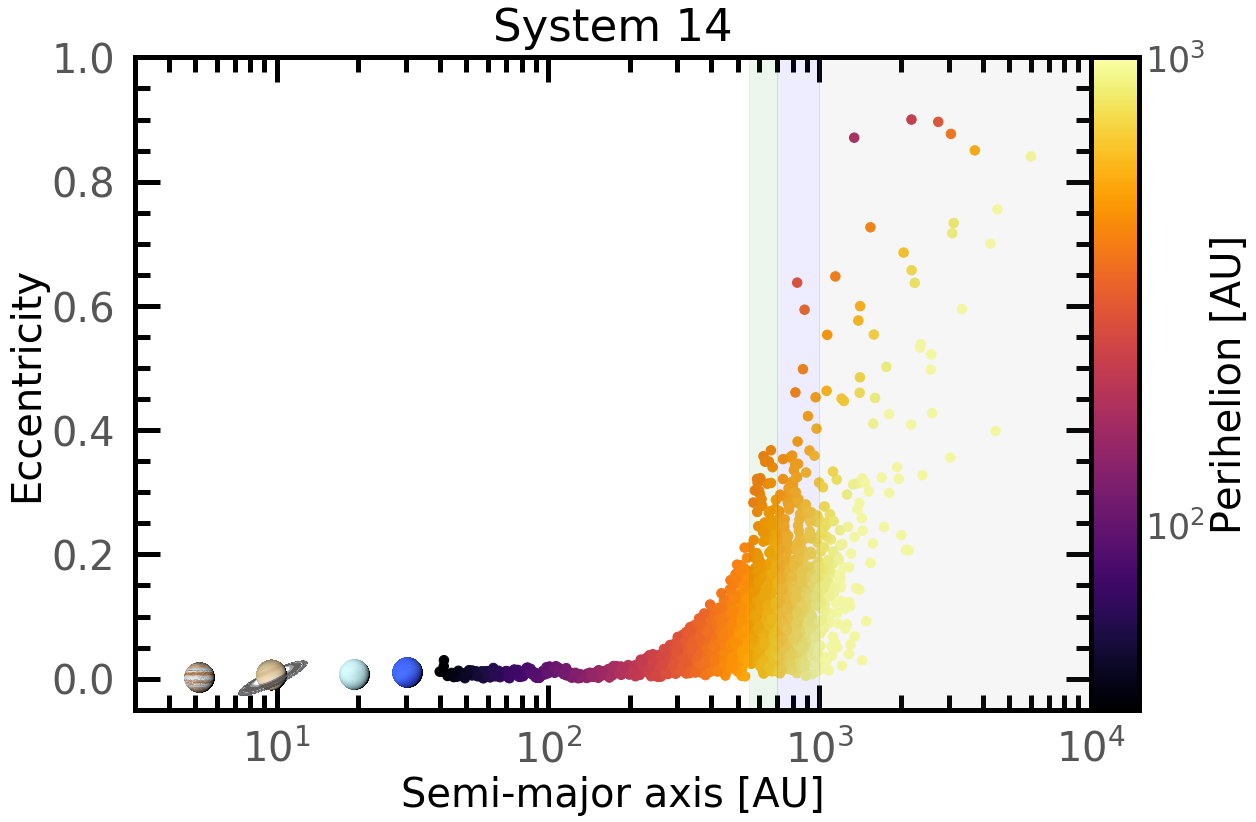}
  \includegraphics[width=0.34\textwidth,trim=0 0 0 0,clip]{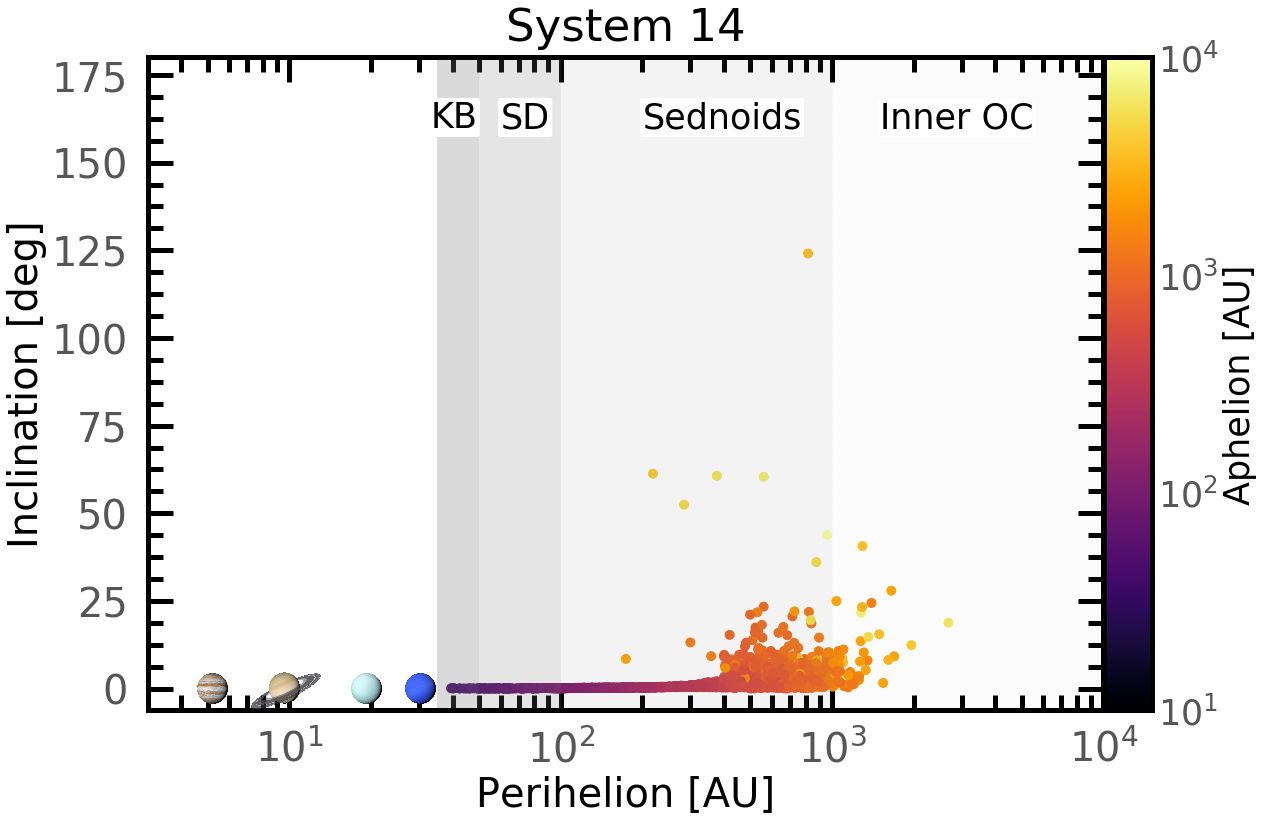}
  \includegraphics[width=0.34\textwidth,trim=0 0 3 0,clip]{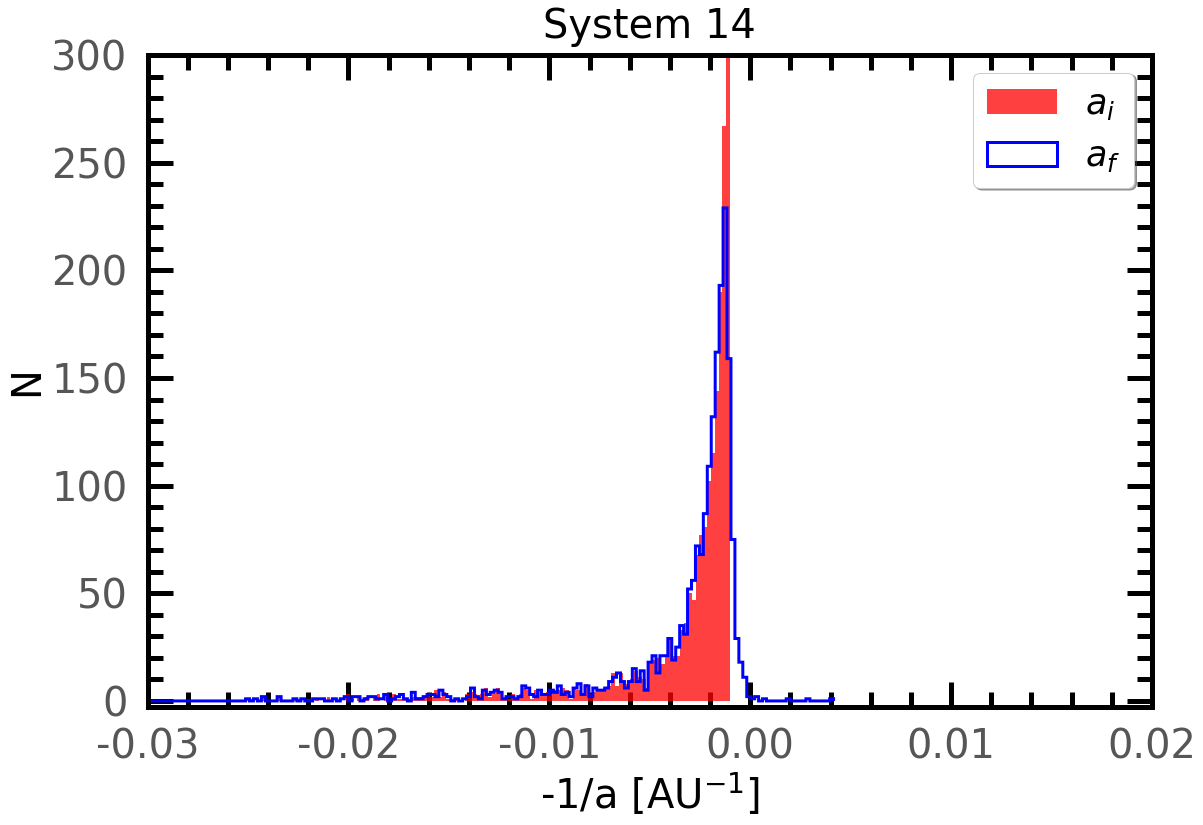}\\
  \includegraphics[width=0.34\textwidth,trim=0 0 0 0,clip]{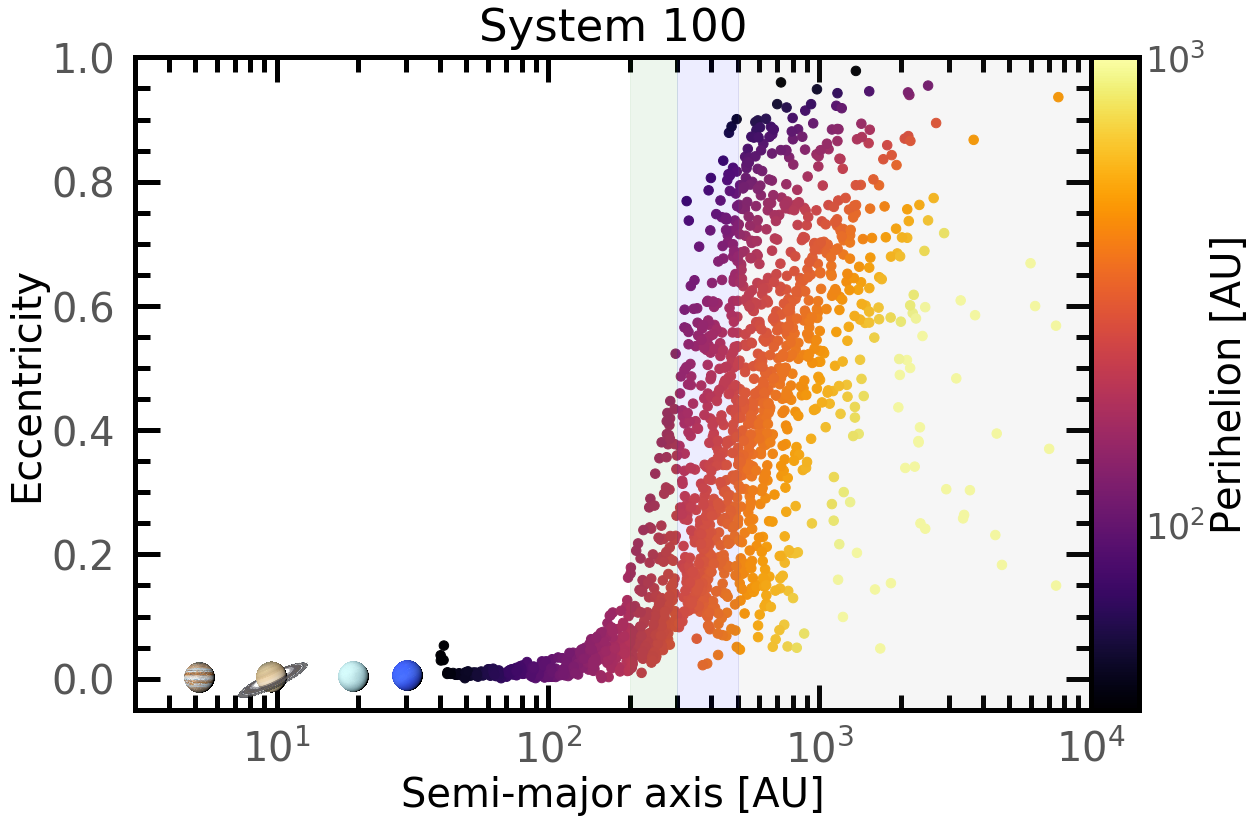}
  \includegraphics[width=0.34\textwidth,trim=0 0 0 0,clip]{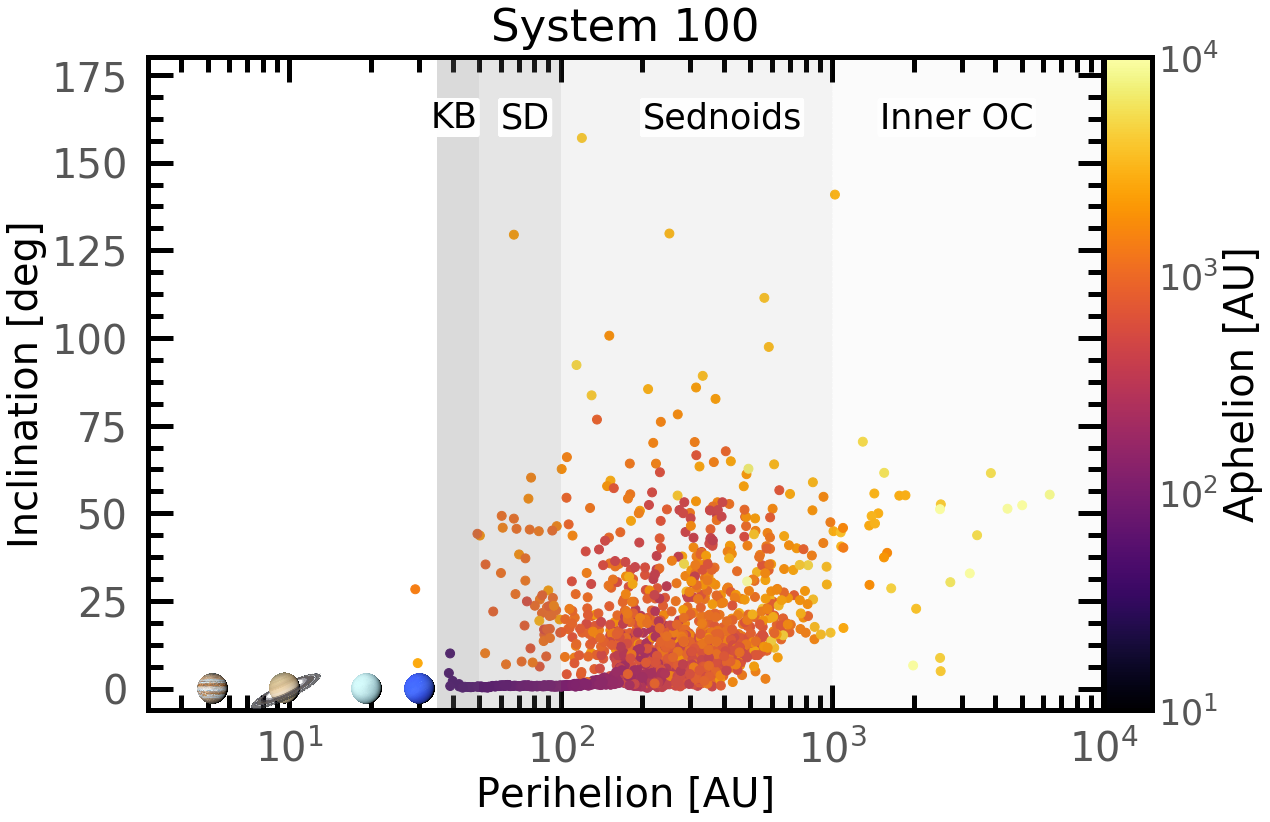}
  \includegraphics[width=0.34\textwidth,trim=0 0 2 0,clip]{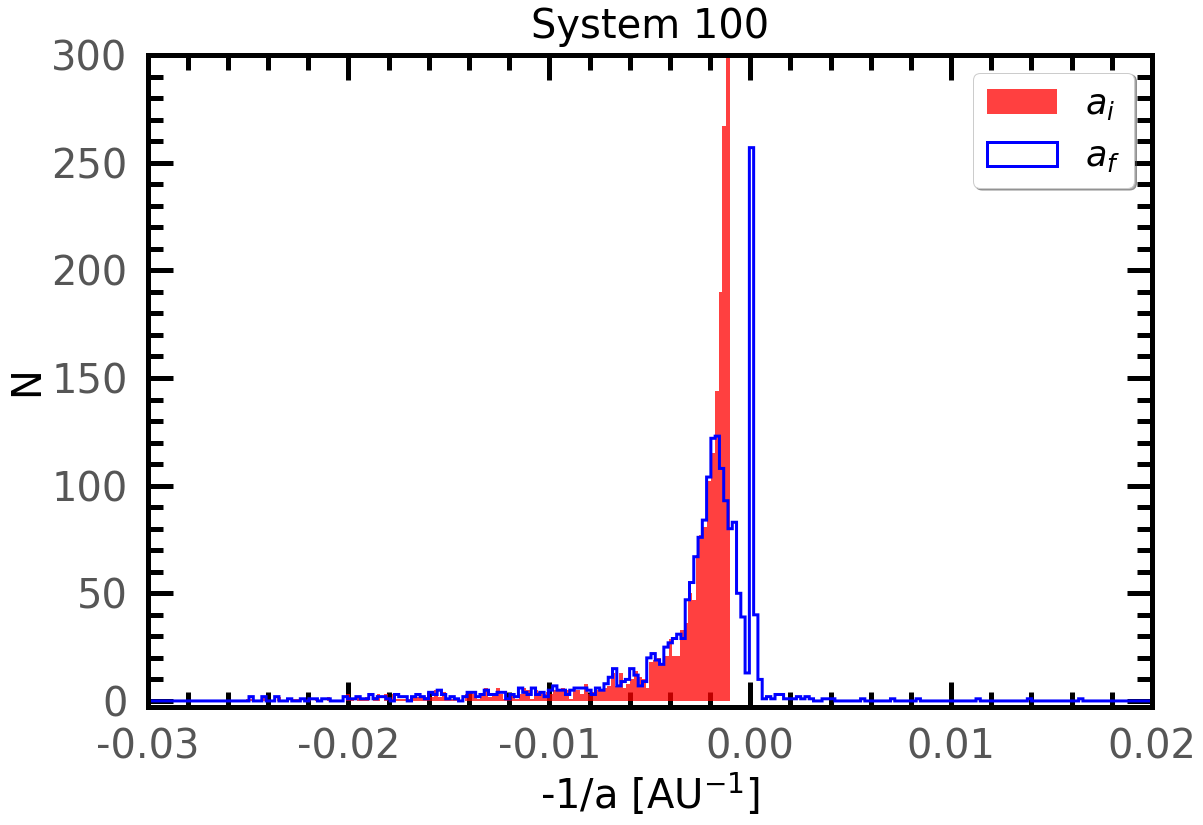}\\
  \includegraphics[width=0.34\textwidth,trim=0 0 0 0,clip]{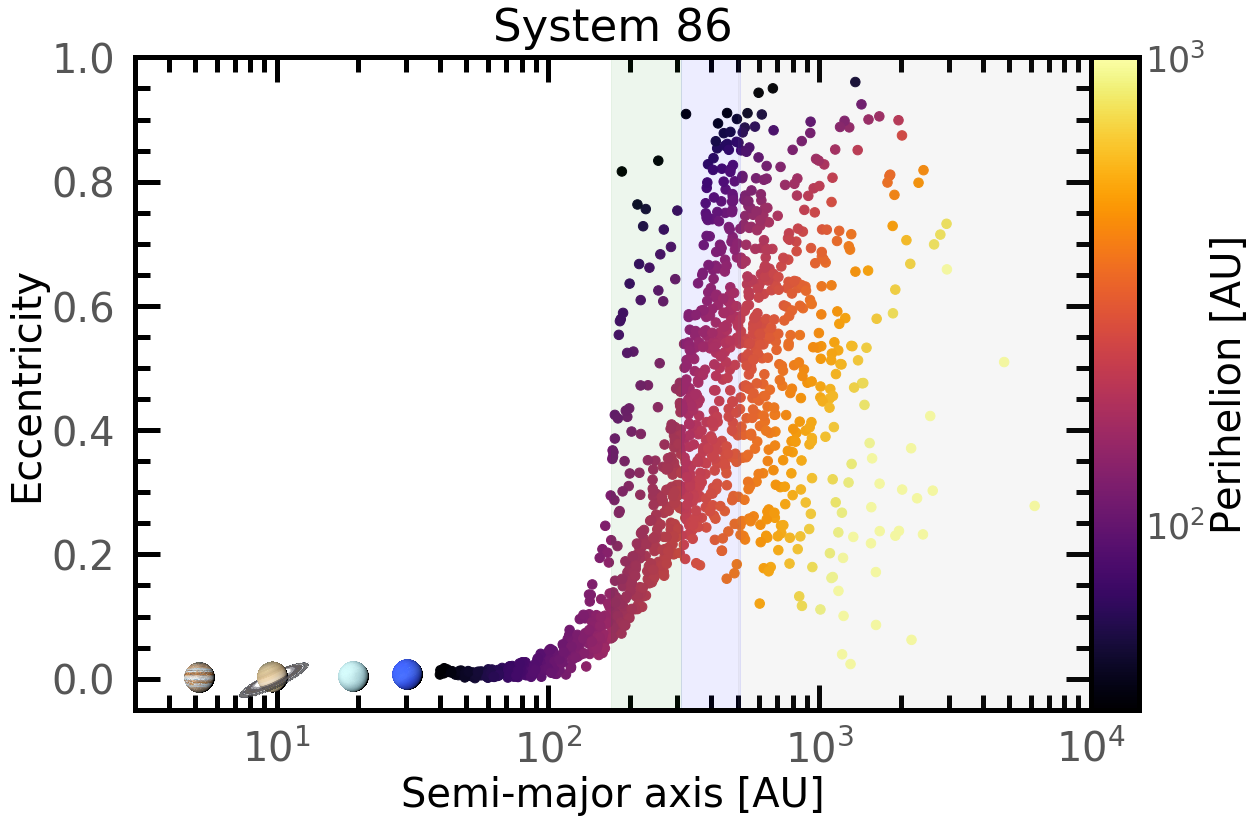}
  \includegraphics[width=0.34\textwidth,trim=0 0 0 0,clip]{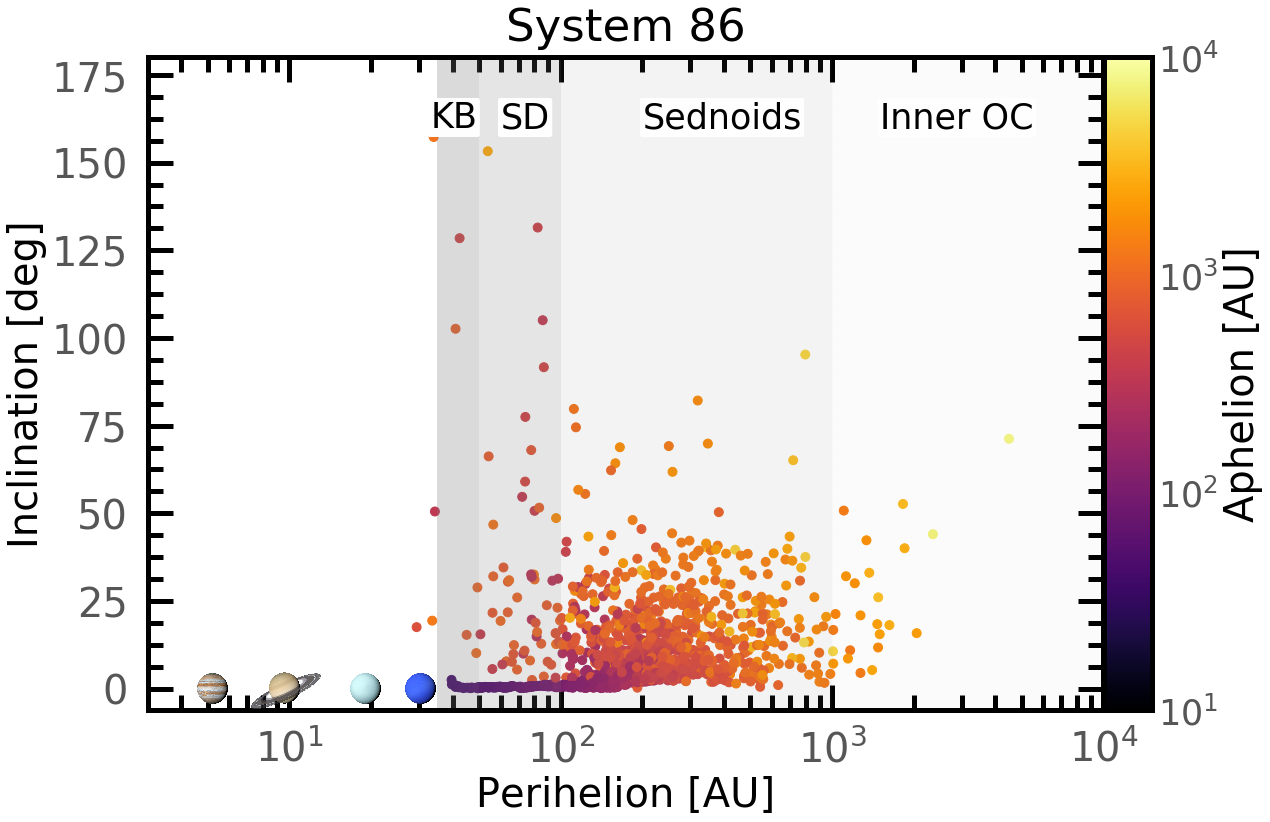}
  \includegraphics[width=0.34\textwidth,trim=0 0 1 0,clip]{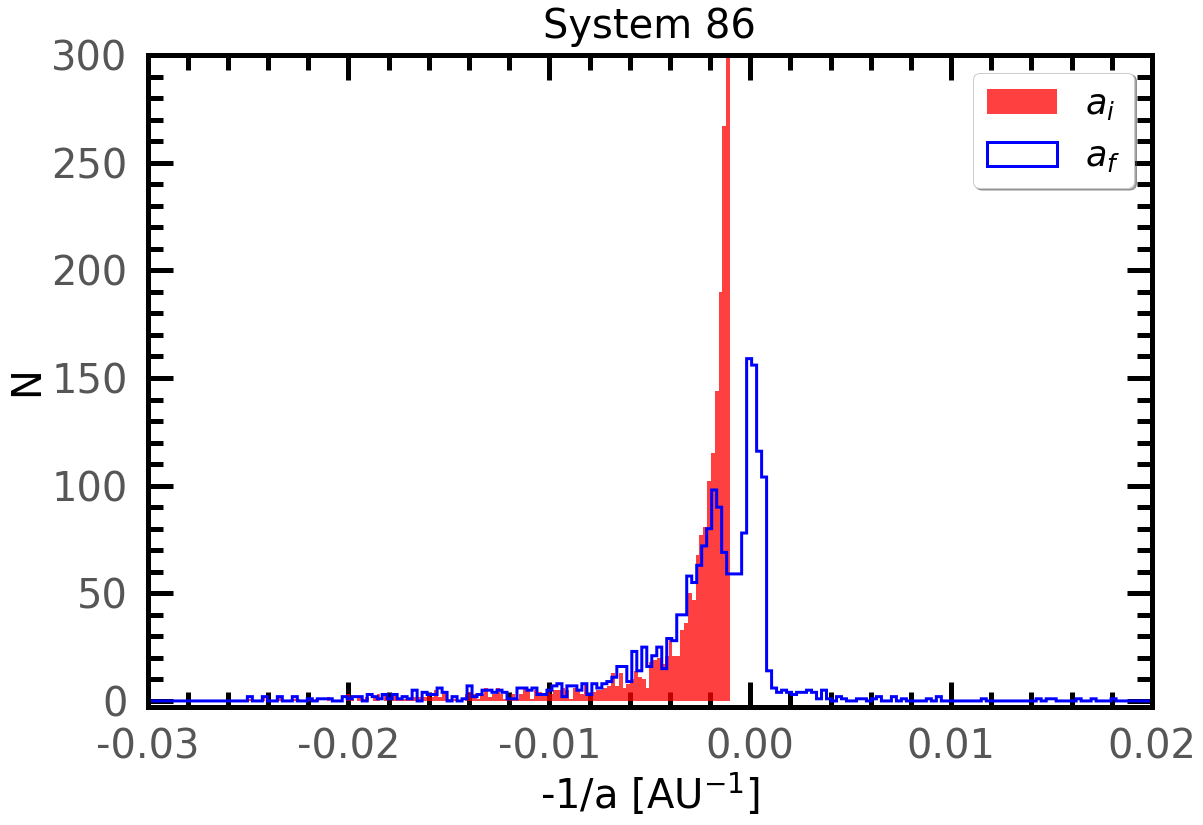}\\
  \includegraphics[width=0.34\textwidth,trim=0 0 0 0,clip]{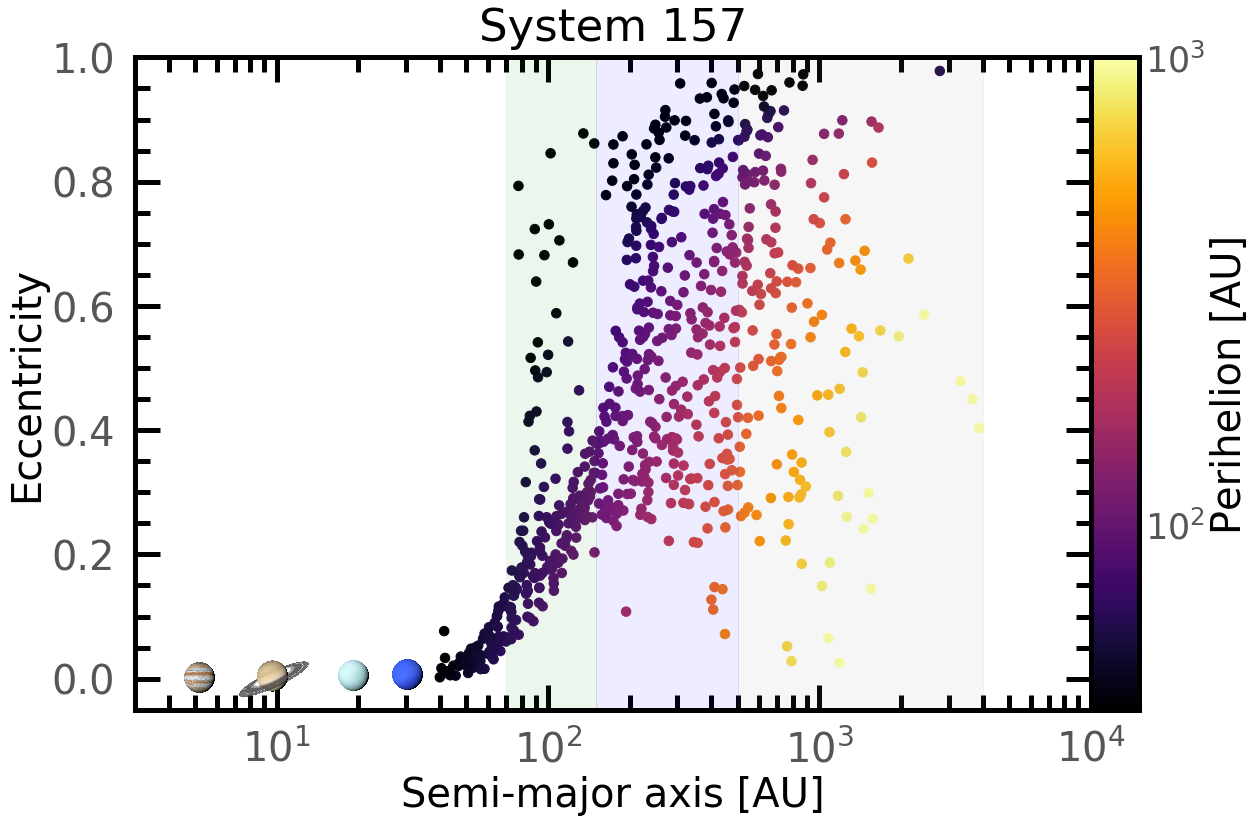}
  \includegraphics[width=0.34\textwidth,trim=0 0 0 0,clip]{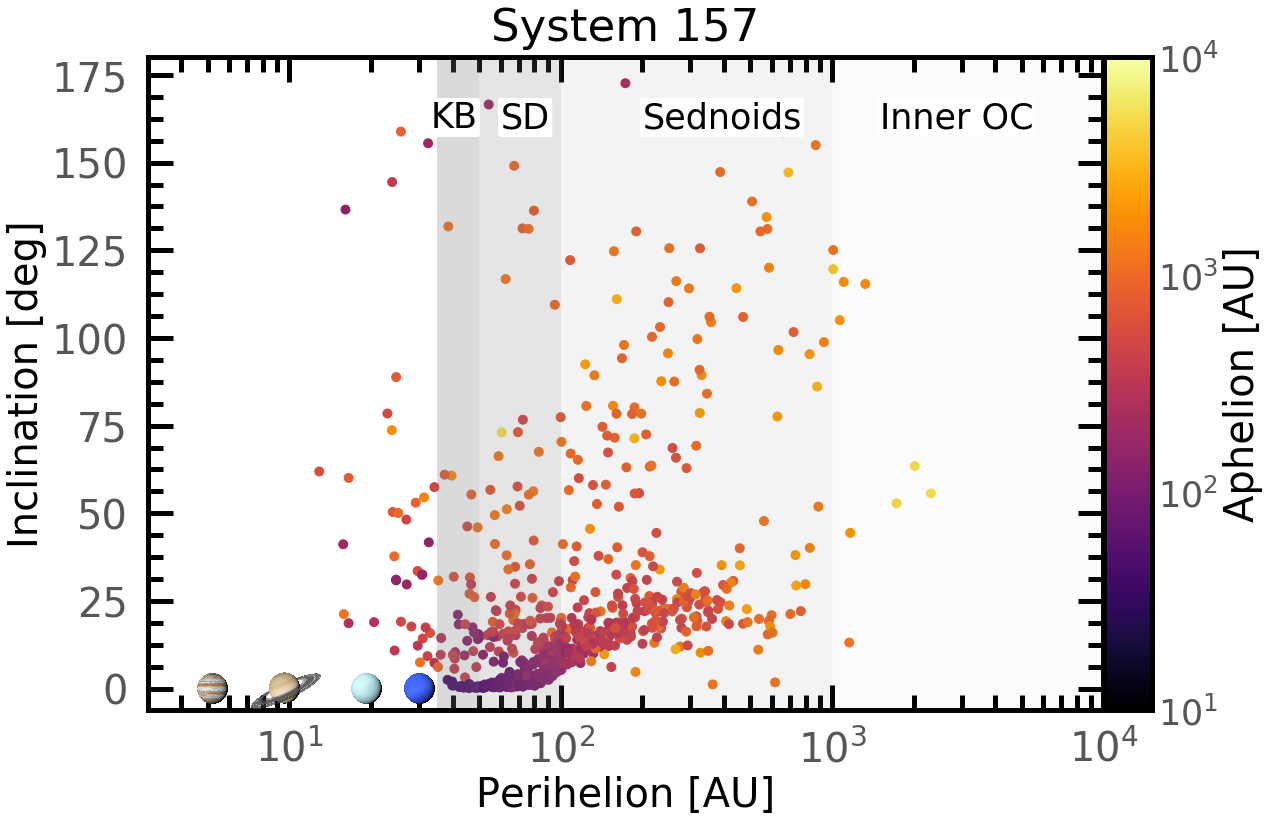}
  \includegraphics[width=0.34\textwidth,trim=0 0 0 0,clip]{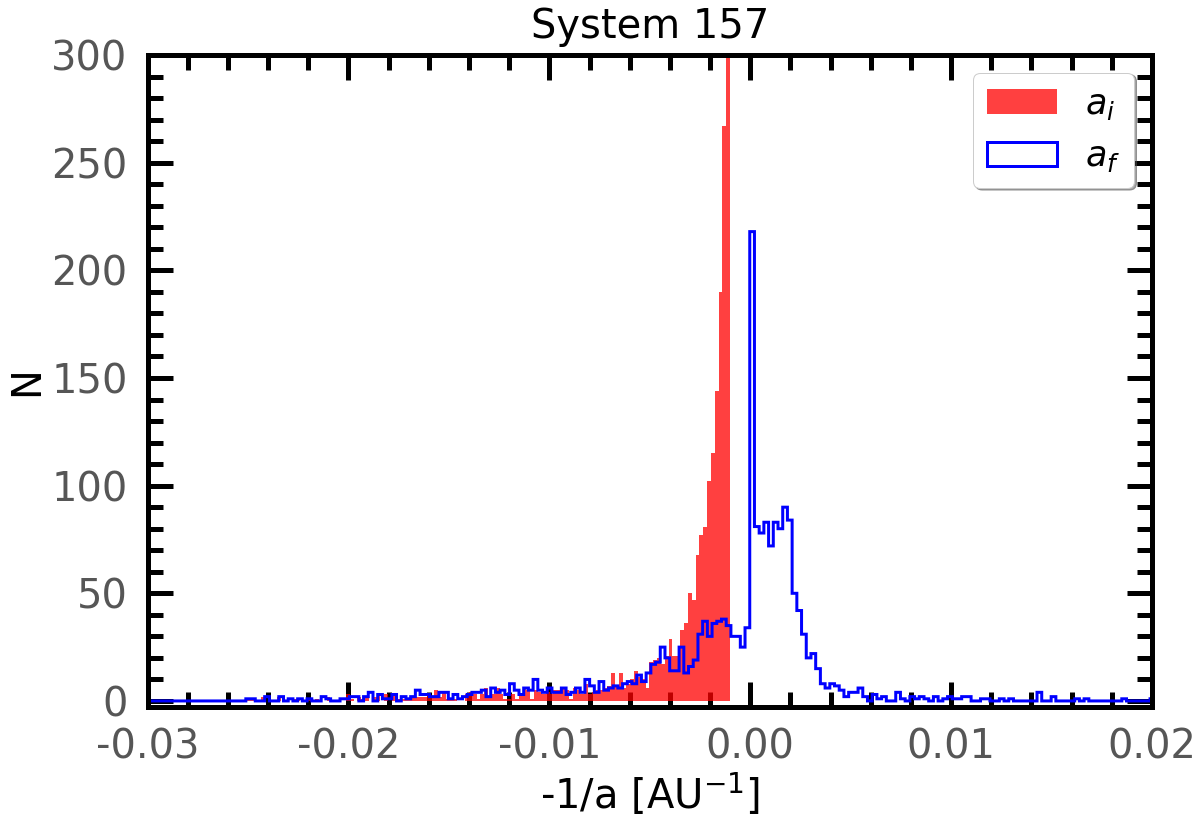}
  \caption{\extn\ model. Orbital elements of particles in the disk after 100 Myr and multiple stellar encounters. First column: Semimajor axis as a function of eccentricity, with particles color-coded by their perihelion distances. The green, blue, and gray shaded regions highlight different populations formed due to stellar encounters. Second column: Perihelion as a function of orbital inclination, with dots color-coded by aphelion distance. The gray shaded areas represent the distinct regions of the Solar System:  Kuiper Belt (KB), scattered disk (SD), Sednoids, and the inner Oort Cloud (OC). Third column:\ Distribution of orbital energy for the particles. The red histograms correspond to the initial energy distribution, and the blue curve represents the final energy distribution. The blue curves with positive values indicate interstellar comets. Each row corresponds to systems numbered 128, 14, 100, 86, and 157, respectively.}
  \label{aeiq_ss} 
\end{figure*}

\begin{figure*} 
  \includegraphics[width=0.34\textwidth]{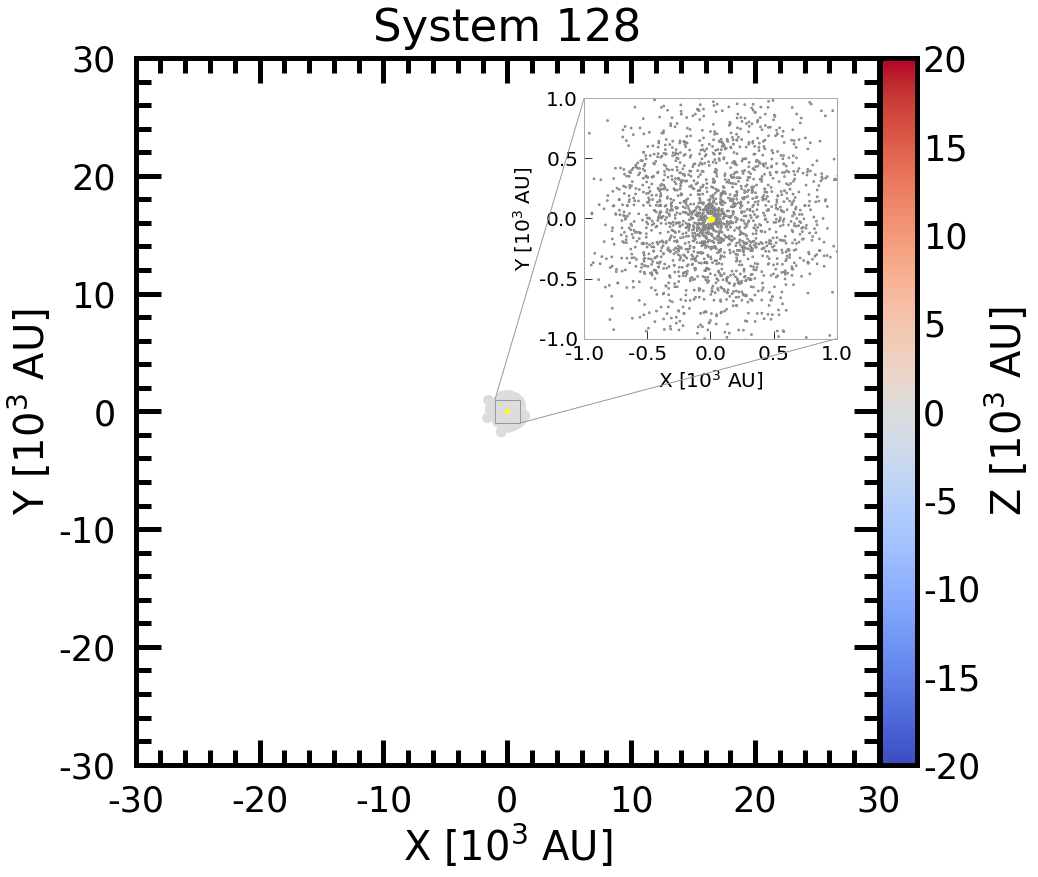}
  \includegraphics[width=0.34\textwidth,trim=0 0 3 0,clip]{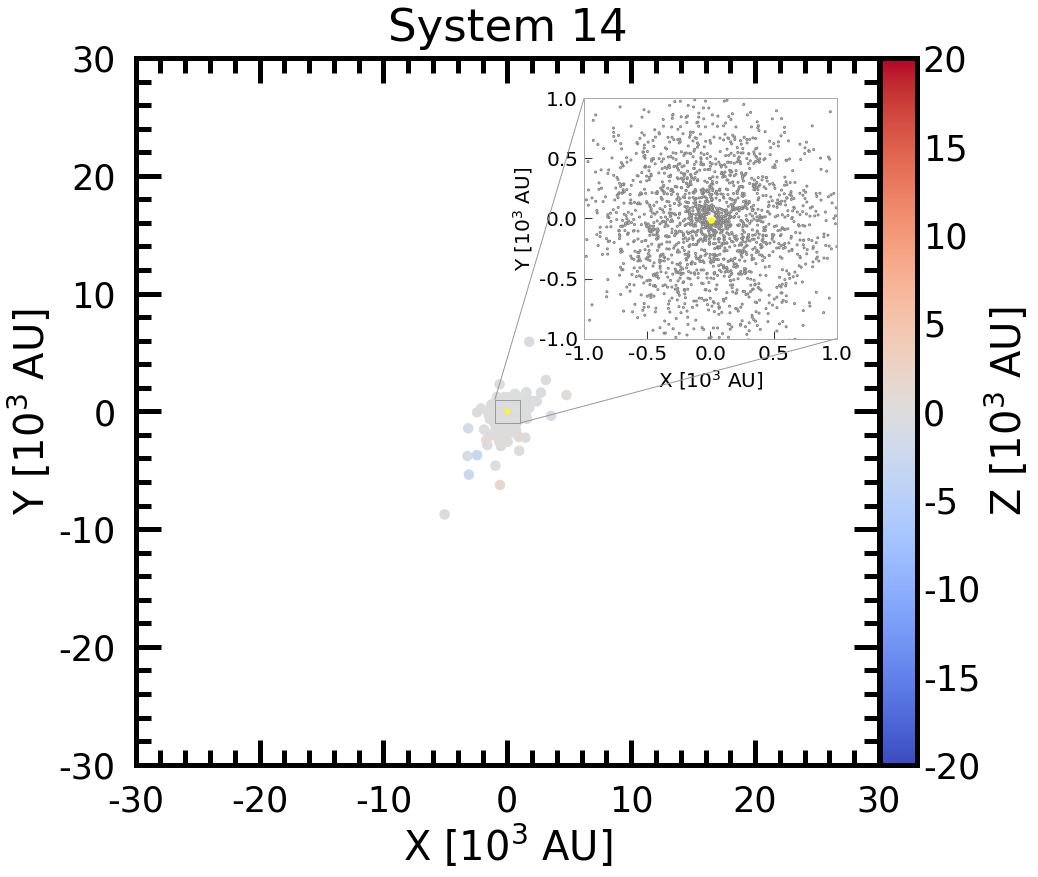}
  \includegraphics[width=0.34\textwidth,trim=0 0 2 0,clip]{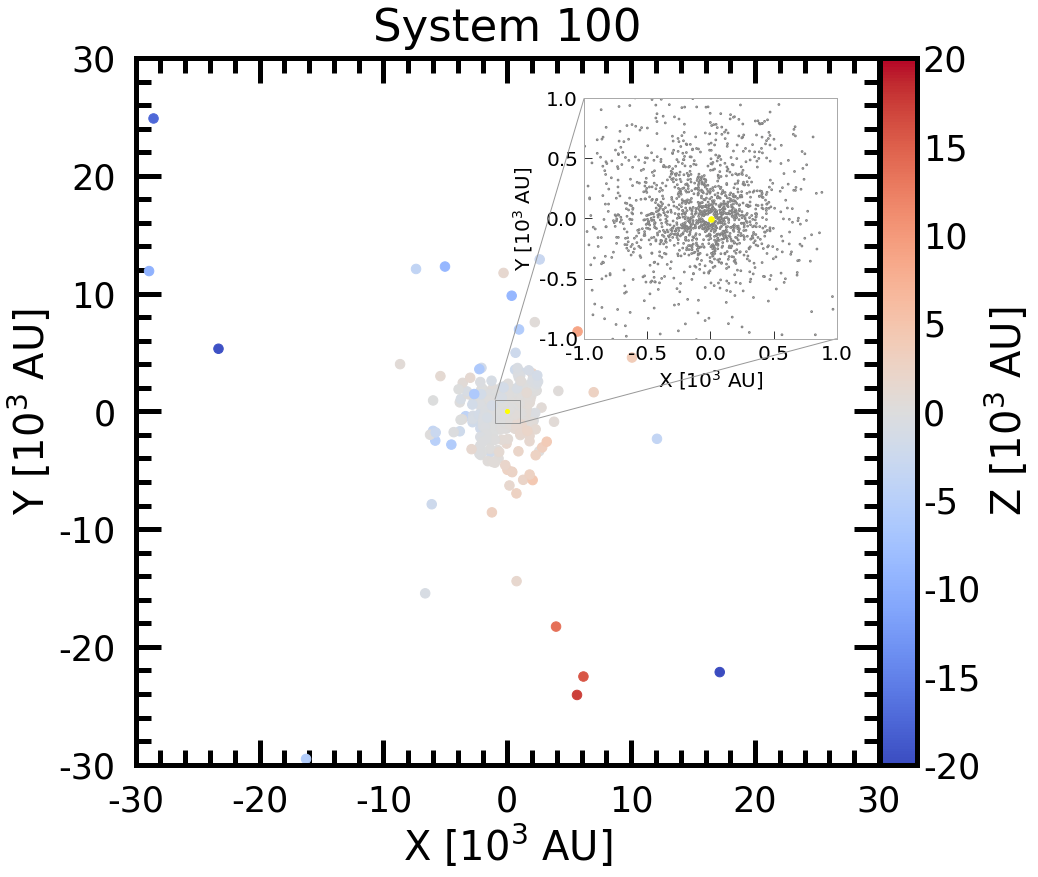}\\
  \includegraphics[width=0.34\textwidth,trim=0 0 1 0,clip]{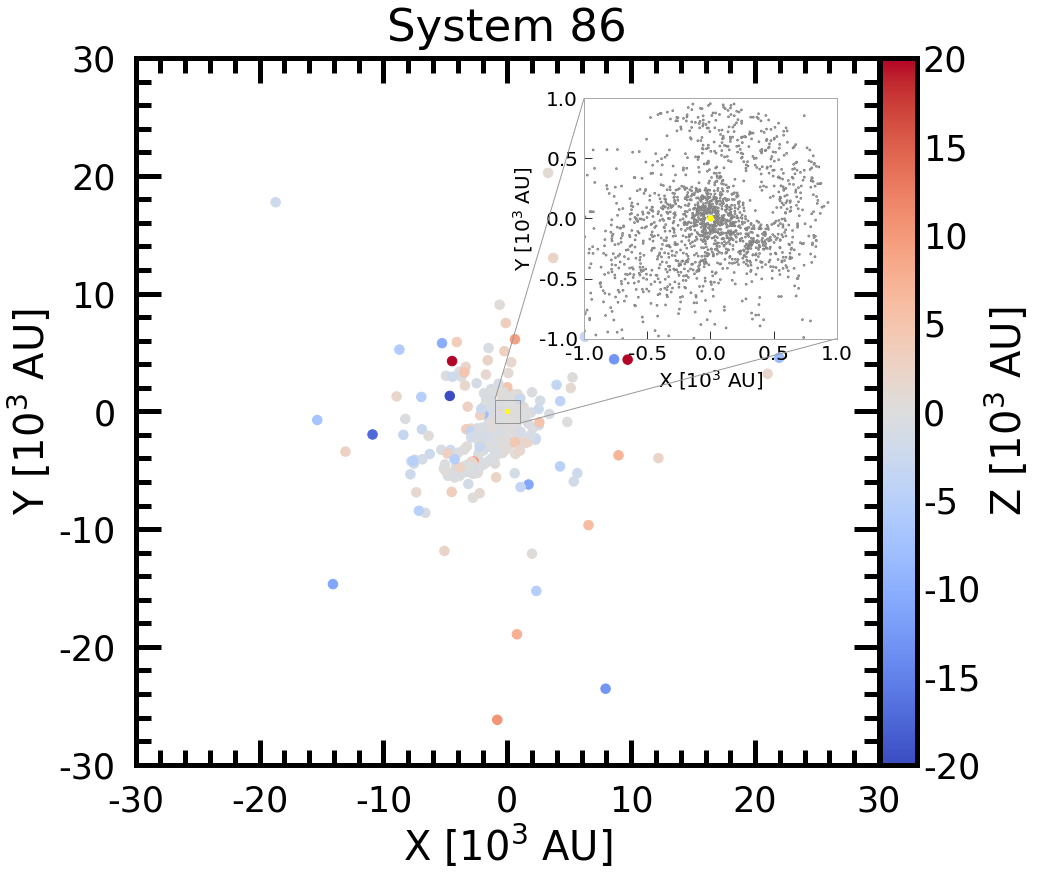}
  \includegraphics[width=0.34\textwidth,trim=0 0 0 0,clip]{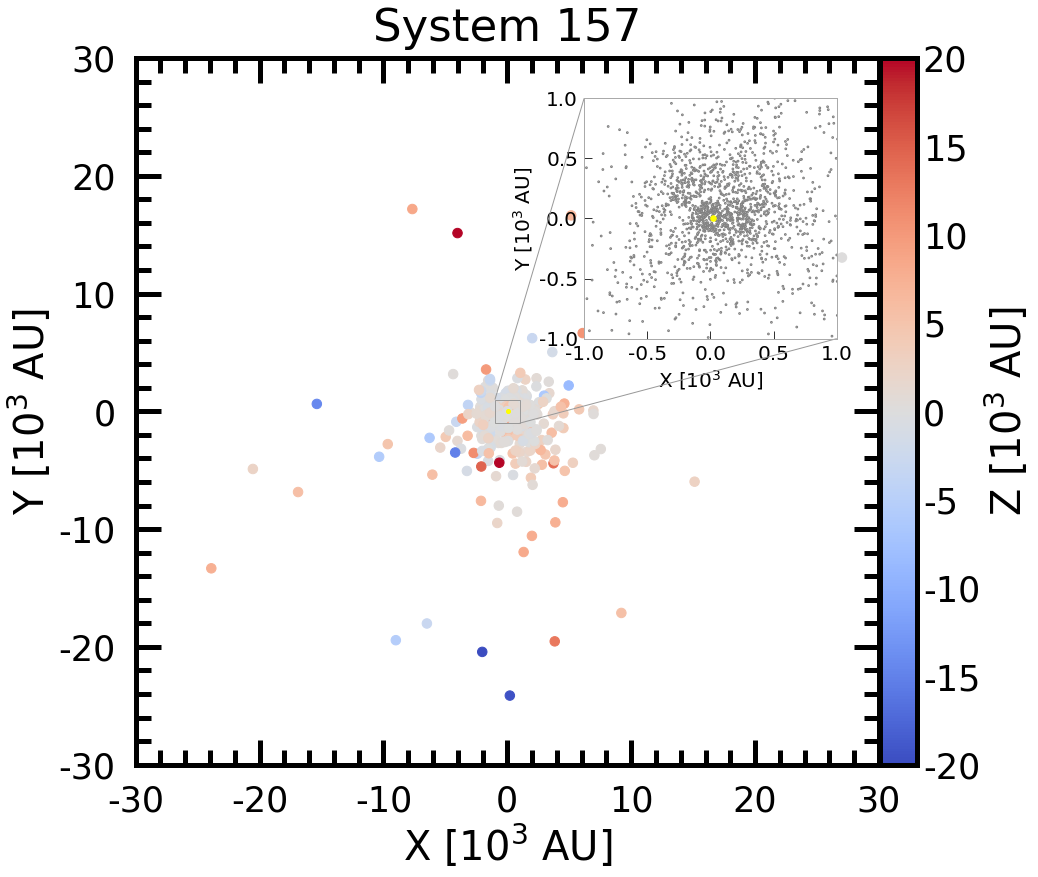}
  \caption{\extn\ model. Phase-space distribution ($X$ vs.\ $Y$) of particles in the disk after $100$~Myr. The color scale indicates the particles' positions along the vertical ($Z$) axis. The panels (from top to bottom) show systems 128, 14, 100, 86, and 157.}
  \label{pos_ss}
\end{figure*}

We now examine in detail a selection of individual systems as shown in \figrefthree{tad_ss}{aeiq_ss}{pos_ss}. From the set of 200 simulations, we selected five representative systems (128, 14, 100, 86, and 157) that experienced a range of stellar encounter intensities, characterized as weak (systems 128 and 14), moderate (system 100), and strong (systems 86 and 157). In Fig. \ref{tad_ss}, we present the evolution of the semimajor axis as a function of time, alongside the stellar encounter distances, highlighting the closest stellar approaches (top panels of Fig. \ref{tad_ss}) for each of the five selected systems. In Fig. \ref{aeiq_ss}, we show the final distributions after $100$\ Myr of the semimajor axis, eccentricity, inclination, perihelion, and aphelion distances, along with their respective particle energy distributions. Finally, in Fig. \ref{pos_ss} we present the resulting phase-space distributions of the particles, highlighting the structures formed within the disk due to stellar encounters.

The most stable system (weakly perturbed) is no.\ 128. All the particles remain in the disk after multiple encounters. The five closest encounters over $100$~Myr range from $1000$ to $5,000$ au (first panel, \figref{tad_ss}) are not close enough to inflict a considerable perturbation. However, the disk is heated somewhat in its outer parts, with a small fraction of particles reaching semimajor axes up to $2,000$~au. In system 14 (second row in Fig. \ref{aeiq_ss}), approximately $99.5$\% of the initial particle population remains bound; however, an early close stellar encounter (second panel, Fig. \ref{tad_ss}) causes $2.1$\% of particles to acquire eccentricities between $0.4$ and $1$, and $7.1$\% to reach semimajor axes greater than $1000$au. System 100 is moderately perturbed (third row, Fig. \ref{aeiq_ss}), roughly $71$\% of particles retain semimajor axes between $40$ and $1000$~au, while $9.6$\% of particles migrate outward, achieving semimajor axes between $1000$ and $5,000$~au and eccentricities in the range $0.1$–$1$. Additionally, most particles ($78.4$\%) maintain orbital inclinations below $50^\circ$, whereas the remaining inclinations span from $50^\circ$ to $150^\circ$. Approximately $17$\% of the particles are ejected from this disk, primarily due to a very close encounter ($\sim300$au) with a low-mass star ($\sim0.2\mathrm{M}_\odot$).

The most strongly perturbed systems are systems 86 (fourth row in Fig. \ref{aeiq_ss}) and 157 (fifth row in Fig. \ref{aeiq_ss}). System 86 experienced two very close stellar encounters: first with a red dwarf star ($0.5~\mathrm{M}_\odot$) at a distance of $488$\ au, followed later by an encounter with a brown dwarf ($0.08\mathrm{M}_\odot$) at $899$\ au. These encounters resulted in the ejection of approximately $13.4$\% of the initial particles, subsequently forming three distinct particle populations. The first population (green shaded region, fourth row of Fig. \ref{aeiq_ss}) comprises particles near $200$\ au with eccentricities ranging from $0.2$ to $1$. The second population (blue shaded region, fourth row of Fig. \ref{aeiq_ss}) occupies semimajor axes between $300$ and $500$\ au, with orbital inclinations predominantly between $0^\circ$ and $20^\circ$. About $23$\% of these particles attained eccentricities between $0.4$ and $1$, while the remainder had eccentricities spanning from $0.2$ to $1$. Finally, the third population (gray shaded region, fourth row of Fig. \ref{aeiq_ss}) consists of approximately $2.1$\% of the initial particles distributed across semimajor axes ranging from $\sim500$ to $6,000$~au, eccentricities from $0.2$ to $1$, and orbital inclinations up to $80^\circ$.

Among all our simulations, system 157 is the most dynamically perturbed and develops orbital distributions closely resembling those observed for trans-Neptunian objects and long-period comets in the Solar System. Two very close stellar encounters at impact parameters of $336$ au and $507$au (Fig. \ref{tad_ss}), involving stars with masses of $0.5\,\mathrm{M}_\odot$ and $0.1~\mathrm{M}_\odot$, respectively, result in the ejection of the system of approximately $57.5$\% of the initial particle population. Similar to the previously discussed systems, three primary populations and multiple secondary branches are subsequently formed due to weaker stellar encounters (fifth panel of Fig. \ref{aeiq_ss}). The first population (green shaded region, fifth row of Fig. \ref{aeiq_ss}) has semimajor axes between $40$ and $100$~au, eccentricities ranging from $0$ to $0.99$, and orbital inclinations from $5^\circ$ to $50^\circ$. Notably, particles with semimajor axes of approximately $40$–$50$\ au enter mean-motion resonances with Neptune. The second population (blue shaded region, fifth row of Fig. \ref{aeiq_ss}) occupies semimajor axes between $100$ and $300$au, eccentricities from $0.3$ to $0.99$, and inclinations up to $20^\circ$. The third and final population (gray shaded region, fifth row of Fig. \ref{aeiq_ss}) spans a broader range of orbital inclinations, from $20^\circ$ to $180^\circ$, with eccentricities ranging between $0$ and $0.99$. Approximately $2.2$\% of the particles reach semimajor axes between $1000$ and $4,000$~au, populating the inner Oort cloud region. Despite the strong encounters experienced by this system, only a small fraction of particles reach the outer Oort cloud region ($a_c > 10,000$~au), primarily because the closest stellar encounters occurred within the inner disk region, resulting in a substantial ejection of particles.

In all five selected systems, a tail emerges in the semimajor axis–eccentricity parameter space due to a combination of strong stellar encounters followed by weaker perturbations. In certain cases (systems 100, 86, and 157), multiple distinct branches appear, primarily resulting from subsequent weaker stellar encounters. Additionally, three characteristic particle populations are clearly identifiable (highlighted in green, blue, and gray shaded regions in the first column of Fig. \ref{aeiq_ss}). The repeated stellar encounters trigger particle ejections extending to the boundary of the present-day inner Oort cloud region ($\sim20,000$\ au). Subsequently, perturbations induced by the Galactic tidal field will circularize the orbits of particles with large semimajor axes \citep[see][]{Brasser2012}, eventually transforming these particles into stable Oort cloud objects. Conversely, particles maintaining smaller semimajor axes, where the Galactic tidal field is ineffective, remain on (highly) eccentric orbits, rendering their orbits dynamically unstable over secular timescales. Thus, while the observed tail structure is transient, it will ultimately evolve into a stable population of objects sharing similar orbital characteristics, analogous to the trans-Neptunian objects and Sednoids observed within our Solar System. The overall dynamical architecture of each system strongly depends on the geometry, stellar mass, and velocity of each encounter, as previously discussed in Sect.\ref{sec3} and illustrated in Fig. \ref{pos_ss}.

Interestingly, after $100$\ Myr, the planets remained in stable orbits in all five of the analysed systems. This result indicates that, even within dense stellar cluster environments, planetary systems may endure multiple close stellar encounters without significant orbital modifications of their planets. 

In summary, the simulations presented in this section illustrate different scenarios of how planetary systems can be dynamically perturbed during their early evolutionary stages. A key condition for long-term stability is that the disk of planetesimals must initially reside sufficiently far from the planets. If the planetesimal disk is too compact or located too close to the planetary region, planet–disk interactions can trigger planetary orbital instabilities, dynamically heating the disk and resulting in the ejection of a large fraction of particles to interstellar distances, thus contributing significantly to the population of interstellar comets. In the scenarios explored here, the formation of Oort cloud-like structures could have been initiated solely by stellar encounters without invoking planetary orbital instabilities. This scenario assumes that the planets already occupied their current positions while the planetary system was still embedded within its natal stellar cluster and that the protoplanetary gas disk had already dissipated \citep[see also][]{Brasser2012}. Subsequent dynamical evolution, driven by perturbations from the Galactic tidal field and occasional additional stellar encounters, will facilitate the gradual transfer of particles from the inner Oort cloud region to populate the outer Oort cloud \citep{PZ_ST_2021}.

\subsection{Compact models}
\label{sec3.2}

For the construction of the \como\ and \comn\ models, we followed the classical Nice model \citep{Gomes2005,Tsiganis2005,Morbidelli2005} for the early Solar System. Although newer versions of the model have been proposed \citep{Nesvorny2012,Morbidelli2019}, \citet{Morbidelli2007} noted that the orbital evolution of the planets remains largely consistent with the original version. The classical Nice model, therefore, serves our purpose by providing a framework to understand the dynamical instability of the giant planets and the subsequent scattering of planetesimals, leading to the formation of Oort cloud-like structures in a compact planetary configuration. In Sect. \ref{1enc_nice} we present our results for a single encounter, while in Sect. \ref{Nenc_nice}, we analyse the outcomes of multiple stellar encounters.

\subsubsection{\como}
\label{1enc_nice}

\begin{figure}[ht!]
  \centering
  \includegraphics[width=0.87\hsize]{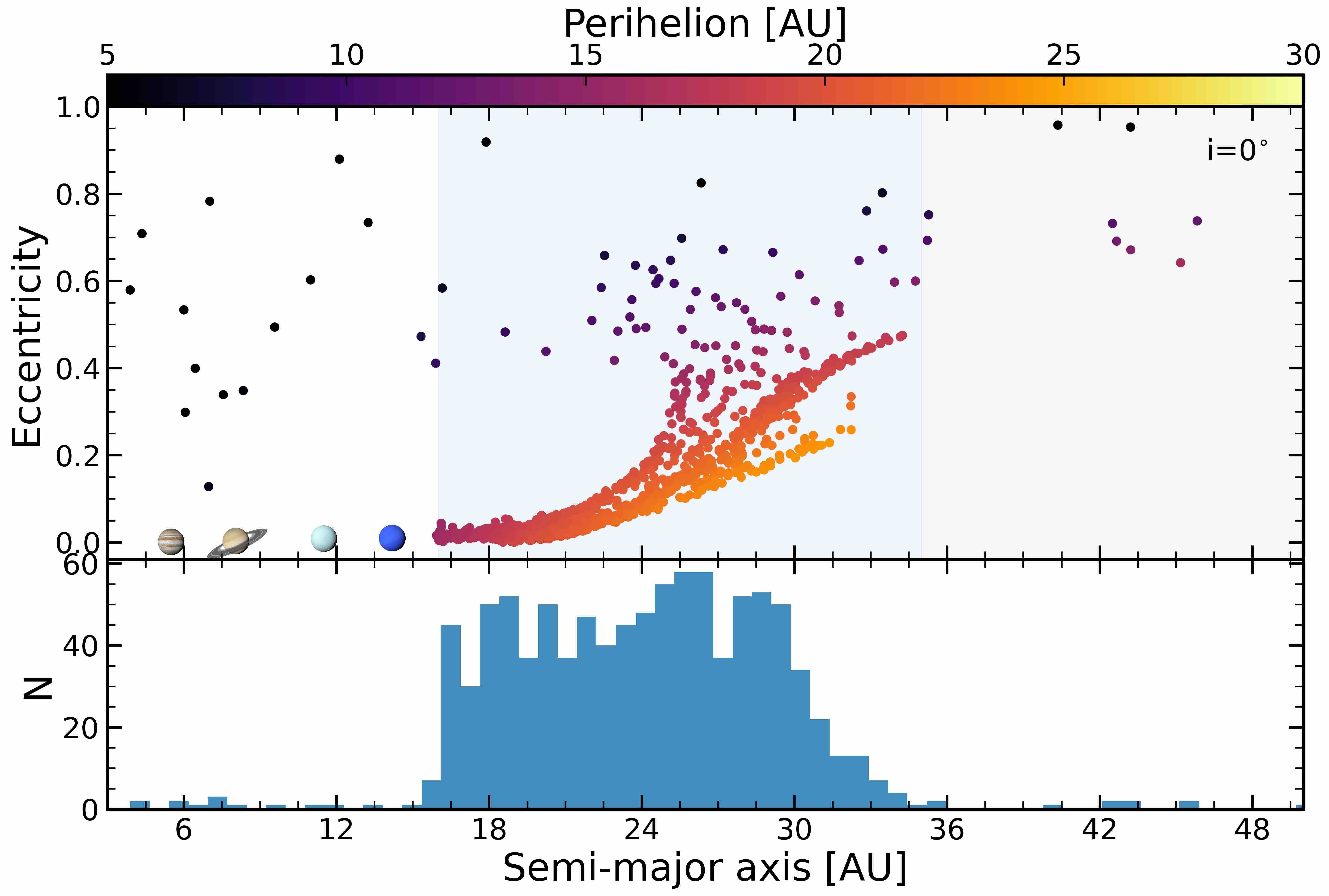}
  \includegraphics[width=0.87\hsize]{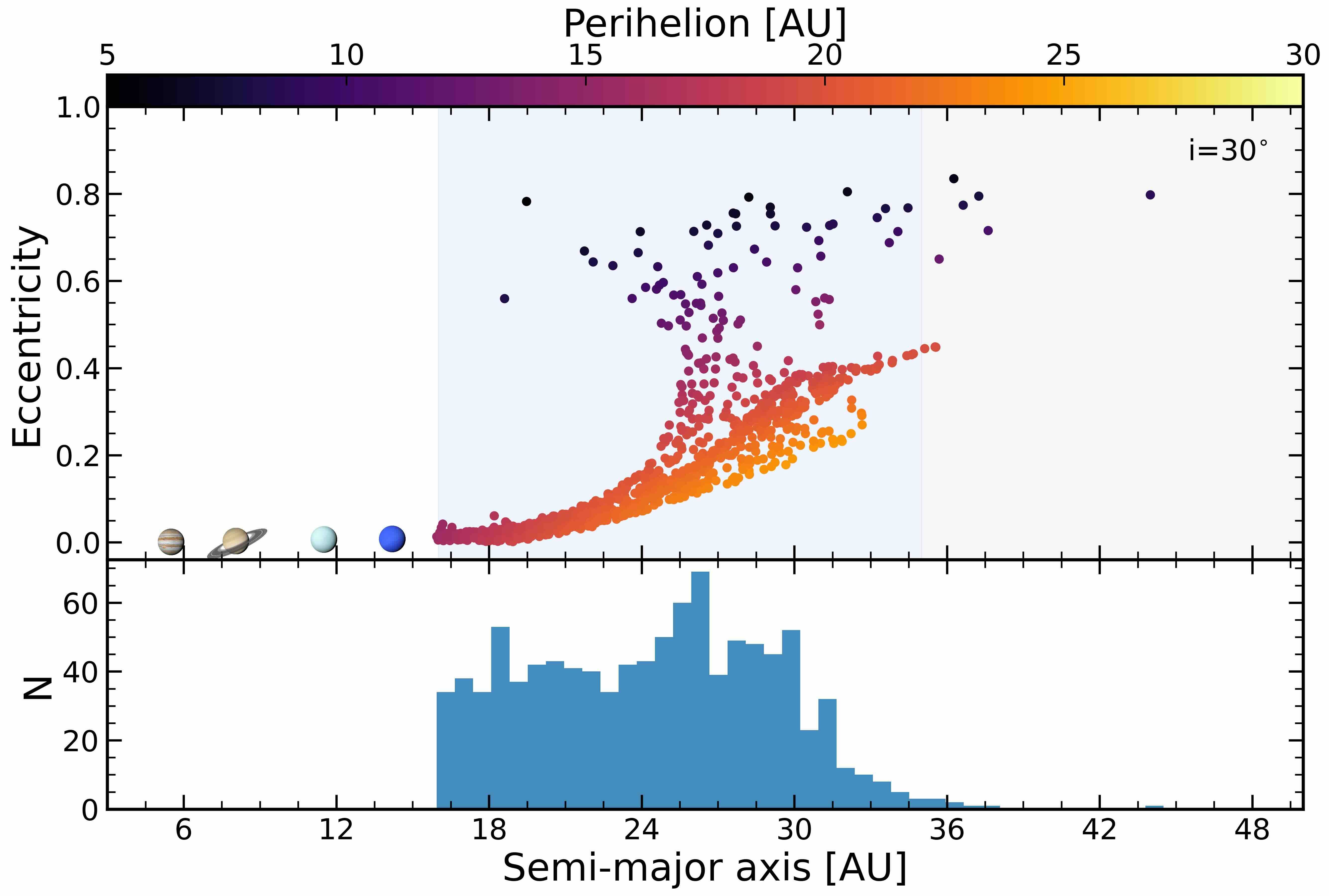}\\
  \includegraphics[width=0.87\hsize]{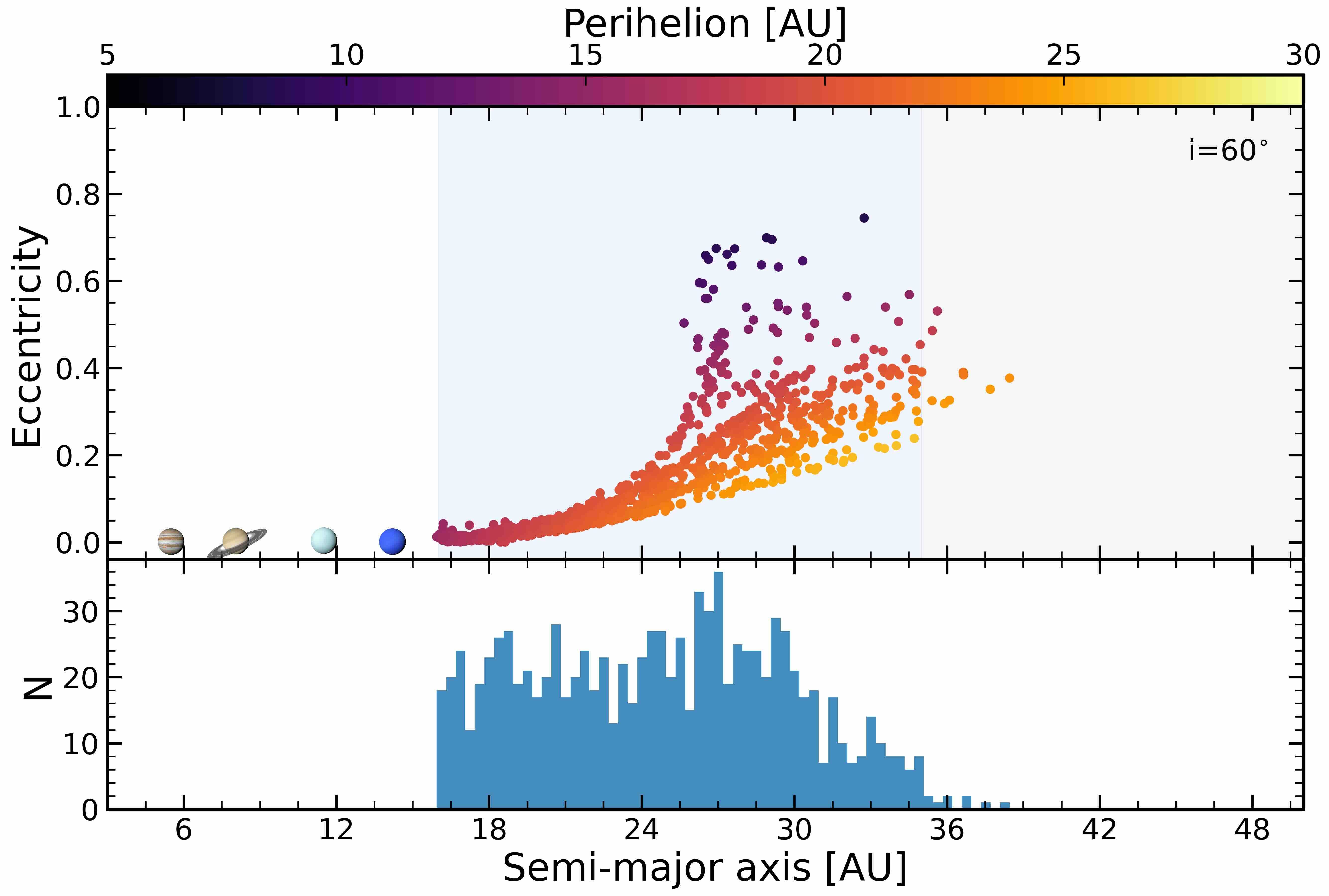}\\
  \includegraphics[width=0.87\hsize]{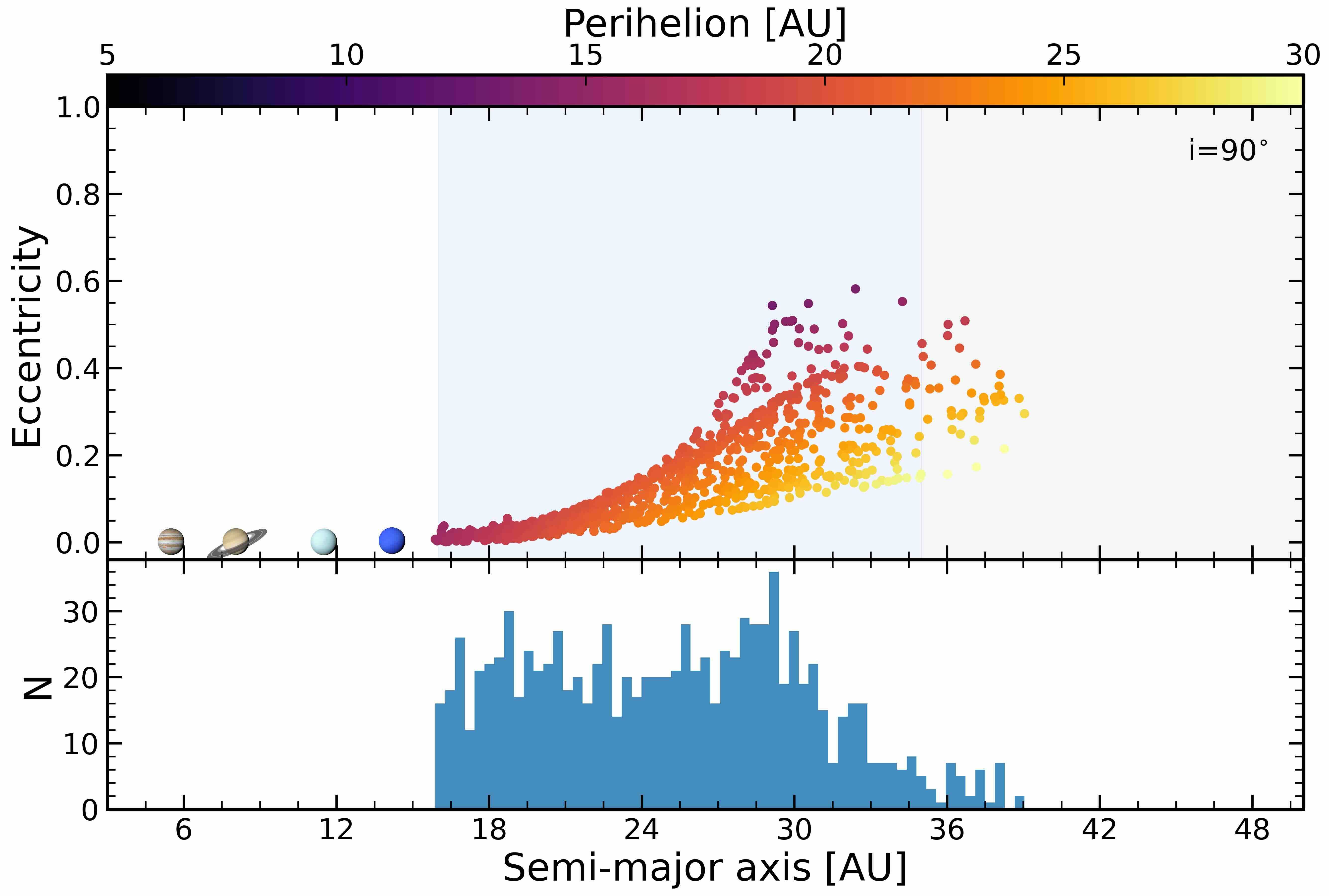}
  \caption{\como\ model. Semimajor axis as a function of eccentricity after an encounter with a $1$~\mSun star at $q_\star\simeq 75$ au. The color bar indicates the perihelion distance of the particles. Each panel corresponds to a different inclination angle of the encounter ($0^{\circ}$, $30^{\circ}$, $60^{\circ}$, and $90^{\circ}$).}
  \label{pss}
\end{figure}

Following the methodology described in Sect.~\ref{sec2} and adopting the initial conditions listed in Table~\ref{table1}, we performed 32 simulations for the \como\ model, varying the inclination of the encounter ($0^\circ$, $30^\circ$, $60^\circ$, and $90^\circ$). We find that the minimum pericenter distance at which the SSA becomes significantly perturbed and begins to produce ISOs is $q_\star \simeq 75$~au. Figure~\ref{pss} shows that the efficiency of disruption is strongly geometry dependent. Coplanar, prograde flybys maximize the energy transfer to disk particles, whereas inclined or polar encounters are comparatively inefficient.

In the coplanar case ($i=0^\circ$), $2.4\%$ of particles are ejected from the system, and the post-encounter disk develops three distinct dynamical families. A small fraction forms a scattered population ($\sim6\%$), characterized by low perihelia and high eccentricities ($e\simeq0.4$–1), with Tisserand parameters relative to Uranus below the scattering boundary ($T_{\rm U}\lesssim2.5$). These values place the particles deep within the Uranus-coupled regime, where high relative velocities lead to strong, chaotic encounters and efficient diffusion in orbital energy and angular momentum \citep{Carusi1987,LevisonDuncan1994}. A second group, the scattered tail ($\sim5.5\%$), occupies a narrow band near the Uranus-scattering boundary ($2.5\lesssim T_{\rm U}\lesssim3$). These objects were shaped by earlier scattering events and remain marginally coupled to Uranus. Although close to the nominal decoupling limit, their relative velocity is sufficiently low to allow continued, though weaker, gravitational encounters \citep{ValsecchiManara1997}, leading to moderate excitation in eccentricity ($e\simeq0.2$–0.5) and inclination. The majority of particles ($\sim88.5\%$) form a lower tail with $T_{\rm U}\gtrsim3$, corresponding to orbits dynamically decoupled from close-encounter scattering with Uranus, though still subject to secular and resonant perturbations. These objects remain only weakly perturbed from their initial configuration. Overall, the wide range of Tisserand parameters ($0.98<T_{\rm U}<8.3$) reflects the coexistence of strongly scattered, marginally coupled, and dynamically decoupled populations produced by the stellar flyby.

The Tisserand parameter, $T_{\rm U}$, serves as a quasi-invariant diagnostic of the restricted three-body problem. It constrains how a particle’s orbital elements, semimajor axis, eccentricity, and inclination evolve during gravitational encounters. Lower values of $T_{\rm U}$ correspond to more chaotic, dynamically heated trajectories \citep[e.g.,][]{MurrayDermott2000}. At higher inclinations, the morphology of the disk changes, but the same three families persist. At $i=60^\circ$, no particles are ejected, the scattered population nearly disappears ($\simeq1.7\%$), and the scattered tail reaches $\simeq6.3\%$, leaving about $\simeq92\%$ of the disk in the lower tail. For $i=90^\circ$, the scattered population vanishes entirely, the scattered tail decreases to $\simeq3\%$, and the majority of the particles in the disk ($\simeq97\%$) survives in a dynamically cold state.

These results confirm that the three-branch tail seen in Fig. \ref{pss} is a hallmark of coplanar or moderately inclined encounters, where gravitational focusing is strongest. In contrast, polar flybys leave the primordial disk largely intact, producing only weak heating without significant ejection.

\begin{figure}    
  \includegraphics[width=\columnwidth]{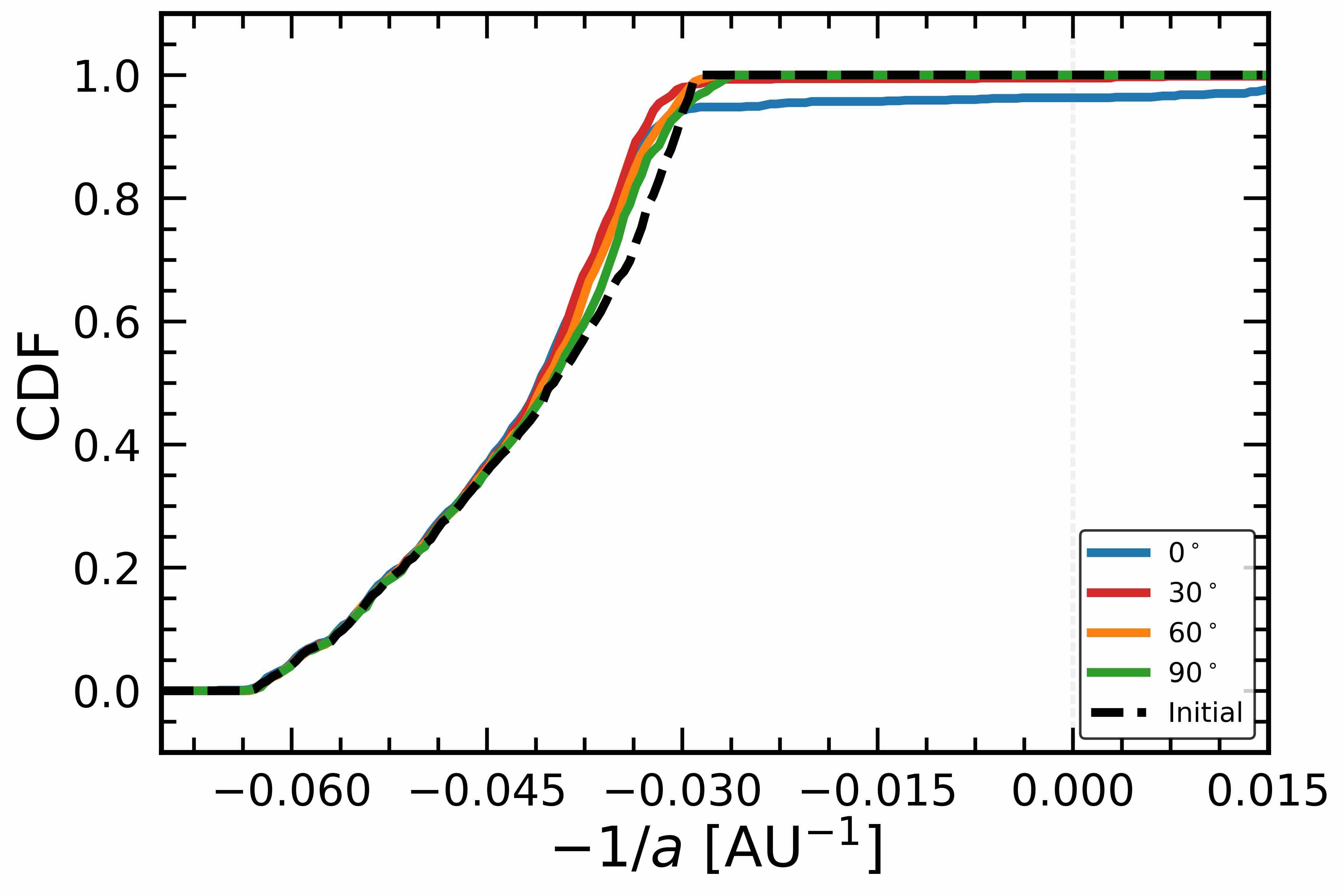}
  \caption{\como\ model. Cumulative distribution of the final orbital energy of disk particles. The colored lines correspond to different inclination angles of the encounter ($0^{\circ}$, $30^{\circ}$, $60^{\circ}$, and $90^{\circ}$), while the black dashed line represents the initial distribution. Negative values ($-1/a<0$) indicate bound orbits, while positive values correspond to ejected ISOs.}
  \label{energy_nice}
\end{figure}

The cumulative energy distributions (Fig. \ref{energy_nice}) confirm these trends. Only the coplanar case ($0^{\circ}$) shows a pronounced unbound tail ($2.4\%$), with a smaller contribution at $30^{\circ}$ ($0.5\%$). At $60^{\circ}$ and $90^{\circ}$, the CDFs overlap almost perfectly with the initial distribution, indicating that such geometries do not efficiently generate interstellar comets. The step-like features in the CDF mirror the three families identified in Fig. \ref{pss}: the scattered population produces the most energetic tail, the scattered tail contributes a secondary rise at intermediate energies, and the lower tail represents the bulk of the disk that remains weakly perturbed and bound. Together, Figs.~\ref{pss} and \ref{energy_nice} show that close solar-mass flybys at $q_\star\sim75$~au can only weakly populate an Oort-like reservoir, and that the efficiency declines sharply with increasing inclination for prograde geometries. 

In contrast, retrograde encounters are expected to be less effective in disrupting the disk.
Because the perturber and disk particles move in opposite directions, their higher relative velocity shortens the interaction time and suppresses gravitational focusing, thereby reducing angular-momentum exchange and limiting both ejection efficiency and dynamical heating of the disk \citep[e.g.,][]{Winter2018}.

\begin{figure}[ht!]
  \includegraphics[width=1.\hsize]{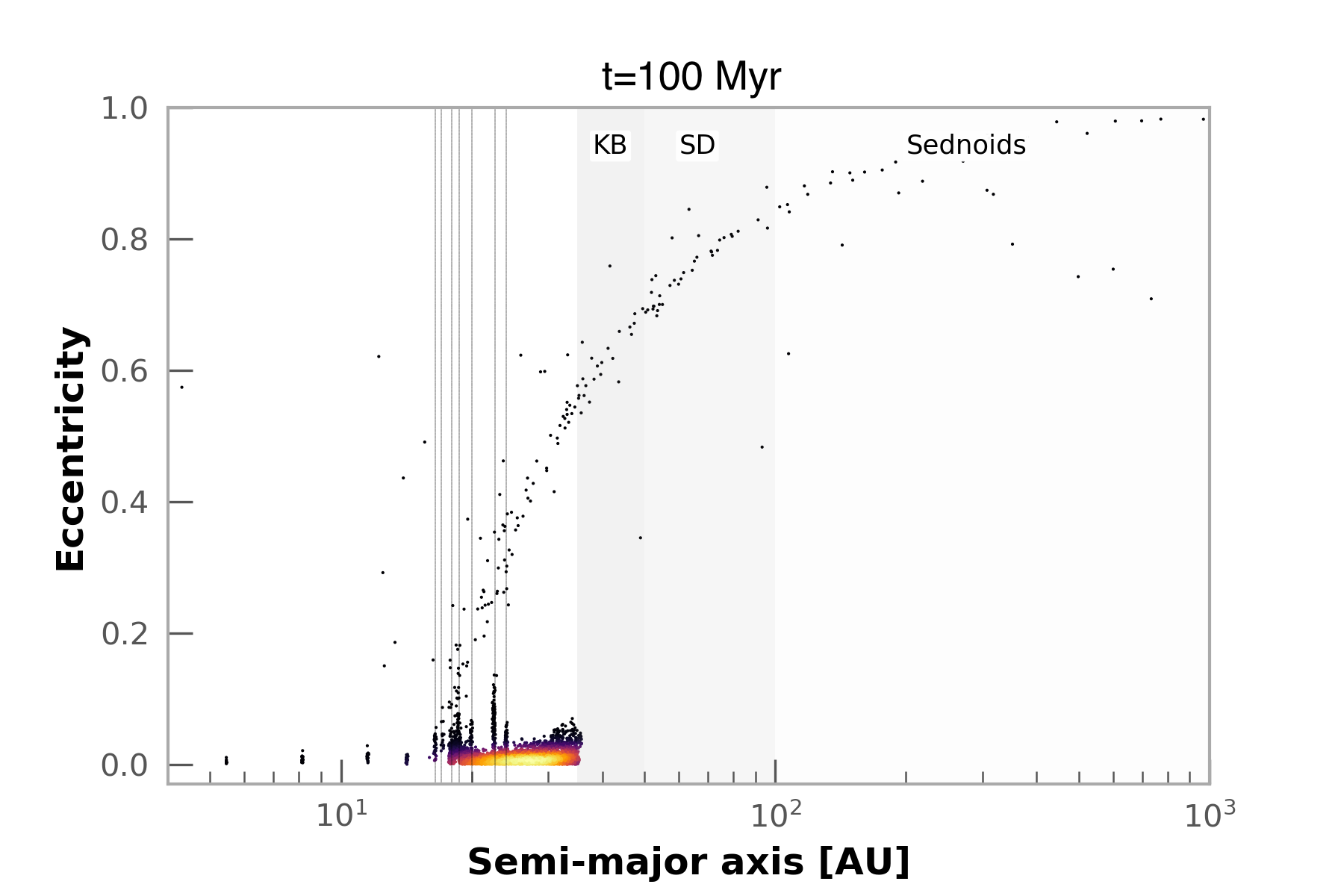}
  \includegraphics[width=1.\hsize]{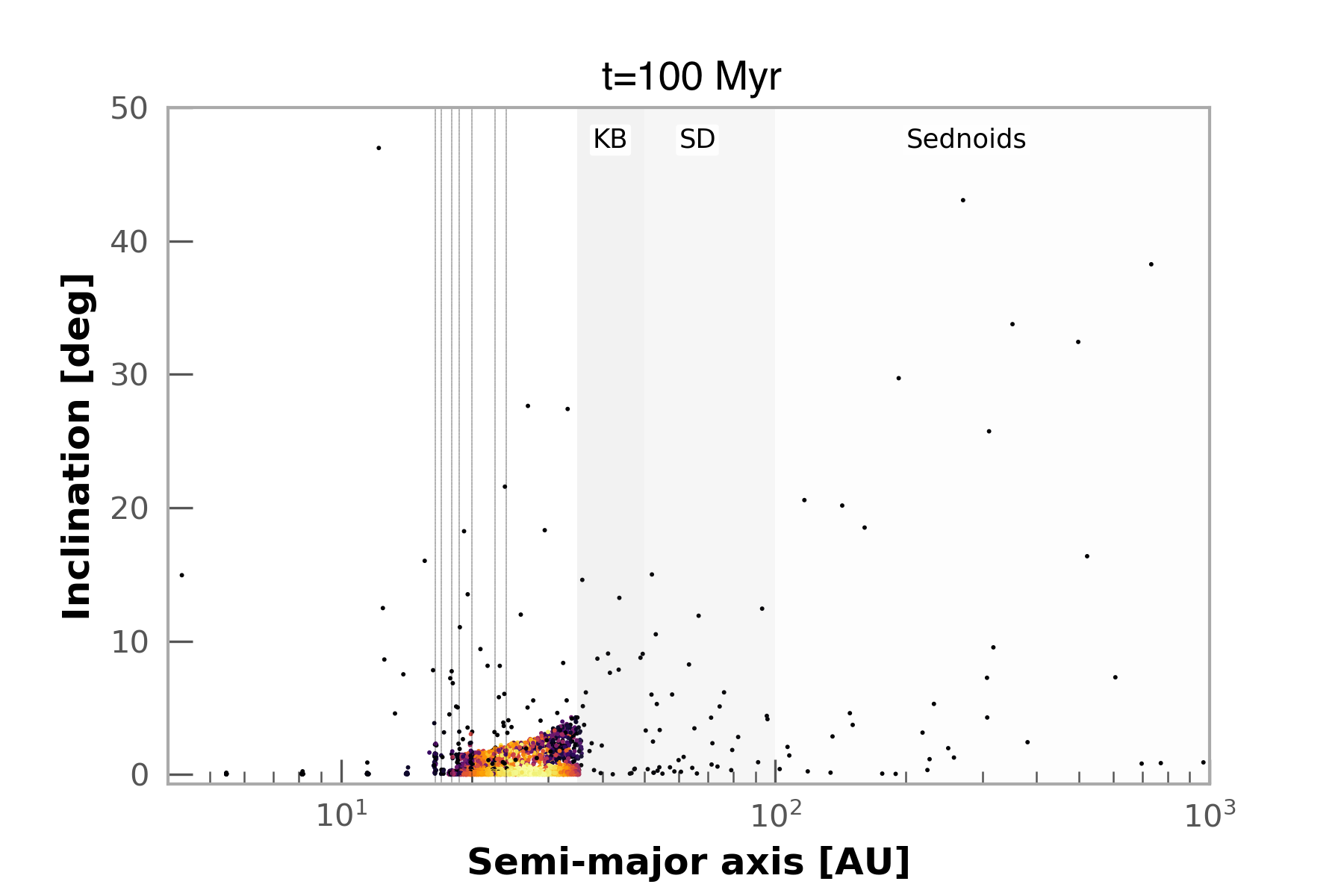}\\
  \caption{\comn\ model. Semimajor axis as a function of eccentricity (top panel) and orbital inclination (bottom panel) for all the particles in the 200 simulated systems. The gray areas represent the different regions of the Solar System:  Kuiper belt (KB), scattered disk (SD), and the Sednoids. The dashed lines show the mean motion resonances with Neptune. The integration time is set to 100~Myr. An animation is available on the \href{https://www.aanda.org/articles/aa/olm/2026/04/aa54978-25/aa54978-25.html}{A\&A website}.} \label{nice_all_eai_p}
\end{figure}

\begin{figure*}[t]      
  \includegraphics[width=0.5\textwidth]{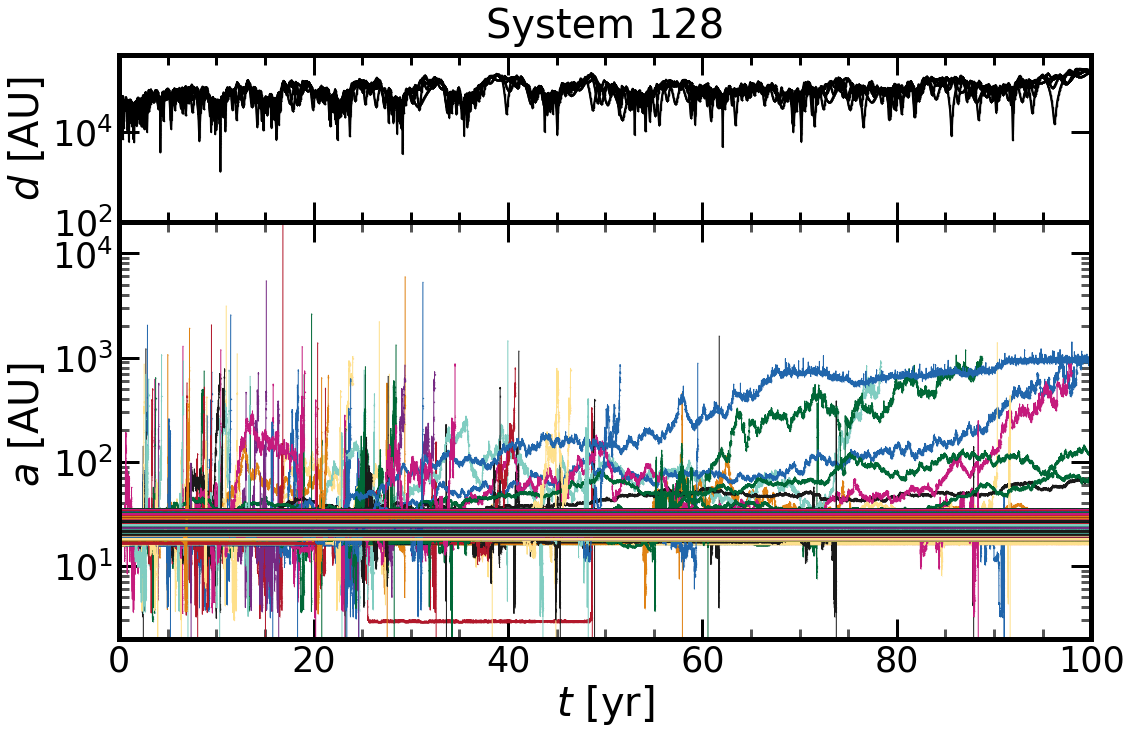}
  \includegraphics[width=0.5\textwidth]{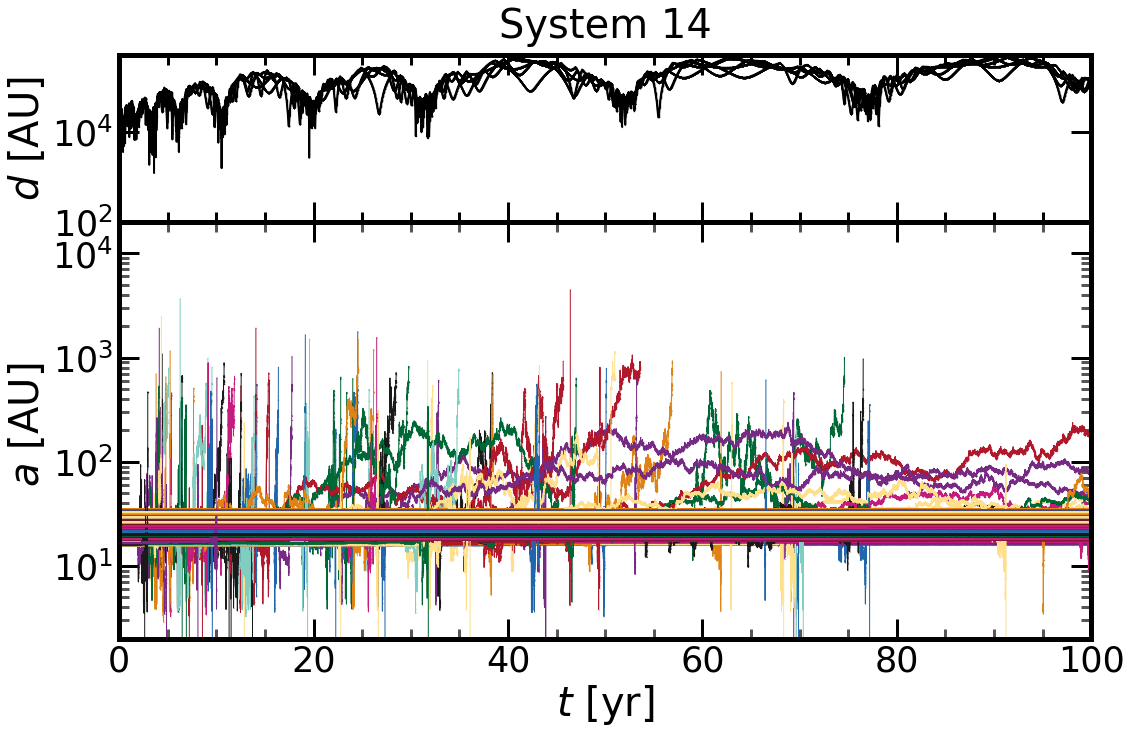}
  \includegraphics[width=0.5\textwidth]{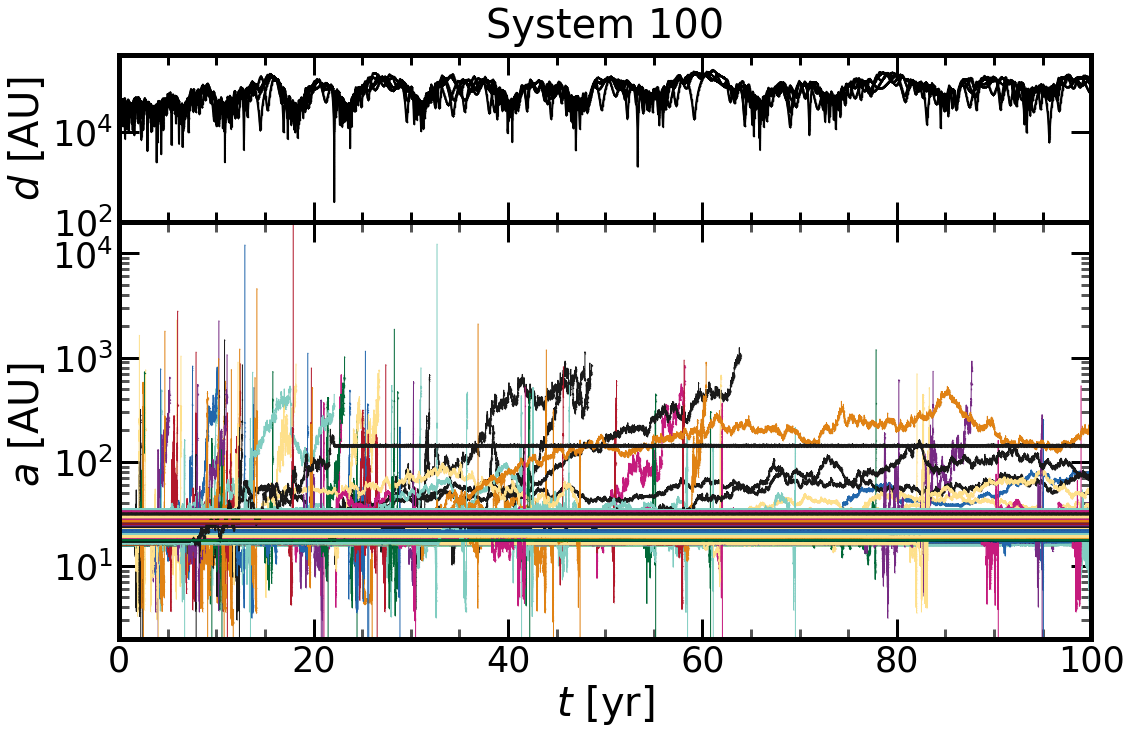}
  \includegraphics[width=0.5\textwidth]{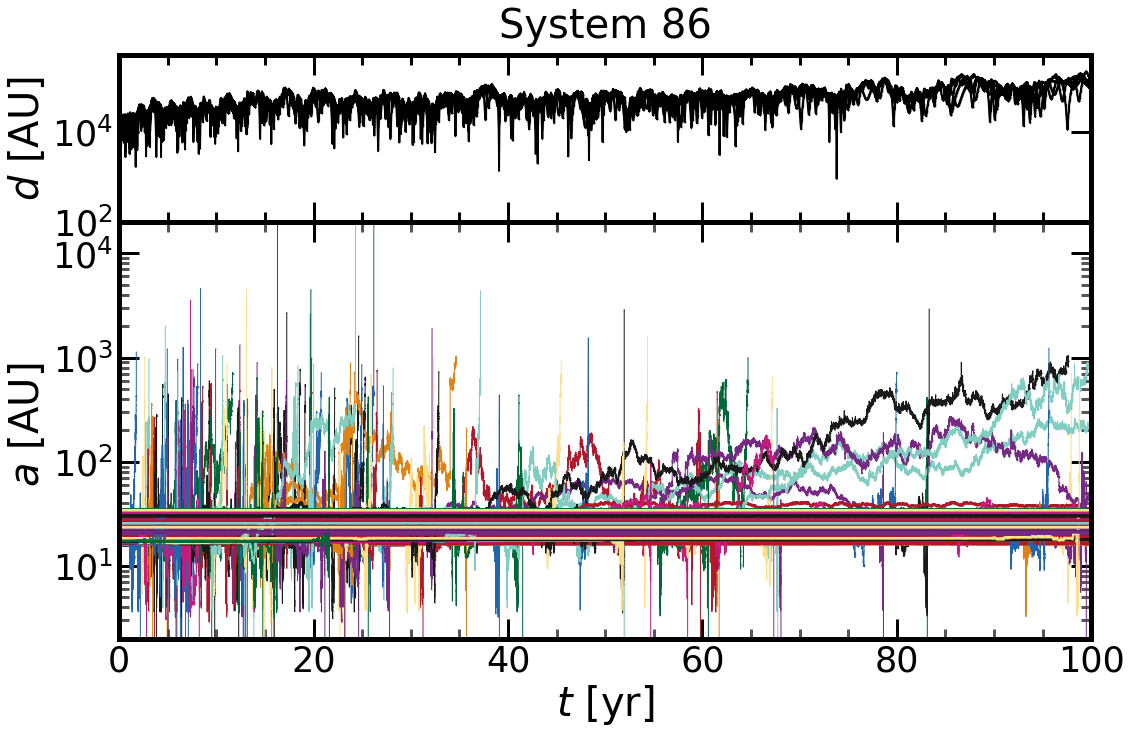}
  \includegraphics[width=0.5\textwidth]{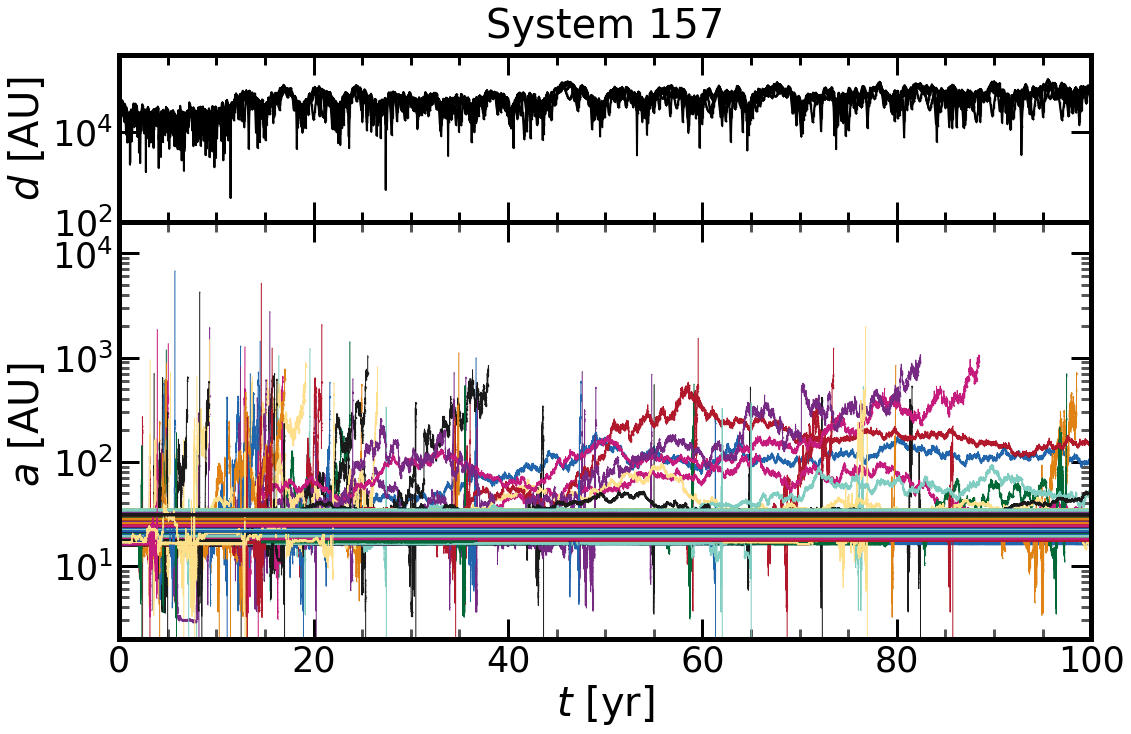}
  \caption{\comn\ model. Orbital evolution of particles in the disk over 100 Myr. The bottom panels in each plot depict the semimajor axis as a function of time, while the top panels show the distance of the perturber. The panels represent  systems 128, 14, 100, 86, and 157 (as labeled). The colored lines correspond to the individual particles within each system.}
  \label{a_t_nice}
\end{figure*}

\begin{figure*}

\includegraphics[width=0.31\textwidth]{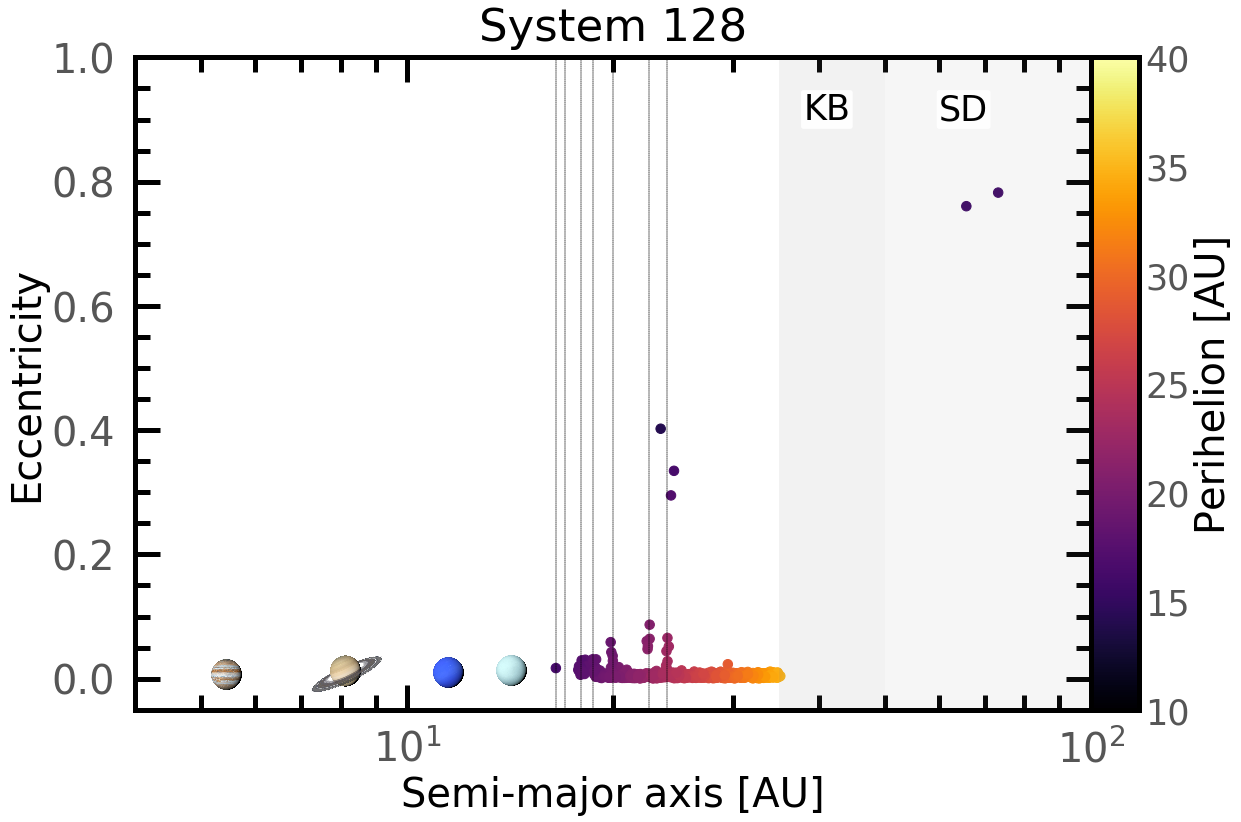}
\includegraphics[width=0.31\textwidth]{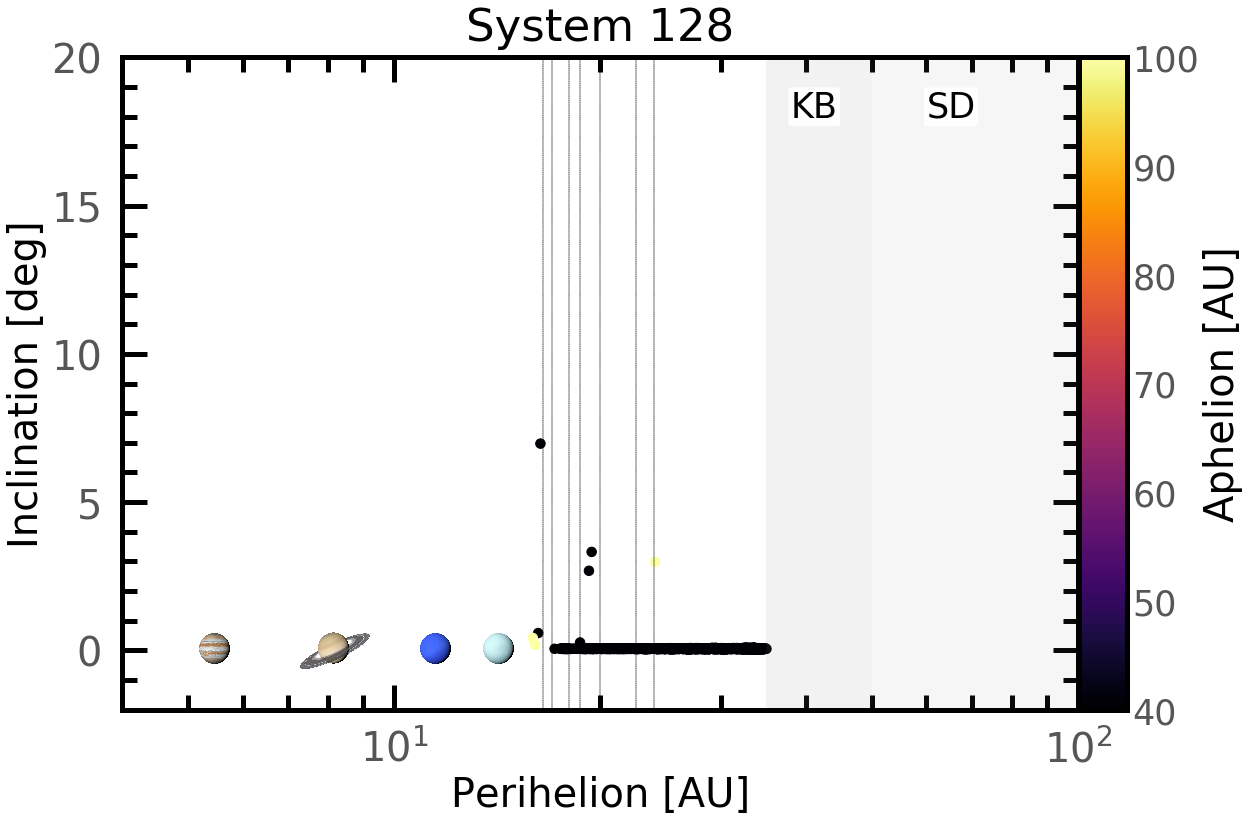}
\includegraphics[width=0.31\textwidth]{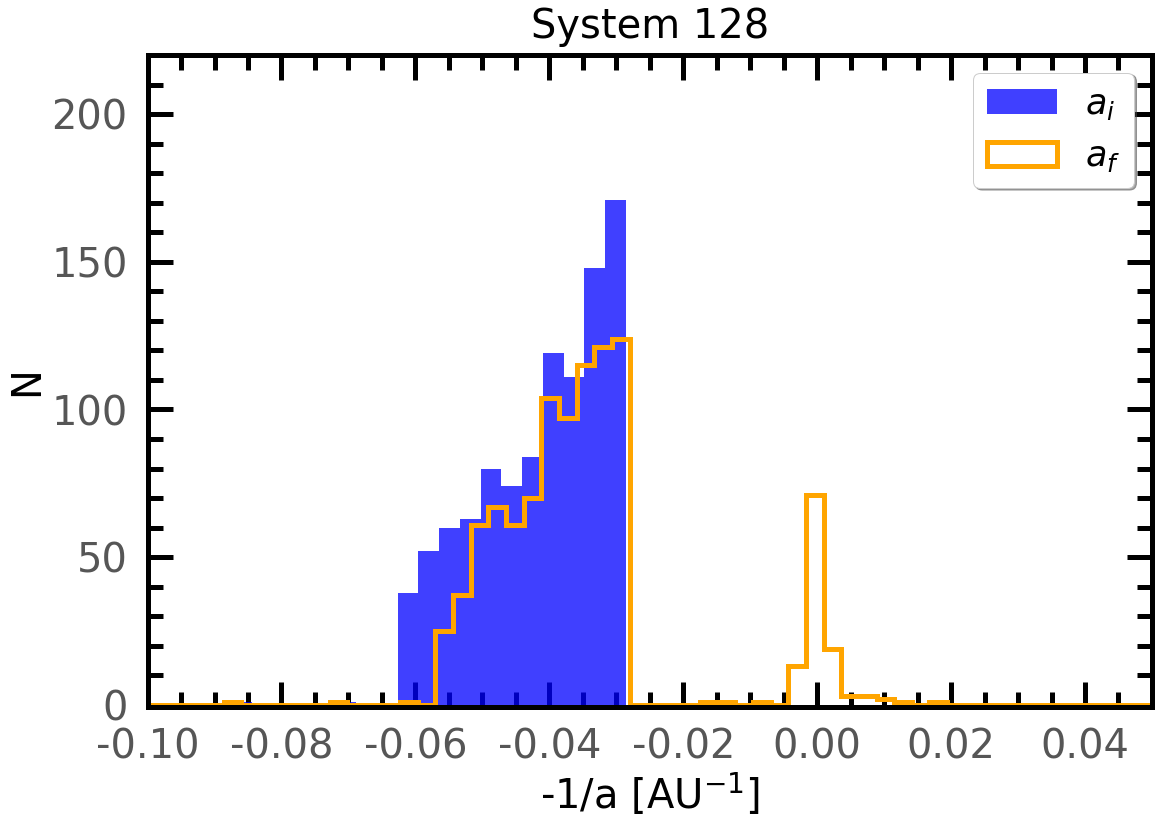}\\

\includegraphics[width=0.31\textwidth,trim=0 0 0 0,clip]{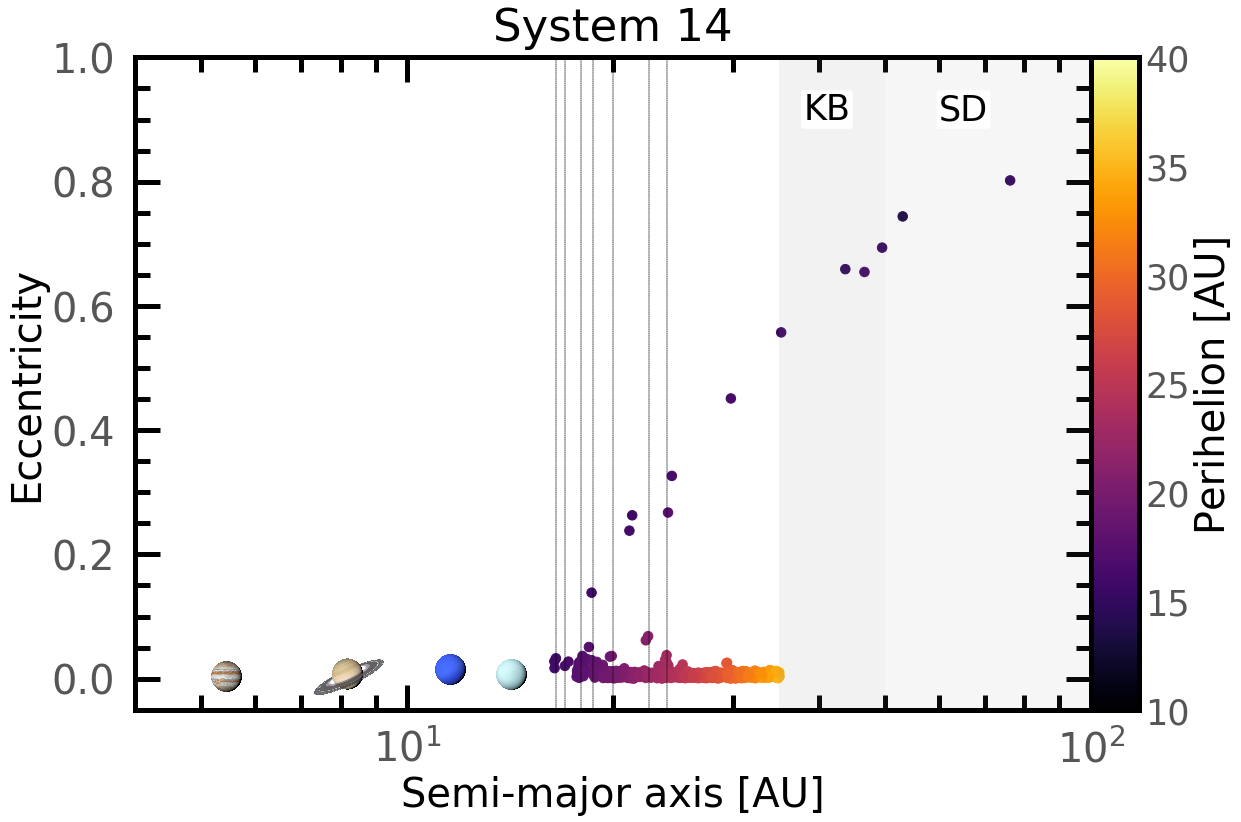}
\includegraphics[width=0.31\textwidth,trim=0 0 0 0,clip]{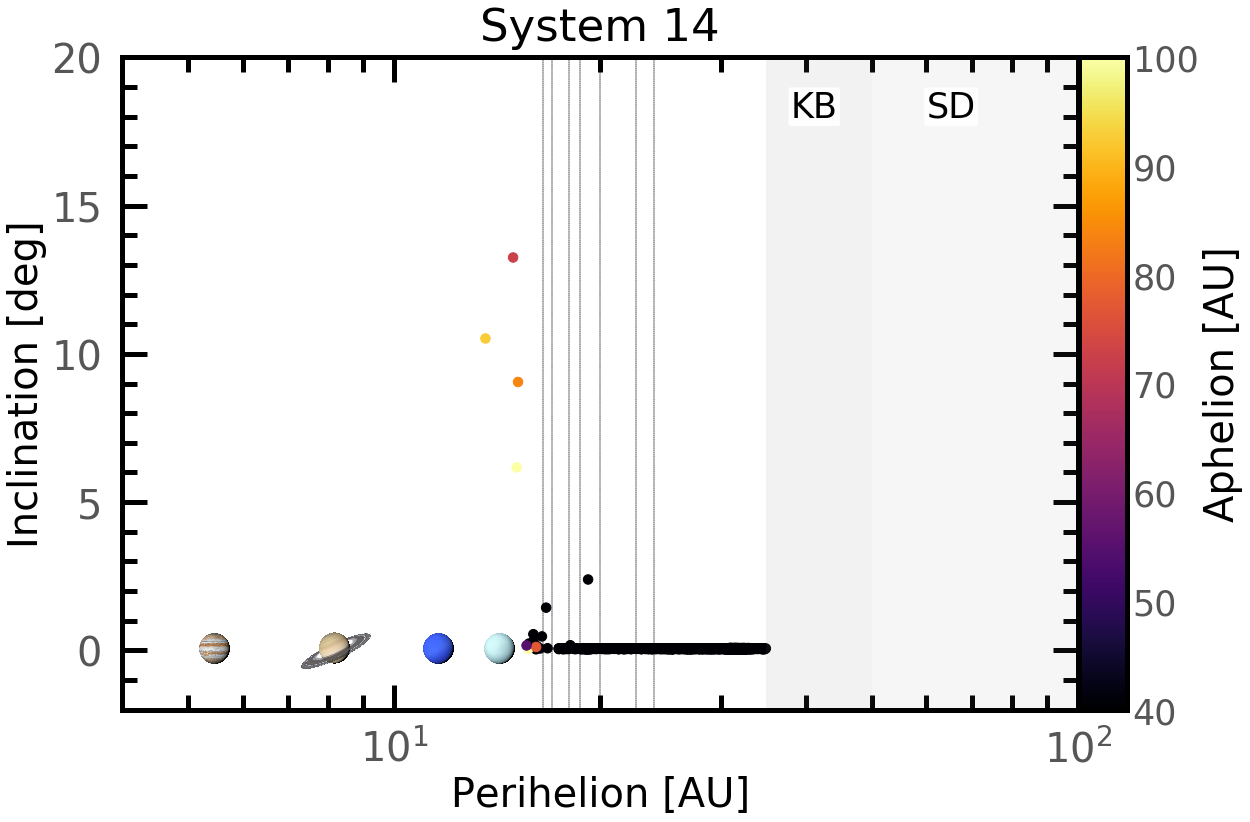}
\includegraphics[width=0.31\textwidth,trim=0 0 3 0,clip]{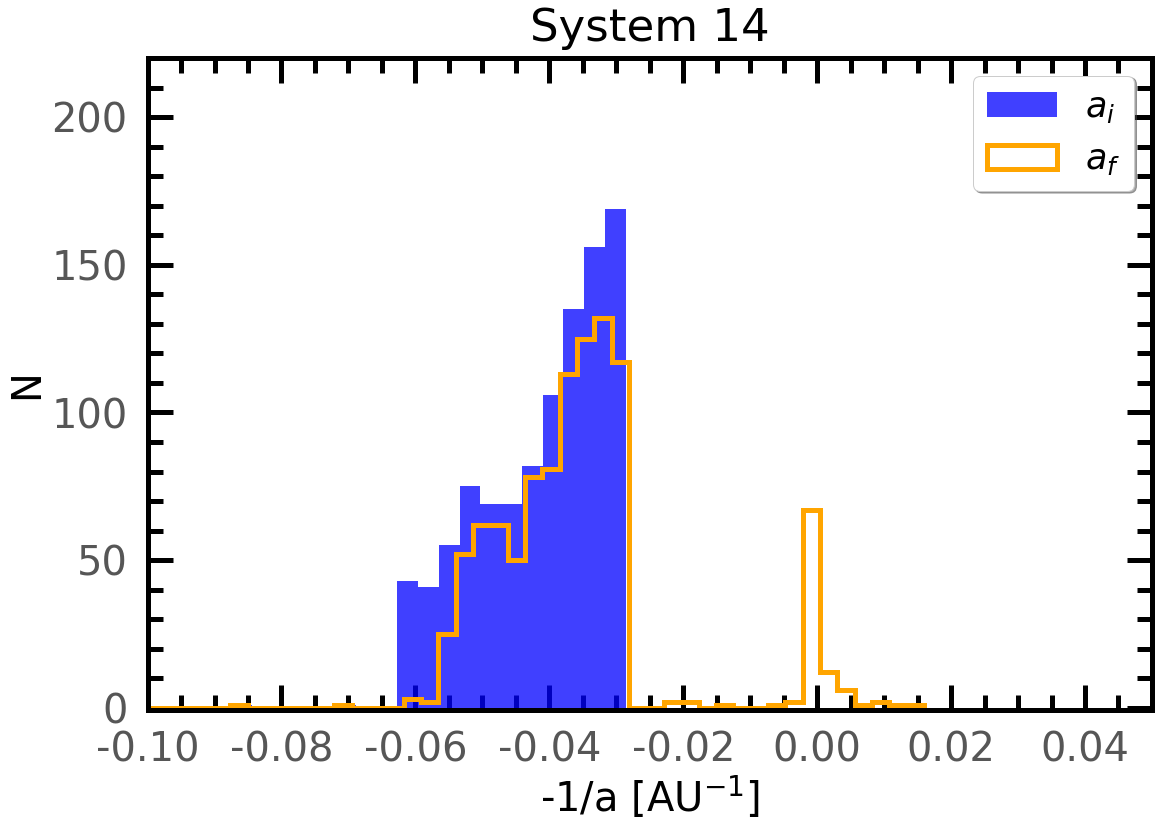}\\

\includegraphics[width=0.31\textwidth,trim=0 0 0 0,clip]{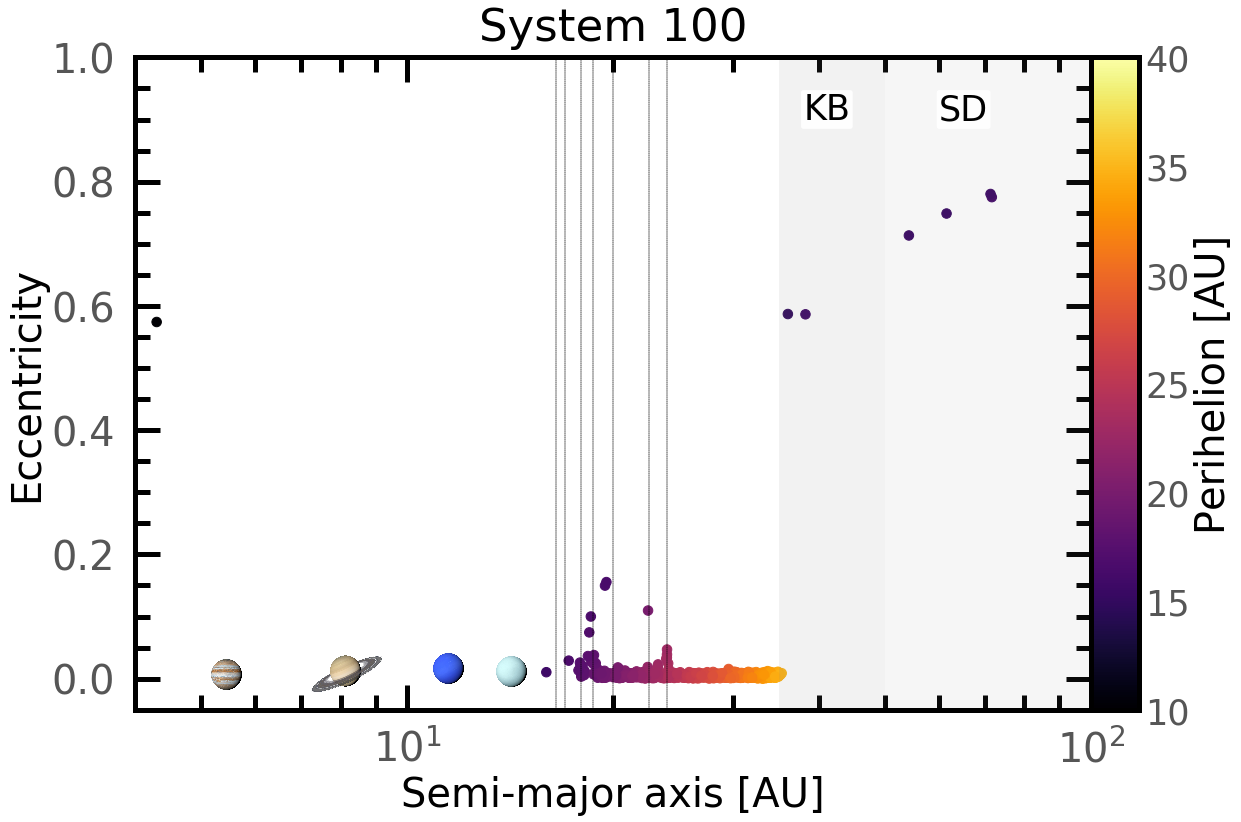}
\includegraphics[width=0.31\textwidth,trim=0 0 0 0,clip]{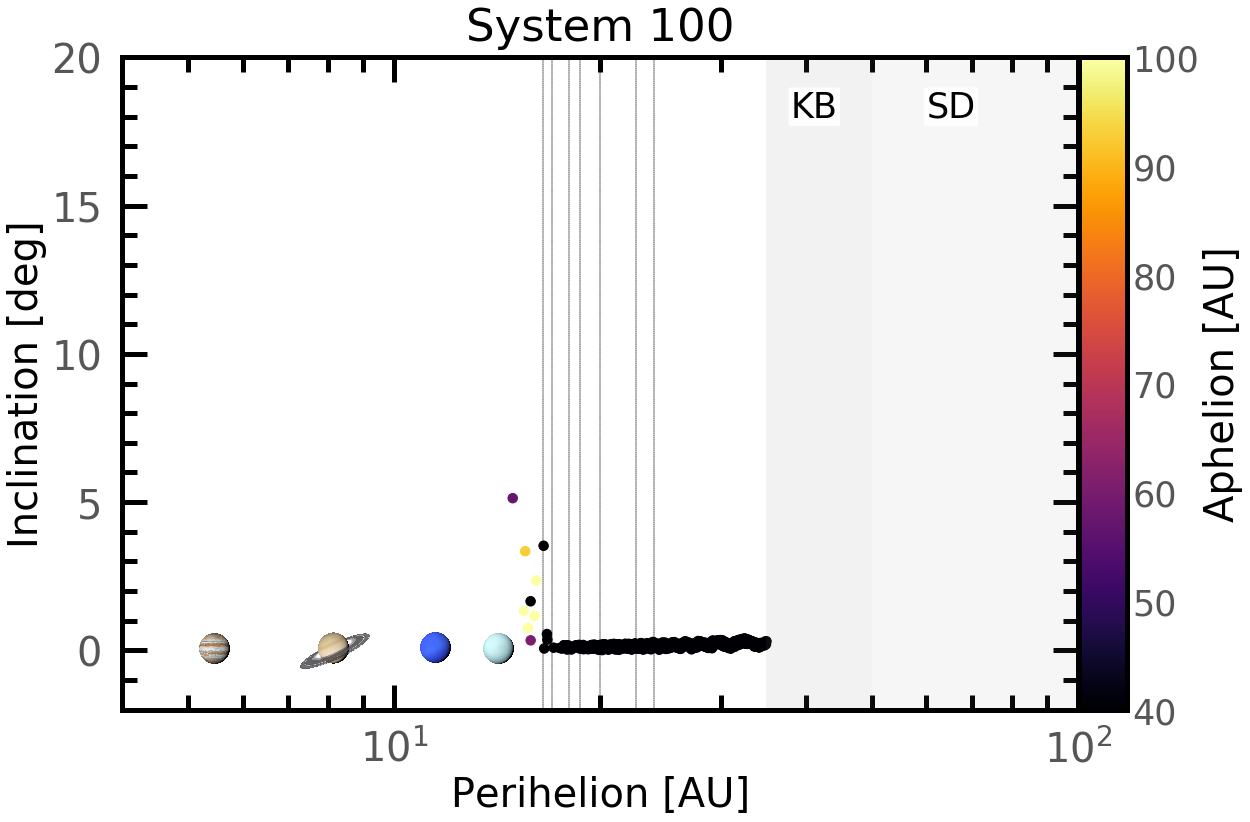}
\includegraphics[width=0.31\textwidth,trim=0 0 2 0,clip]{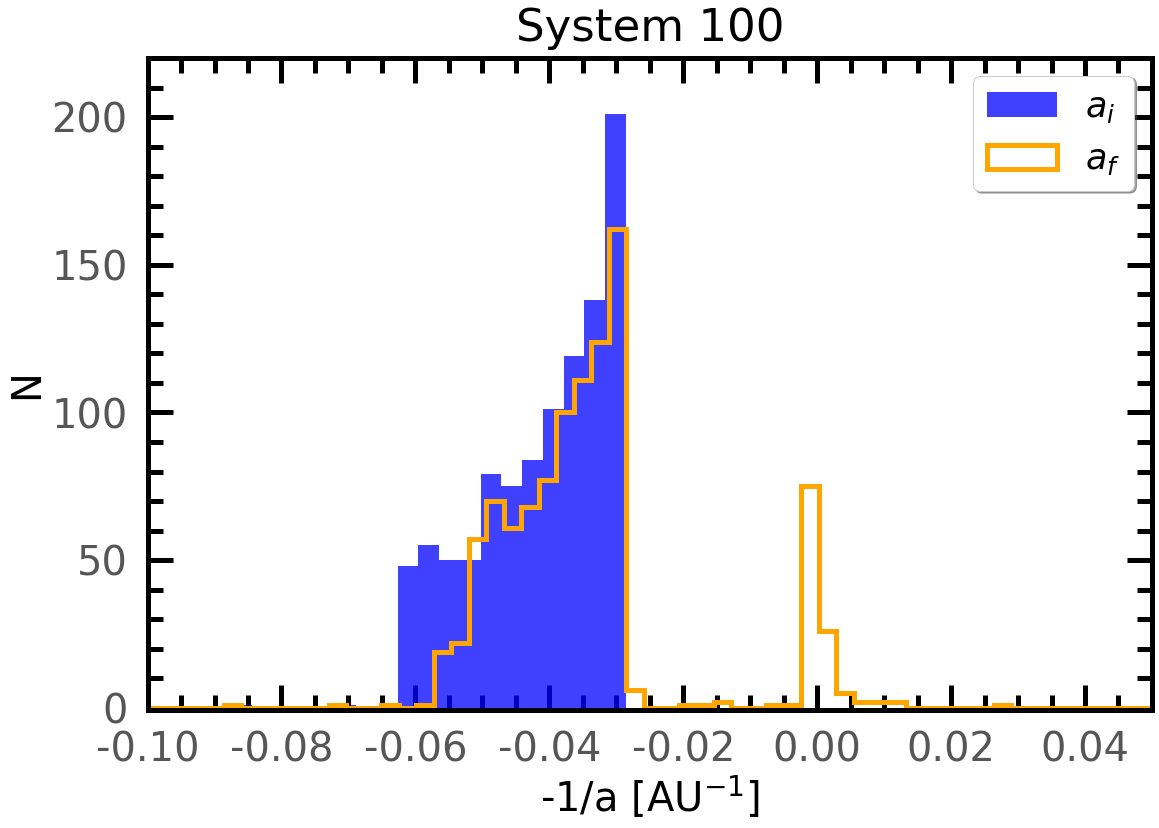}\\

\includegraphics[width=0.31\textwidth,trim=0 0 0 0,clip]{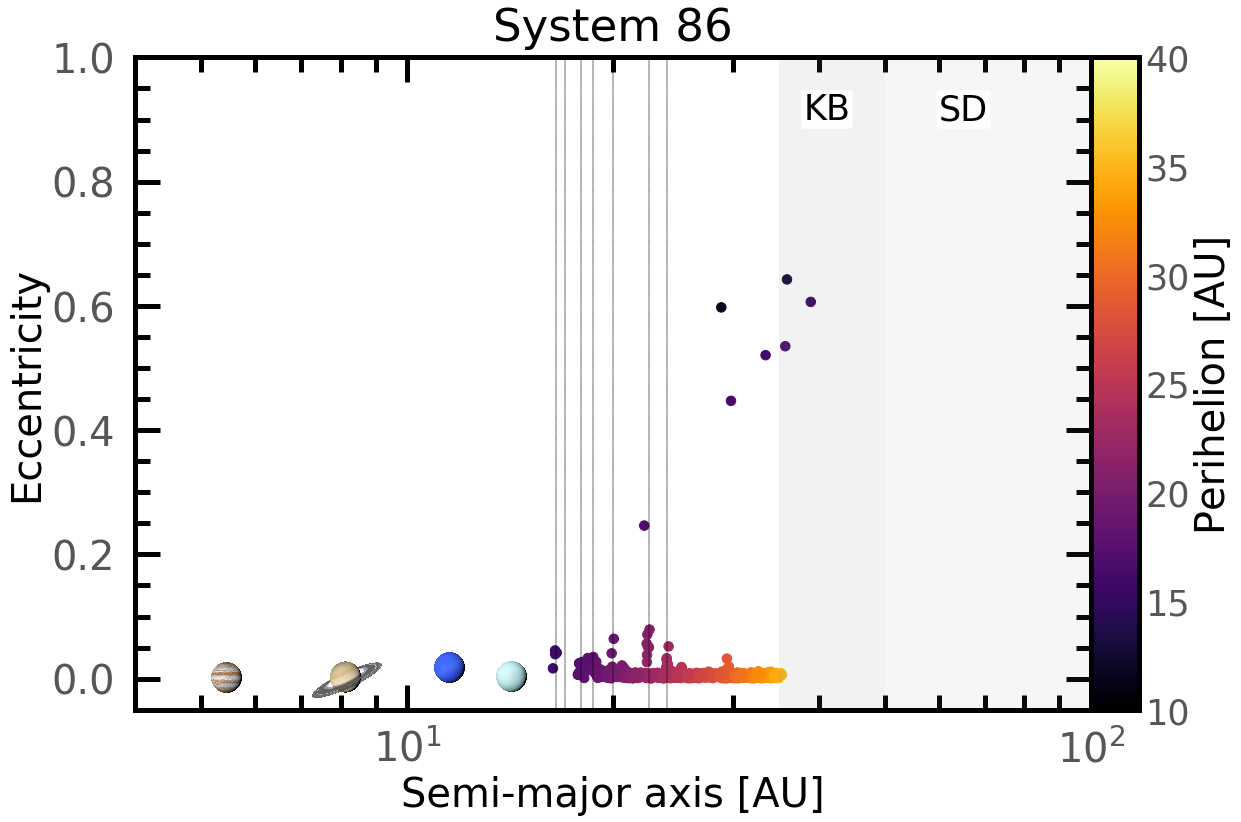}
\includegraphics[width=0.31\textwidth,trim=0 0 0 0,clip]{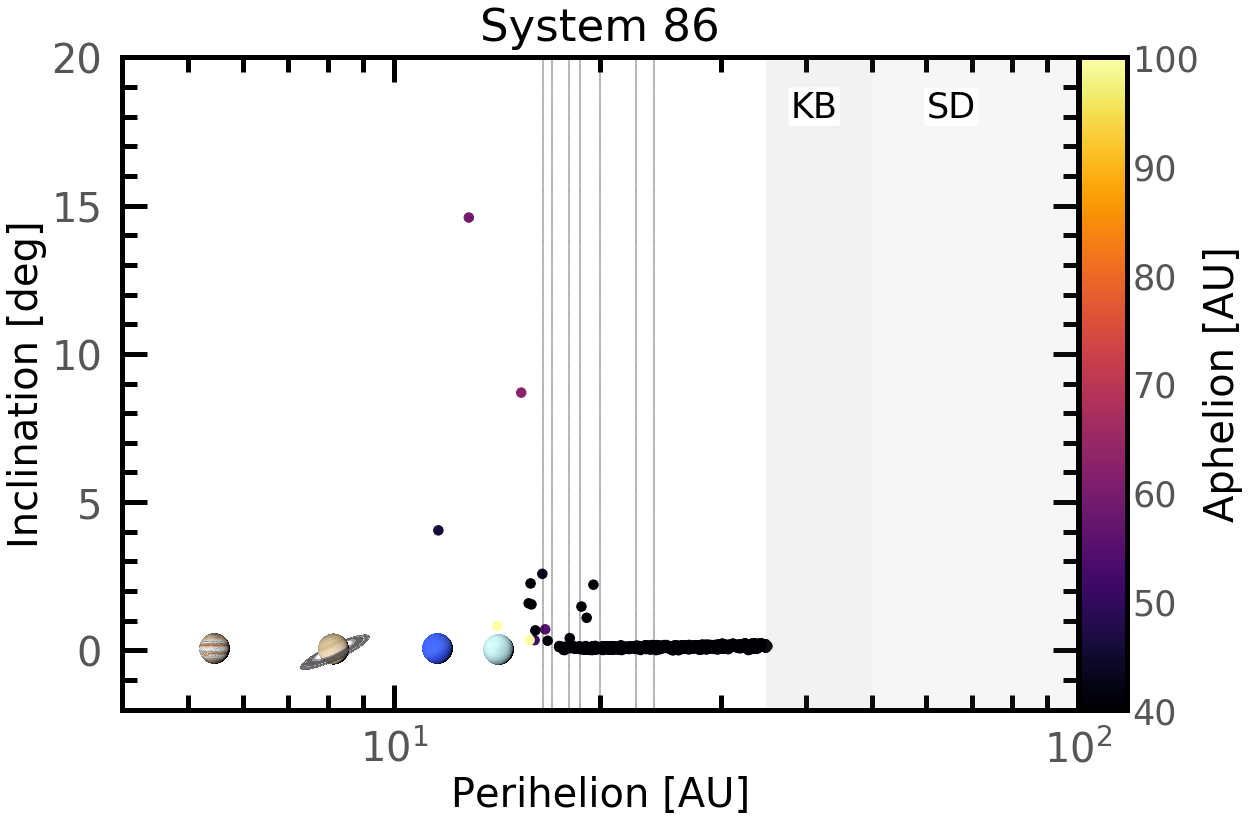}
\includegraphics[width=0.31\textwidth,trim=0 0 1 0,clip]{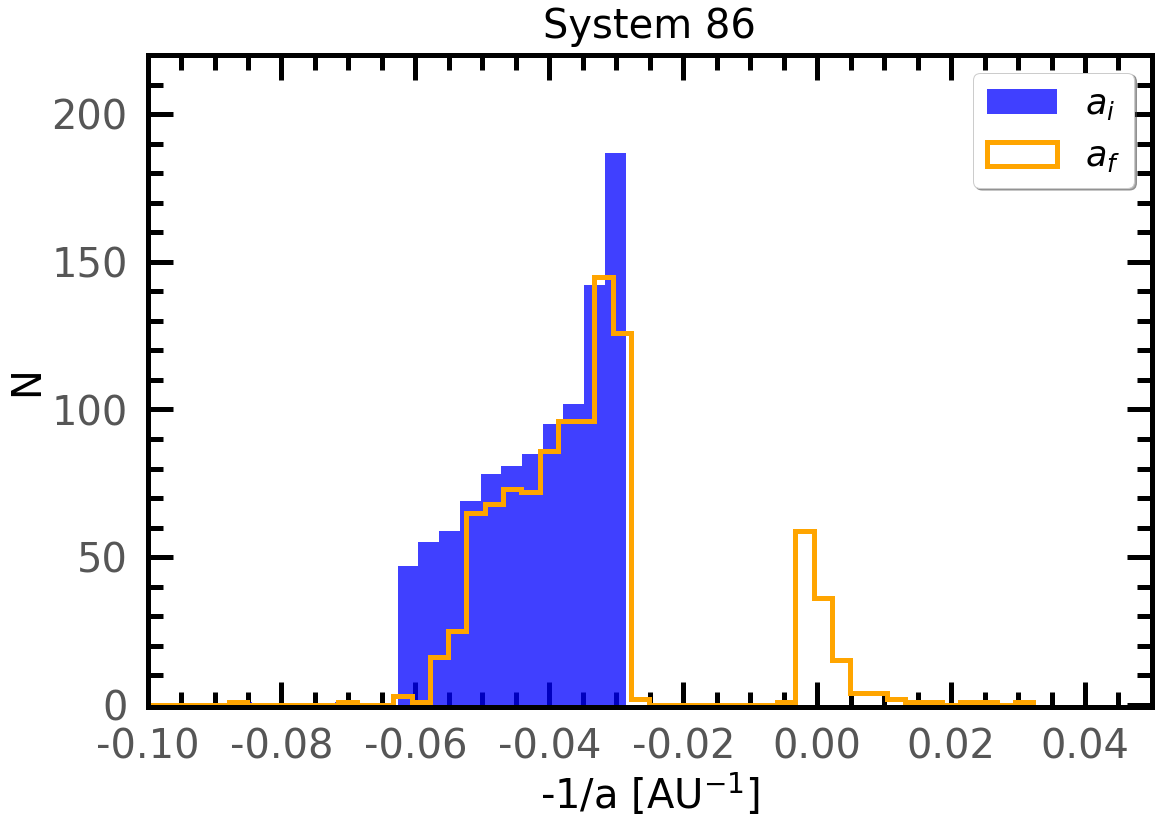}\\

\includegraphics[width=0.31\textwidth,trim=0 0 0 0,clip]{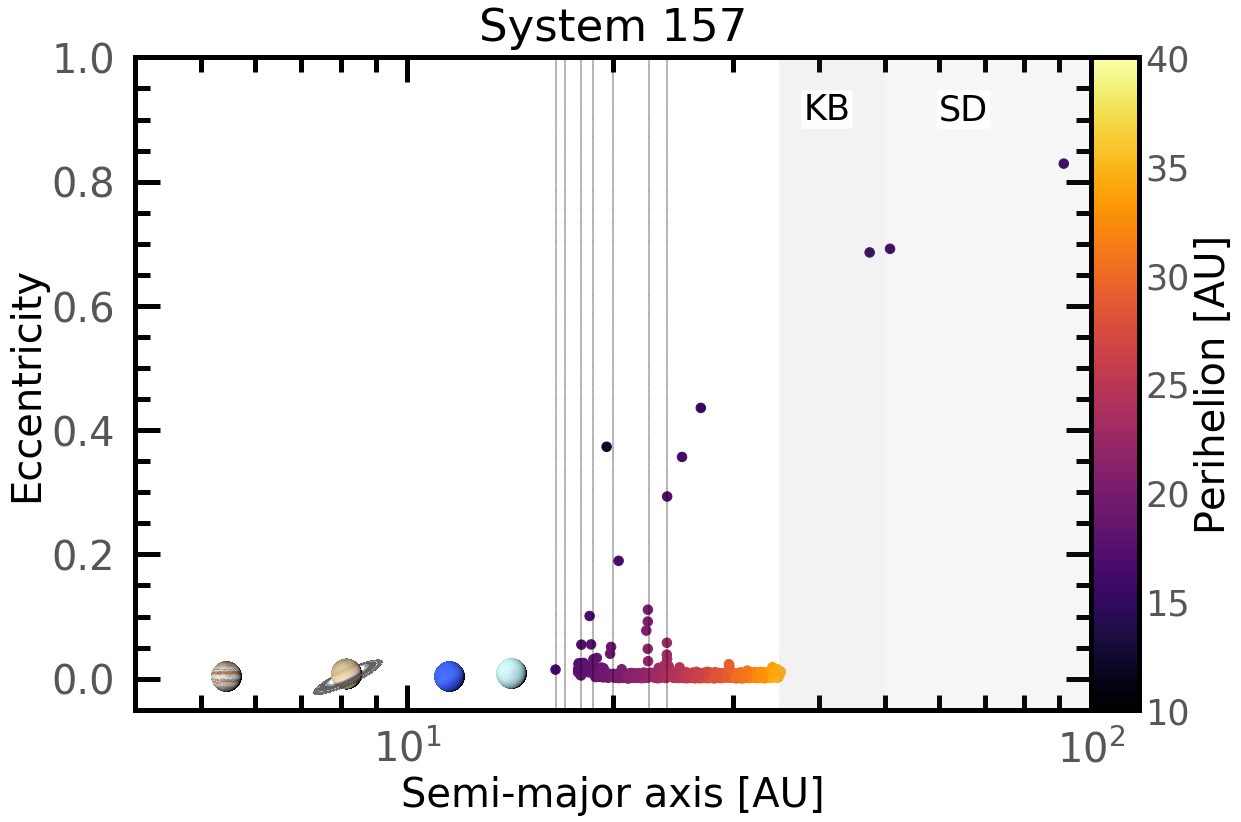}
\includegraphics[width=0.31\textwidth,trim=0 0 0 0,clip]{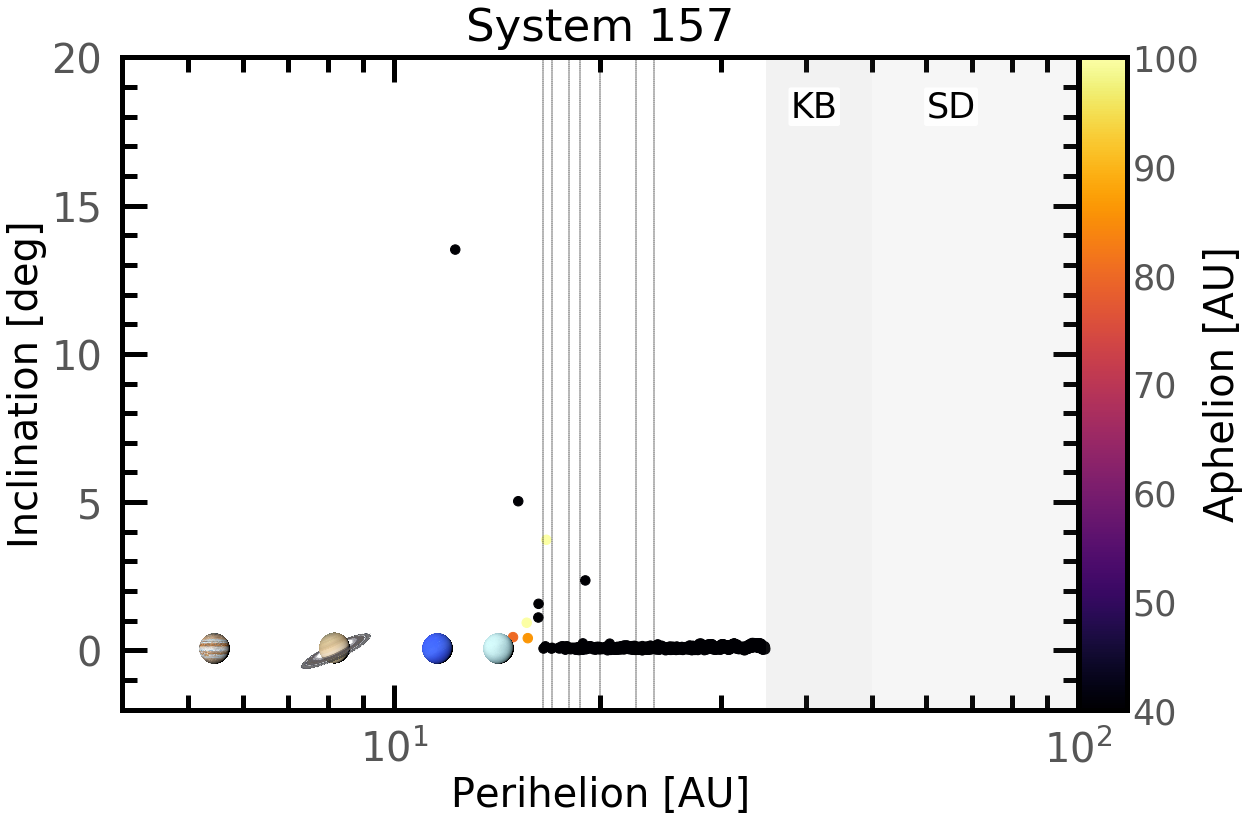}
\includegraphics[width=0.31\textwidth,trim=0 0 0 0,clip]{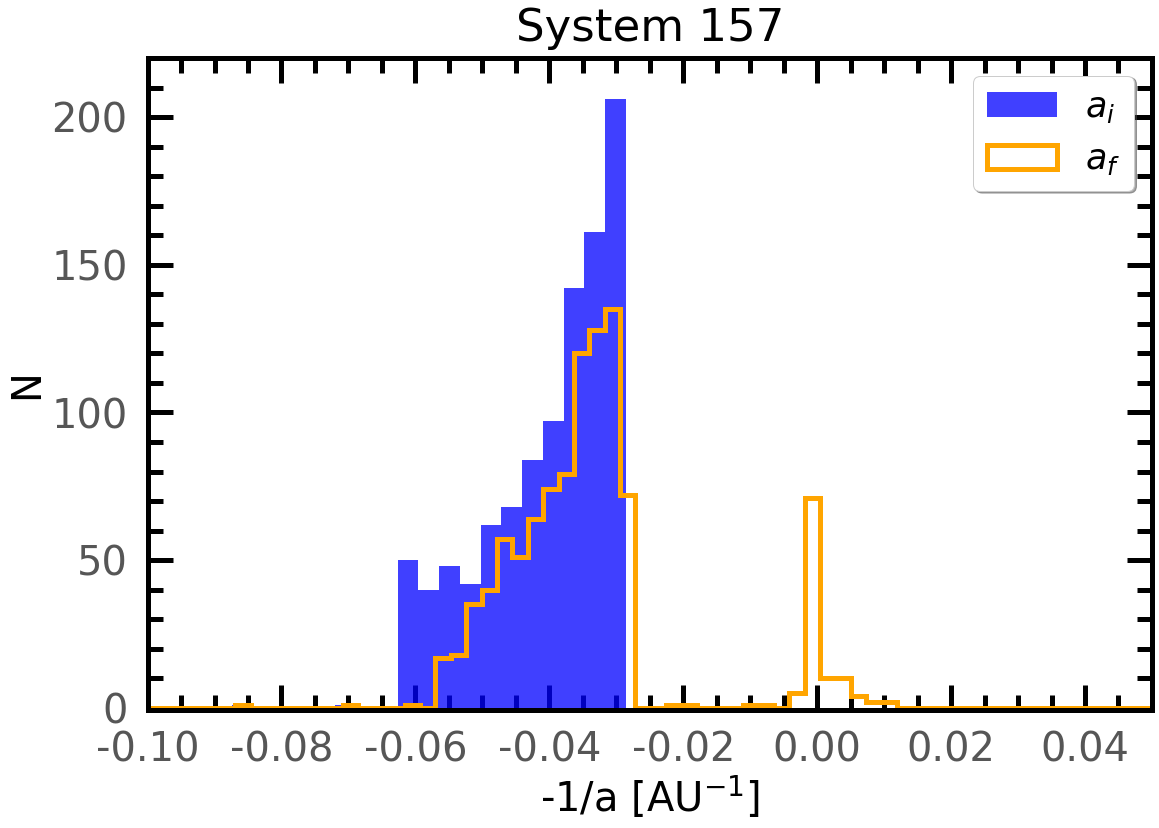}

\caption{\comn\ model. Orbital elements of particles in the disk after 100 Myr and multiple stellar encounters. First column: Semimajor axis as a function of eccentricity. The particles are color-coded by their perihelion distances. The green, blue, and gray shaded regions highlight different populations formed due to stellar encounters. Second column: Perihelion as a function of orbital inclination. The dots are color-coded by aphelion distance. The gray shaded areas represent distinct regions of the Solar System: Kuiper Belt (KB) and  scattered disk (SD). Third column: Distribution of orbital energy for the particles.  The red histograms correspond to the initial energy distribution, and the blue curve represents the final energy distribution. The blue curves with positive values indicate interstellar comets. The rows (from top to bottom) correspond to systems 128, 14, 100, 86, and 157.}
\label{aeiq_nice}
\end{figure*}

\begin{figure*} 

  \includegraphics[width=0.34\textwidth]{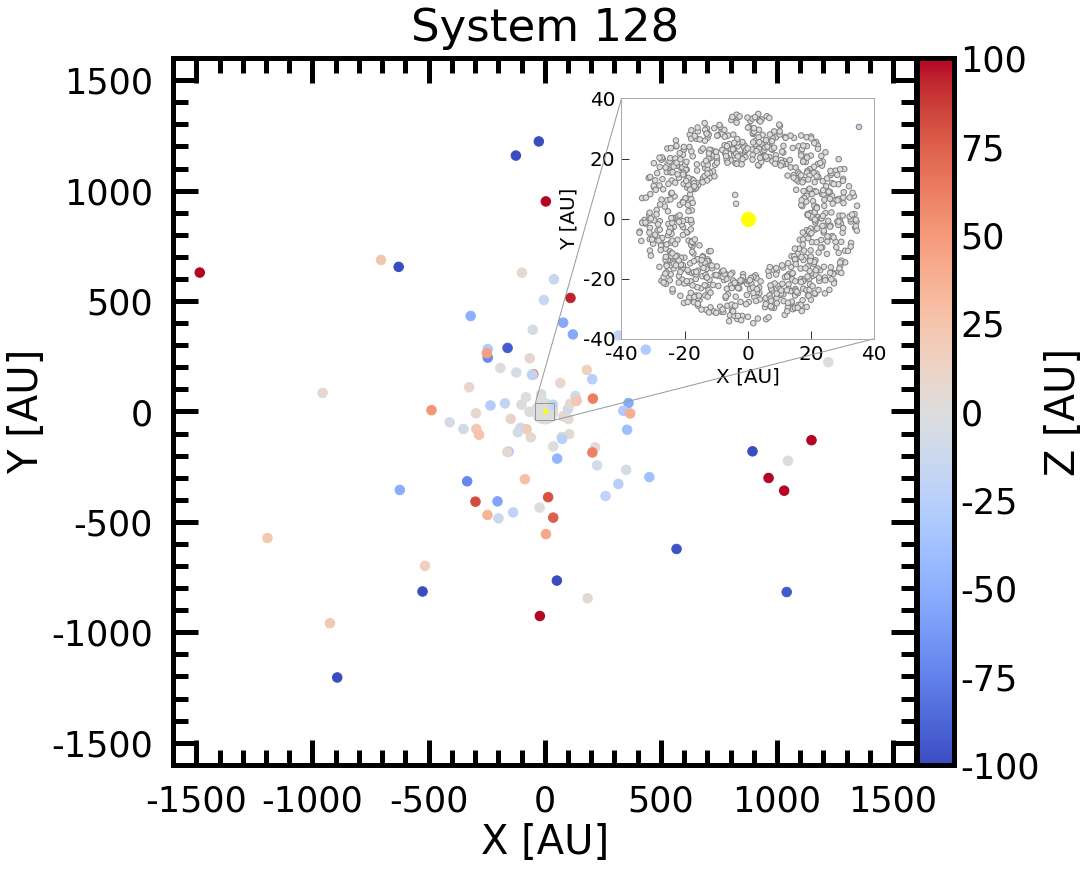}
  \includegraphics[width=0.34\textwidth,trim=0 0 3 0,clip]{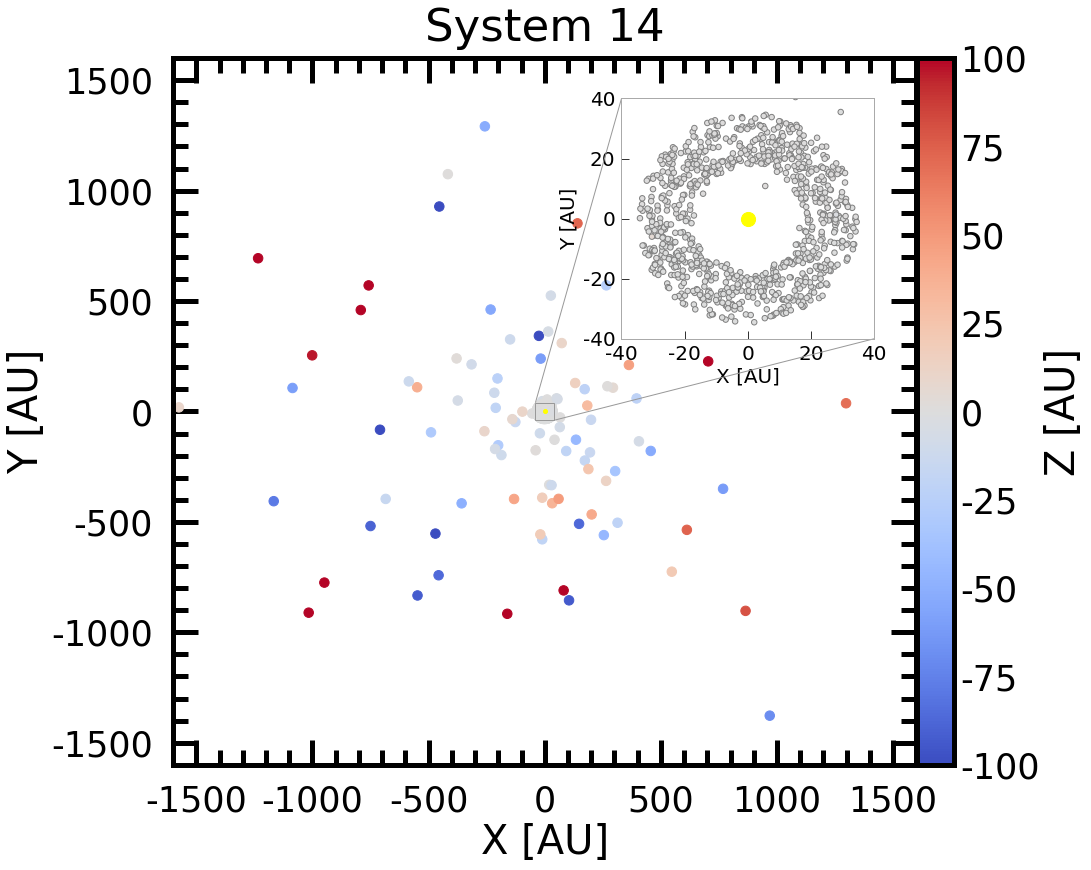}
  \includegraphics[width=0.34\textwidth,trim=0 0 2 0,clip]{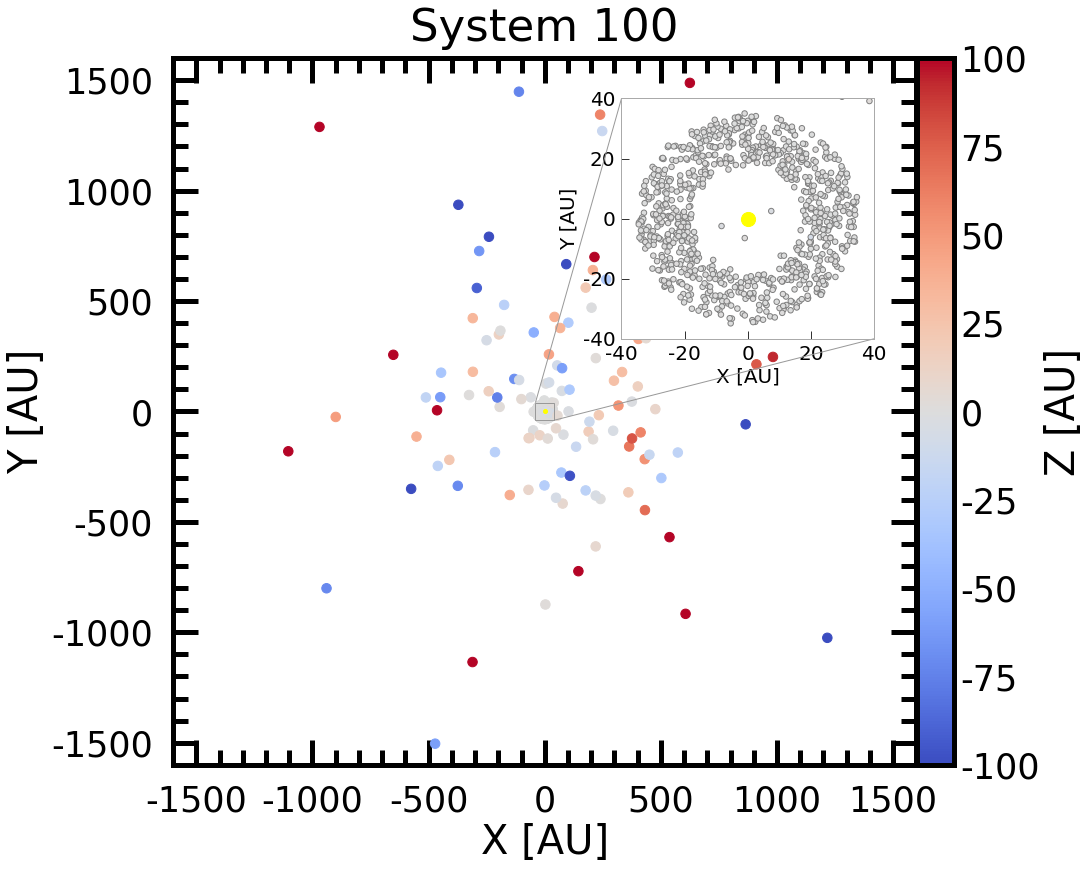}\\
  \includegraphics[width=0.34\textwidth,trim=0 0 1 0,clip]{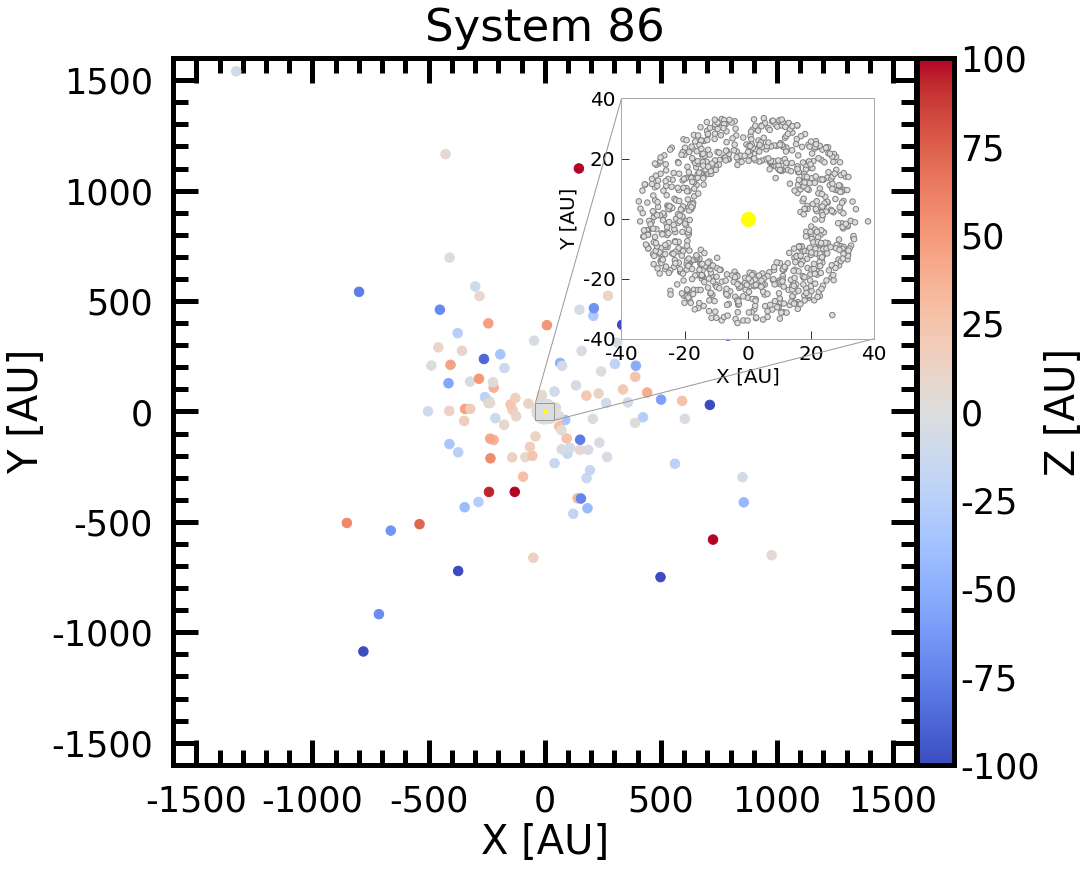}
  \includegraphics[width=0.34\textwidth,trim=0 0 0 0,clip]{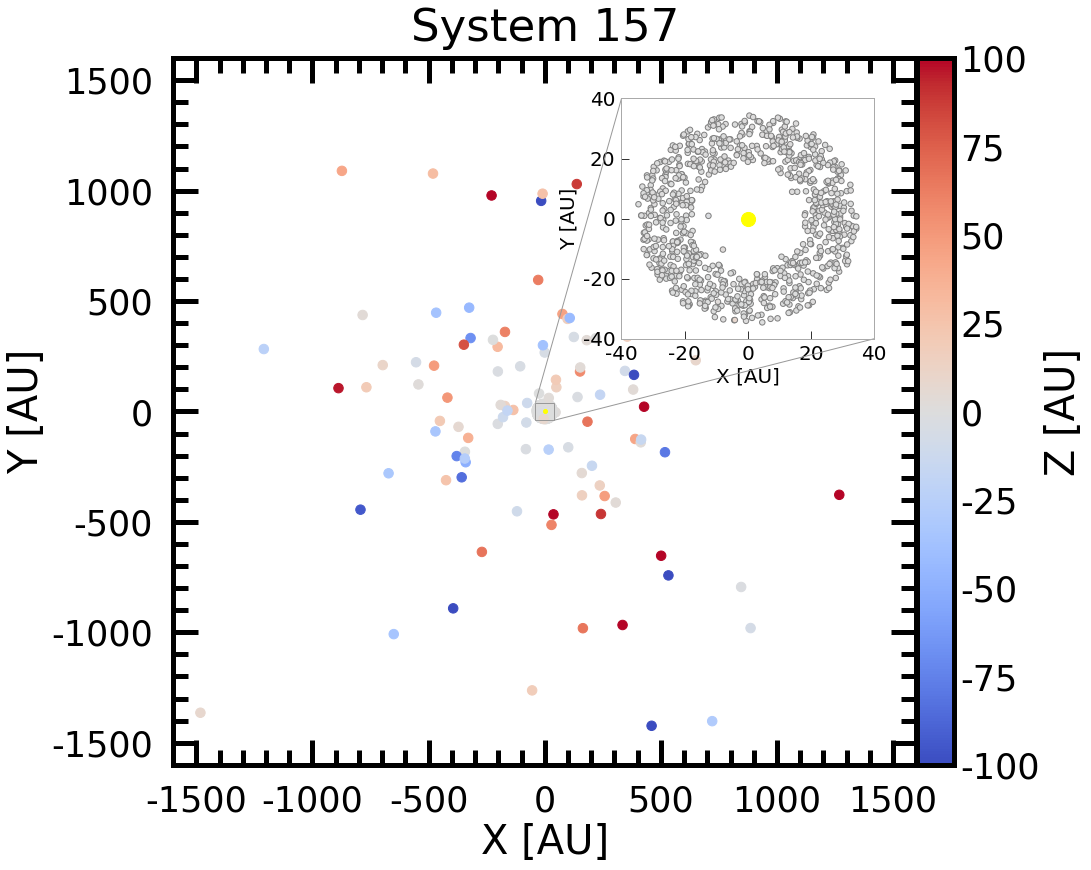}
  \caption{\comn\ model. Phase-space distribution ($X$ vs.\ $Y$) of particles in the disk after $100$~Myr. The color scale indicates the particles' positions along the vertical ($Z$) axis. The panels (from top to bottom) show systems 128, 14, 100, 86, and 157.}
  \label{pos_nice}
\end{figure*}

\subsubsection{\comn}
\label{Nenc_nice}

As described in Sect. \ref{sec2}, we performed a set of 200 simulations for the \comn\ model, set up analogously to the \extn\ models. Each system evolved for 100~Myr, accounting for multiple stellar encounters throughout the evolution. Figure \ref{nice_all_eai_p} shows the final orbital distribution of all particles. Particles with semimajor axes between $15$ and $20$ au are found in mean-motion resonance with Uranus, resembling Kuiper Belt objects in the Solar System. The outer part of the disk experiences only mild perturbations due to the stellar encounters. Approximately 0.8\% of particles are scattered into the Kuiper Belt region, 0.45\% into the scattered disk, and 0.40\% into the Sednoid region. Notably, none of the particles reach the Oort Cloud region. Overall, the strongest perturbations experienced by the particles arise from interactions with the giant planets, particularly Uranus. After 100 Myr, the orbital architecture of the giant planets remains stable. No planetary migration is observed in any of the simulations, primarily due to the distant nature of the encounters. Around 12.5\% of the particles acquire eccentricities between 0.1 and 0.7, while approximately 7\% are ejected from the system, forming interstellar comets.

In \secref{Nenc_ss} we studied the five most interesting cases in our \extn\ simulations. In this section, for comparison, we analyse the same systems, i.e.,\ numbers 128, 14, 100, 86, and 157. In \figrefthree{a_t_nice}{aeiq_nice}{pos_nice} we show the evolution of the orbital elements of the disk particles, the final orbital elements and the energy distribution, and the position of the particles after $100$~Myr, respectively. System 128 (first row in \figref{aeiq_nice}) has an early close encounter ($1000$~au), which in combination with the interaction between the giant planets and the disk and the following encounters leads to the ejection of $6.5$\% of the initial particles. The remaining particles concentrate in the original disk ($95$\%) with eccentricities near zero, while the rest reach the scattered disk region with eccentricities between $0.2$ and $0.6$. System 14 (second row in \figref{aeiq_nice}) does not suffer an important change in the orbital parameters and remains mostly stable during $100$~Myr. A small fraction ($1.5$\%) of particles reach high eccentricities ($0.1< e < 0.6$), and approximately $6$\% are ejected mainly because of the interaction with Uranus. The particles between $16$ and $20$ au enter in mean-motion resonance with Uranus. Systems 100 (third row in \figref{aeiq_nice}) and 86 (fourth row in \figref{aeiq_nice}) retain $94.5$\% of the initial particles.

The most significant perturbation is due to Uranus, which perturbs all the particles between $16$ and $20$~au (as in systems 14 and 128). Such particles become resonant bodies with Uranus. In systems $100$ and $86$, about $5.5$\% of the particles end up on hyperbolic orbits, while the two systems have a similar disk structure after $100$~Myr (\figref{pos_nice}). However, system 100 suffered only one close encounter with a very low-mass star ($0.3$~\mSun) at $273$~au at an early stage ($\sim20$~Myr,), while system 86 faced two close encounters, the first one with a red dwarf star with a mass of $0.5$~\mSun at $488$~au and later on with a brown dwarf ($0.08$~\mSun) at $899$~au. The most perturbed system is $157$. Approximately $7$\% of the particles are ejected from the system, attaining hyperbolic orbits. This system faced two very close encounters at $336$~au and $507$~au, with masses of $0.5$~\mSun and $0.1$~\mSun, respectively. Both encounters
took place at an early stage ($10$ and $27$~Myr). Even though system 157 faced two close encounters, the particles with semimajor axes between $20$ and $36$~au do not suffer an important change. 

The dynamical evolution of the particles in a compact disk depends primarily on the evolution of the giant planets. However, close encounters define their further evolution depending on the mass of the perturber. In the five systems presented here, we examined different scenarios. For stellar encounters, we find that in all the cases, a small fraction of the particles were ejected from the systems ($\sim5$--$7$\%) in hyperbolic objects. The number of objects ejected not only depends on the perturbation due to the planet closest to the disk, but also depends on the proximity of the encounter. This suggests that the SSA faced at least one encounter in its early evolution, which partly determined the evolution of the inner regions. It is essential to highlight that in all our experiments, the inner SSA, particularly the planets, do not get perturbed even when a close encounter occurs. The scenario presented here is a simplification of the classic Nice model, which, however, gives us an idea of the importance of the giant planets in forming and shaping a Kuiper belt-like structure in planetary systems.

\subsection{Extended versus Compact models}
\label{nice_ss}

In Sects. \ref{sec3.1} and \ref{sec3.2} we presented our results for the numerical simulations for the Extended and Compact models, respectively. First, we focused on the effects of a single but close encounter (Sects. \ref{sec3.1.1} and \ref{1enc_nice}) and then studied the effect of multiple encounters (Sects. \ref{Nenc_ss} and \ref{Nenc_nice}) in a star cluster environment, over a time span of $100$~Myr. For the \exto\ model, we found that a close encounter with the present-day planet configuration of the SSA and a disk extending up to 200 au can produce different structures and populations in the disk, but the planets remain in stable orbits. On the other hand, for the \como\ model, a close encounter triggers the migration and excitation of the planets and the debris disk. The original disk gets perturbed, but most of the particles remain in the system. If we compare Figs. \ref{e_ss} and \ref{energy_nice}, we see that the production of unbound particles is higher for the extended case ($5.6$\% of the original disk) than for the compact model ($1.4$\%). It is important to highlight that even when a perturbing star comes close ($\sim300$~au) to a planetary system, the evolution of the planetesimal disk will be determined by its size. If the disk is compact, the perturbations due to the planets will dominate over the effect of the stellar encounter. If the disk is large, the planets will not suffer any orbital variation, while the particles in the disk can be heavily perturbed.

When considering multiple stellar encounters (Figs. \ref{all_eai_p} and  \ref{nice_all_eai_p}) the effect on the disk for the case of the \extn\ model is similar to the case of one encounter. However, the multiple effects of the passing stars create several populations of particles. These structures can be associated with different families in the Solar System, particularly the Sednoids region (\citealt{Jilkova2016a,Pfalzner2018a}). This region ($100 < a < 1000$~au; Fig. \ref{all_eai_p}) is populated with large numbers of particles with eccentricities ranging from $0.2$ to $0.9$ with orbital inclinations from $25$ to $\sim150$ degrees. When we compare this to the \comn\ model, we can see that the disk is barely perturbed due to the passing stars. The major perturbations are because of the interaction with the giant planets, which produces at least six resonances with the particles with semimajor axes between $16$ and $20$~au (Fig. \ref{nice_all_eai_p}). 

Overall, our simulations indicate that stellar encounters, whether single or multiple, distant or close, represent the most efficient mechanism for populating the inner Oort cloud-like region in planetary systems. Such perturbations result in a significant and rapid outward transport of particles from the planetesimal disk, typically on timescales of a few megayears. Conversely, giant planets primarily influence and sculpt the inner small SSA regions over comparatively longer periods, on the order of tens of megayears. Our results further suggest that Kuiper belt-like structures originate predominantly from dynamical instabilities and gravitational interactions with giant planets, in agreement with various realizations of the Nice model \citep[e.g.,][]{Morbidelli2007}. Extending these findings, the overall architecture and orbital configuration of planetary systems are ultimately determined by their birth environment, initial disk size, dynamical evolution, and planetary configurations. Given that planet formation is considered a common process around stars, structures analogous to the Oort cloud and Kuiper belt should be universal \citep[see also][]{Hands2019}. These features are shaped primarily by stellar birth environments, which consequently dictate their long-term stability and survivability. Furthermore, this scenario implies that the ejection of interstellar comets is likely a widespread phenomenon, resulting in interstellar space being populated by numerous objects expelled from their parent planetary systems.

Both the Extended and Compact models show that the disk’s vertical structure is governed primarily by the encounter geometry. Coplanar and moderately inclined flybys ($i\lesssim30^\circ$) efficiently excite eccentricities and semimajor axes while preserving a flattened configuration with inclinations below $\sim40^\circ$. In contrast, high-inclination and polar encounters ($i\gtrsim60^\circ$) inject vertical angular momentum, broadening the inclination distribution and producing thicker, more isotropic remnants. These results explain how early stellar flybys can create dynamically excited yet geometrically thin disks, while repeated encounters and planet–disk interactions progressively isotropize the outer regions and leave the inner disk dynamically cold (see Figs.~\ref{all_eai_p}, \ref{aeiq_ss}, \ref{nice_all_eai_p}, and \ref{aeiq_nice}). This mechanism supports the idea that the inner Oort cloud might have a disk-like origin \citep[e.g.,][]{Fouchard2018,Fouchard2020,Fouchard2023}, while the outer Oort cloud will become isotropic due to repeated Galactic tides and stellar perturbations \citep{PZ_ST_2021}.

\section{Interstellar comets}
\label{sec4}

\begin{figure}[t]
  \includegraphics[width=\columnwidth]{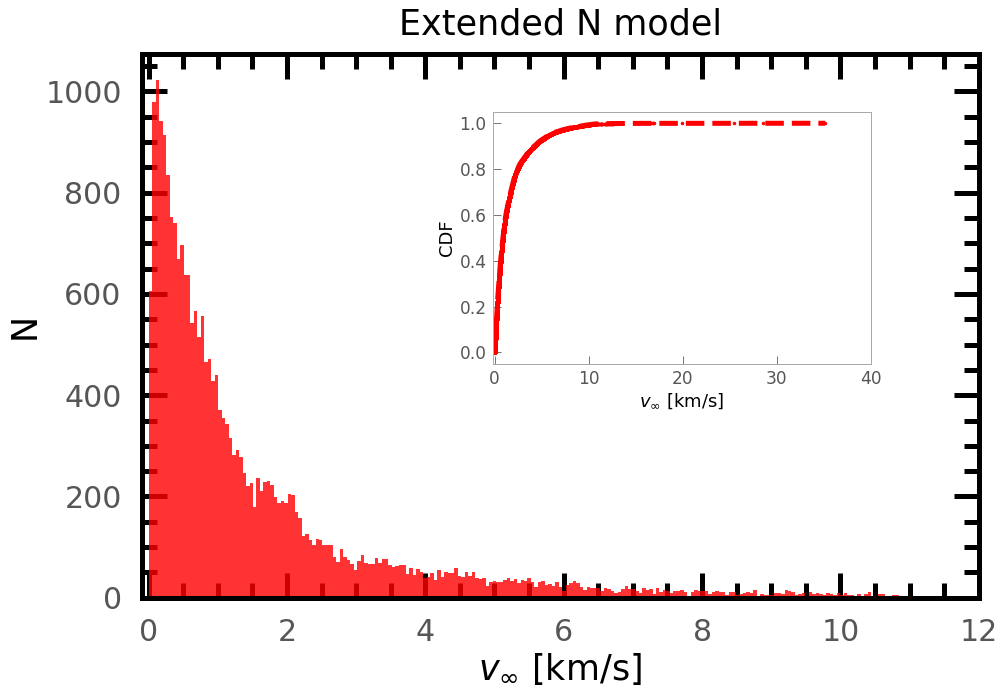}\\
  \includegraphics[width=\columnwidth]{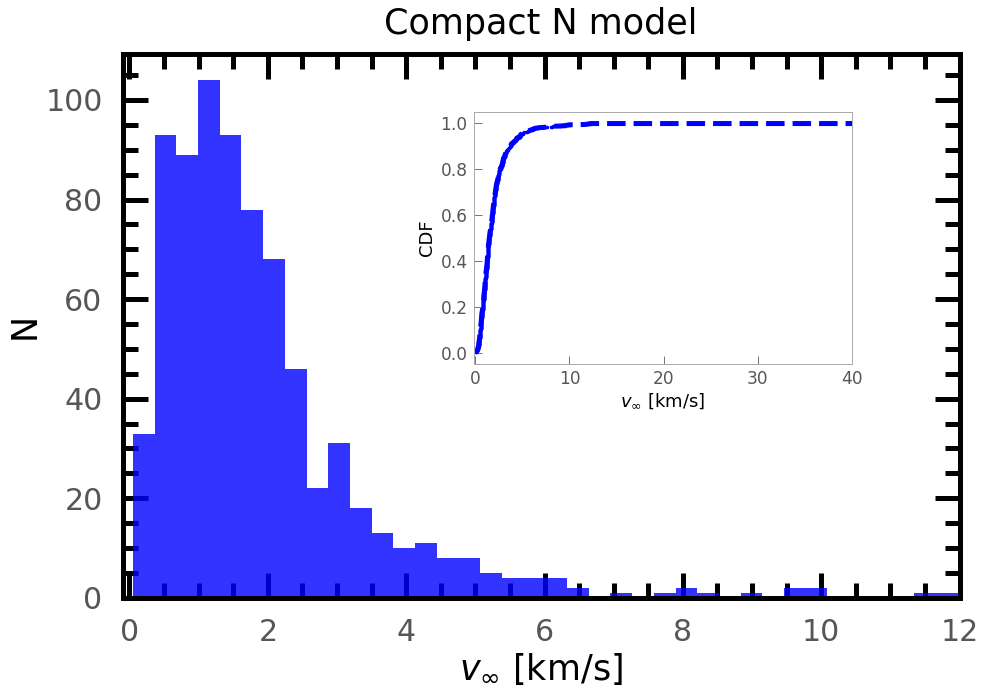}
  \caption{Distribution of hyperbolic excess velocities ($v_{\infty}$) of unbound objects from all simulated systems, comparing the Extended (top panel, red) and Compact (bottom panel, blue) models. The histograms represent the particle distributions, while the insets display their cumulative distributions.}
  \label{ios_ss_nice}
\end{figure}

The discovery of the first interstellar comet, 1I/'Oumuamua \citep[e.g.,][]{Chambers2016,Meech2017}, and the subsequent detections of 2I/Borisov \citep{Borisov2019,Guzik2019} and 3I/Atlas \citep{Seligman2025}, have opened a new era in the exploration and understanding of the role of comets in the evolution of planetary systems. The origin of such ISOs, whether asteroidal or cometary, remains an active area of research, with various formation scenarios and dynamical evolution pathways proposed \citep[see, e.g.,][]{PZ_Torres2018,Torres2019,Hands2019,OumuamuaISSI2019,Pfalzner2021,Jewitt2024,XabiTorres2025}. Among these scenarios, the most extensively studied involve gravitational perturbations caused by massive bodies, such as nearby planets, stellar encounters, or combinations thereof. For example, \cite{Hands2019} demonstrated that planetesimals can escape from stellar clusters with terminal velocities ($v_{\infty}$) on the order of a few~km\,s$^{-1}$, implying the existence of numerous objects similar to 'Oumuamua, traversing interstellar space after being ejected from their birth clusters. Furthermore, \cite{Torres2019} showed that comets located in the outermost regions of planetary systems may become TIOs, objects dynamically perturbed onto interstellar trajectories by secular gravitational interactions with distant passing stars. Such objects can pollute the solar neighborhood and, in some cases, potentially collide with nearby planetary systems \citep{Torres2023}. In this section we analyse the unbound particles produced by planet–disk and stellar–disk interactions in our simulations, focusing on both the \extn\ and \comn\ models described in Sects. \ref{sec3.1} and \ref{sec3.2}, respectively.

Our simulations indicate that the hyperbolic ejection velocities ($v_{\infty}$) of particles escaping from the simulated SSA for the \extn\ and \comn\ models typically range between $1$ and $3\ km/s$ (Fig. \ref{ios_ss_nice}). Particles in the \extn\ model acquire greater kinetic energy due to strong gravitational perturbations caused by passing stars, achieving ejection velocities ranging from $\sim 0.5$ to approximately $40\ km/s$ (inset, top panel of Fig. \ref{ios_ss_nice}). Approximately $36$\% of the total particle population across all 200 simulated systems become interstellar comets (Fig. \ref{ios_ss_nice}). Conversely, in the \comn\ model, planet–disk interactions exert weaker perturbations compared to those caused by stellar encounters, resulting in roughly $7$\% of particles being ejected with velocities between $0.5$ and $40\ km/s$ (inset, bottom panel of Fig. \ref{ios_ss_nice}). Overall, the ejection of asteroid- and comet-like objects into interstellar space within a cluster environment is predominantly driven by gravitational interactions with passing stars, especially when encounter angles are within $0^\circ$ to $30^\circ$ (Fig. \ref{e_ss}). If planetary systems near the disk undergo instability, inducing significant planetary migration via planet–disk interactions, the efficiency of ISO production due to planetary perturbations may surpass that of stellar encounters \citep{Raymond2010}. However, in systems with stable planetary orbits, only a small fraction ($\sim7$\%) of the particles is typically ejected into interstellar space.

The subsequent evolution of interstellar comets will primarily be governed by dynamical interactions with other stars within their birth cluster, and those objects that ultimately escape the cluster will further experience perturbations induced by the Galactic tidal field. These long-term gravitational effects will increase their hyperbolic excess velocities from a few $km/s$ up to tens of $km/s$, similar to the observed velocity of `Oumuamua. Estimates for the local density of ISOs currently vary significantly, ranging from approximately $10^{14}$~pc$^{-3}$ \citep{PZ_Torres2018} to $8\times10^{14}$~pc$^{-3}$ \citep{Jewitt2017}, and up to $2\times10^{15}$~pc$^{-3}$ \citep{Do2018}. Additionally, \cite{OumuamuaISSI2019} estimated the underlying mass density of ISOs, considering scenarios involving planetary instabilities. Their results suggest that the total mass of ISOs ejected due to planet–disk interactions ranges from approximately $0.004$ to $3$~M$_\oplus$,pc$^{-3}$. However, given the current lack of observational data on ISOs, these estimates remain highly uncertain. Nevertheless, the findings presented in these previous studies, as well as those discussed in this work, clearly indicate that planetary systems can eject substantial fractions of their asteroid and comet populations, implying that numerous ISOs populate interstellar space.

We estimate the number of ISOs generated in a single close stellar flyby by combining a size–distribution model of the primordial disk with the ejection fractions $f_{\rm ej}$ measured in our simulations of single encounters with the compact disk (Sect.~\ref{sec3.2} and Appendix~\ref{MS_compact}). The latter are defined as the fraction of test particles that become unbound from the Sun, those with $e>1$ and positive orbital energy, by the end of each simulation. These values represent the per-encounter efficiency of ejection and scale linearly with the disk mass. We assume a differential size distribution,
\begin{equation}
n(D)\,dD = k\,D^{-q}\,dD,
\end{equation}
between diameters $D_{\min}$ and $D_{\max}$, with bulk density $\rho$. Here $k$ is a normalization constant that fixes the absolute number of bodies, determined by the total disk mass. For a collisional cascade $q\simeq3.5$ \citep{Dohnanyi1969}, close to values inferred in debris disks and the Kuiper belt \citep{Wyatt2008,Pan2012}. The total disk mass is
\begin{equation}
M_{\rm d} = \int_{D_{\min}}^{D_{\max}} \frac{\pi\rho}{6}\,D^3\,n(D)\,dD 
          = \frac{\pi\rho}{3}\,k\left(\sqrt{D_{\max}}-\sqrt{D_{\min}}\right),
\end{equation}
where the prefactor $\pi\rho/6$ corresponds to the mass of a single spherical body of diameter $D$. This expression fixes the normalization constant as
\begin{equation}
k= \frac{3M_{\rm d}}{\pi\rho\left(\sqrt{D_{\max}}-\sqrt{D_{\min}}\right)}.
\end{equation}
With this normalization, the cumulative number of bodies larger than $D_0$ is
\begin{equation}
N(>D_0) = \int_{D_0}^{D_{\max}} n(D)\,dD 
        \;\simeq\; \frac{k}{q-1}\,D_0^{\,1-q}.
\end{equation}
This approximation holds for $q>1$ and when $D_{\max}^{\,1-q}\ll D_0^{\,1-q}$, i.e., when the contribution of the largest objects to the integral is negligible compared to that of $D_0$ sized bodies. 

A stellar flyby that ejects a fraction $f_{\rm ej}=N_{\rm ej}/N_{\rm tot}$ therefore produces
\begin{equation}
N_{\rm ISO}(>D_0) = f_{\rm ej}\,N(>D_0).
\label{Nisos}\end{equation}

Adopting values of $q=3.5$, $\rho=10^3~{\rm kg\,m^{-3}}$, $D_{\min}=0.1$~km, 
$D_{\max}=100$~km, and $D_0=1$~km, we obtain
\begin{equation}
N(>1~{\rm km}) \simeq 2.35\times10^{11}
\left(\frac{M_{\rm d}}{M_\oplus}\right),
\end{equation}
so that
\begin{equation}
N_{\rm ISO}(>1~{\rm km}) \simeq 2.35\times10^{11}\,
\left(\frac{M_{\rm d}}{M_\oplus}\right)\,f_{\rm ej}.
\end{equation}

For a $20\,M_\oplus$ primordial disk \citep[e.g.,][]{Gomes2005}, we estimate the number of ISOs produced during a single stellar flyby by scaling the ejection fractions $f_{\rm ej}$ measured from the \como\ model. These values correspond to coplanar, prograde encounters at the specified periapsis distances and thus represent upper limits on the ISO yield from individual encounters.
\[
\begin{aligned}
\text{G-type }(1\,M_\odot,\ f_{\rm ej}=0.024):\;& N_{\rm ISO}\!\sim\!1.1\times10^{11},\quad v_\infty\simeq2.5~{\rm km\,s^{-1}},\\
\text{A-type }(2\,M_\odot,\ f_{\rm ej}=0.025):\;& N_{\rm ISO}\!\sim\!1.2\times10^{11},\quad v_\infty\simeq5~{\rm km\,s^{-1}},\\
\text{B-type }(5\,M_\odot,\ f_{\rm ej}=0.201):\;& N_{\rm ISO}\!\sim\!9.5\times10^{11},\quad v_\infty\simeq7~{\rm km\,s^{-1}},\\
\text{B-type }(9\,M_\odot,\ f_{\rm ej}=0.409):\;& N_{\rm ISO}\!\sim\!1.9\times10^{12},\quad v_\infty\simeq9~{\rm km\,s^{-1}}.
\end{aligned}
\]
The resulting asymptotic velocities are comparable to the internal velocity dispersions of young stellar clusters ($v_\infty\simeq2$--10~km\,s$^{-1}$). This implies that the ejected particles are rapidly mixed into the local stellar environment and contribute efficiently to the interstellar reservoir. Because $N_{\rm ISO}\propto M_{\rm d}\,f_{\rm ej}$, rare but massive perturbers dominate early ISO production by ejecting orders of magnitude more material than solar-type encounters acting on disks of similar mass. Each value of $N_{\rm ISO}$ in our examples corresponds to the total number of $\gtrsim1$~km bodies ejected during a single stellar flyby, assuming a $20\,M_\oplus$ primordial solids disk. For instance, a solar-type perturber ($1\,M_\odot$) expels $\sim10^{11}$ objects per encounter, whereas a $9\,M_\odot$ B-type star can unbind nearly $2\times10^{12}$ bodies in a single pass. These encounters imply that even a few close interactions during a cluster’s lifetime can substantially enrich the local interstellar medium with cometary debris.

Although in the \como\ model the inferred ISO yields per encounter and stellar type are large, they correspond to the total number of kilometer-scale objects ejected during a \emph{single, very close} stellar flyby ($q_\star \lesssim 300$~au). In our simulations, such encounters probe the inner regions or outer edge of the primordial planetesimal disk (see Table~\ref{table1}, \como\ model) and occur during the early cluster phase, when disk masses and object numbers are significantly higher than in a fully formed Oort cloud. By contrast, the present-day Oort cloud represents only a small residual fraction of the original planetesimal disk population \citep{Francis2005}. As a result, the number of ISOs produced in an Oort cloud-like structure is expected to be significantly lower. Consequently, the production of ISOs depends sensitively on the encounter impact parameter, the density distribution of particles, and whether the source population is disk-like (early phases) or Oort cloud-like \citep[late phases; e.g.,][]{Engelhardt2017,Pfalzner2021}.

Our results are consistent with theoretical expectations that close encounters and planet–planet scattering efficiently eject particles into interstellar space \citep[e.g.,][]{Raymond2018,Moro-Mart2018,Torres2019,PortegiesZwart2021a}. Although low-mass stars dominate the IMF, their flybys contribute comparatively little. In contrast, encounters with $5$--$9\,M_\odot$ stars, though rare, eject orders of magnitude more bodies, and may thus represent a major source of the first generation of interstellar comets in dense birth clusters

\section{Summary and conclusions}
\label{summary}

In this work, we investigated the dynamical evolution of a Solar System analogue (SSA), focusing on the gravitational interactions among four giant planets, a non-self-gravitating particle disk, and passing stars within a dense stellar environment (Sects.~\ref{sec3.1} and \ref{sec3.2}). Our goal was to understand how stellar perturbations reshape the architecture of outer planetary systems, driving the formation of populations analogous to those observed in the outskirts of the Solar System, such as resonant objects, Sednoids, Kuiper Belt and scattered disk objects, Oort cloud-like comets, and ISOs. We explored two disk models: a Compact and an Extended disk, each subjected to either a single close encounter or a sequence of multiple stellar flybys, following the initial conditions summarized in Table~\ref{table1}. The main results of this work are summarized below.

\begin{itemize}

\item Disk evolution and cluster environment.  
In both the Compact and Extended models (Sects.~\ref{sec3.1} and~\ref{sec3.2}), the morphology and dynamical heating of the particle disk depend on the encounter geometry and the mass of the perturber. Repeated close encounters with low-mass stars (K and M-type stars) efficiently reproduce structures analogous to the Kuiper belt, scattered disk, and inner Oort cloud (Figs.~\ref{all_eai_p} and~\ref{nice_all_eai_p}). These cumulative perturbations drive progressive disk heating and outward migration, raising perihelia and populating long-lived eccentric orbits. Compared to single close encounters, which can strongly reshape or even truncate a disk in one event, the cumulative effect of weaker encounters leads to a gradual redistribution of material rather than its complete removal. As a result, multiple low-mass encounters collectively build up extended scattered populations while preserving the inner regions of the planetary system. For encounters with massive stars, compact disks are comparatively resilient. A-type stars eject only a few per cent of particles, whereas B-type stars truncate the disk at $\sim$20--25~au and excite large eccentricities and inclinations (Fig. \ref {fig:hist_final}). Extended disks, however, are far more fragile; even $2\,M_\odot$ perturbers remove over half the population, and $5$--$9\,M_\odot$ stars can strip 60--75\% of bodies while seeding Oort-like orbits (Fig. \ref{fig:hist_kde_AB5_ext}).

\item Encounter geometry and disk populations.
We find that the efficiency of disk disruption is strongly geometry-dependent. Coplanar, prograde encounters maximize the transfer of angular momentum and energy to disk particles, whereas inclined or polar encounters are comparatively inefficient (Fig. \ref{pss}). In the closest encounter we studied (\como\ model), a solar-type perturber at $q_\star\simeq75$~au ejects $\sim$2.4\% of particles for $i=0^\circ$, giving rise to three distinct dynamical families: a scattered population with low Tisserand parameters ($T_{\rm U}\lesssim2.5$) and small perihelia; a scattered tail near the Uranus-scattering boundary ($2.5\lesssim T_{\rm U}\lesssim3.0$); and a dominant lower tail that remains dynamically cold at larger perihelia. At higher inclinations ($i=60^\circ$--$90^\circ$), the scattered population nearly disappears and most of the disk survives only mildly perturbed, confirming that the three-branch morphology is a hallmark of prograde or moderately inclined encounters. A similar behavior is found in the \exto\ models (Sect.~\ref{sec3.1.1}). In coplanar encounters, the disk remains largely flattened, whereas polar flybys redistribute angular momentum vertically, producing the isotropic outer populations of an emerging Oort cloud. These results provide a framework for interpreting the architecture of the outer Solar System. When the Sun was still embedded in its birth cluster, repeated stellar encounters with low-mass stars likely drove the outward migration and eccentricity excitation of bodies in its primordial disk, thereby populating the outer regions of the Solar System \citep[see, e.g.,][]{PZ_ST_2021}. Particles scattered onto sufficiently large semimajor axes become susceptible to Galactic tides, which gradually circularize their orbits, while those at smaller $a$ remain on eccentric, long-lived trajectories or become dynamically unstable on secular timescales \citep[e.g.,][]{Heisler1986a,Fouchard2006,Dones2015,Nesvorny2018}. The tail-like structure observed in semimajor axis–eccentricity space (Figs. \ref{pss_ss},~\ref{aeiq_ss},~\ref{pss},~and~\ref{nice_all_eai_p}) is consistent with this picture, linking the dynamical families formed in our simulations to the seeds of the observed populations of the outer Solar System in its early stages of evolution.

\item Implications for the formation of interstellar objects in dense environments. 
The unbound populations produced in our simulations exhibit characteristic hyperbolic velocities of $v_\infty\!\sim\!1$--$3~\mathrm{km\,s^{-1}}$, with a high-velocity tail extending up to $\sim40~\mathrm{km\,s^{-1}}$ across both disk models (Fig. \ref{ios_ss_nice}). These velocities are comparable to the velocity dispersion of stars in open clusters, implying that ejected bodies are rapidly mixed into the local stellar environment. By coupling our kinematic results with the measured ejection fractions from our single-encounter compact-disk simulations (Sect.~\ref{1enc_nice}), we estimated the production of kilometre-sized ISOs during individual stellar flybys (Eq.~\ref{Nisos}). For a $20\,M_\oplus$ primordial disk, a solar-type encounter ejects about $10^{10}$ objects per flyby, whereas A- and B-type stars can unbind $10^{11}$--$10^{12}$ bodies at typical asymptotic velocities of 2--10~km\,s$^{-1}$.

Although these numbers appear large, they represent the total number of bodies ejected from a single planetary system and are therefore strongly diluted when distributed over parsec-scale cluster volumes. The Galactic ISO population must instead arise from the cumulative contribution of planetary systems, mainly from dense stellar environments, where the probability of close encounters is high. Over time, the ISOs will gradually blend into the interstellar medium as clusters dissolve and stars exchange material through mutual encounters \citep{Jilkova2016a}. Because $N_{\mathrm{ISO}}$ scales directly with disk mass and ejection efficiency, massive stars, though rare, dominate the overall ISO production, while the more frequent flybys of low-mass stars provide a steady but modest background. These results highlight stellar flybys in dense environments as an efficient mechanism for enriching the Galaxy with ISOs.
\end{itemize}

Our results highlight the key role of stellar encounters in dense environments in shaping the early architecture of planetary systems, seeding distinct dynamical families, and producing ISOs rather than truncating them. This work provides a link between the internal evolution of planetary systems and their external environment. Future studies combining high-resolution $N$-body simulations with cluster-scale dynamics, stellar evolution, and Galactic dynamics models will be essential for tracing the long-term evolution of SSAs and quantifying how the population of ISOs in dense environments enriches the Galaxy.

\begin{acknowledgements}

We thank the referee for their suggestions and comments, which helped us improve the quality and clarity of the paper. ST thanks Ylva Götberg, Maxwell Cai, Diptajyoti Mukherjee, Simon Portegies Zwart, and Anthony Brown for their valuable feedback and comments. ST acknowledges the funding from the European Union’s Horizon 2020 research and innovation program under the Marie Sk\l{}odowska-Curie grant agreement No 101034413.
\end{acknowledgements}

\bibliographystyle{aa}
\bibliography{biblio_AA}

\begin{appendix}

\section{\textbf{The effect of A- and B-type stars}}
\label{MS_ss}

In Sect.~\ref{sec3} we examined the dynamical evolution of the SSA under full cluster dynamics. Because of the adopted IMF (see Sect.~\ref{sec2}), the vast majority of encounters involved low-mass stars. However, previous studies have shown that massive perturbers can strongly affect the evolution and morphology of planetary systems, and in particular Oort cloud-like structures \citep[e.g.,][]{fouchard2011}. To complement the cluster-based results, we perform dedicated single-encounter simulations with A- and B-type stars, designed to probe the impact of these rare but potentially encounters.

We modeled encounters with stars of $2\,M_\odot$, $5\,M_\odot$, and $9\,M_\odot$ for both the Compact (Appendix~\ref{MS_compact}) and Extended (Appendix~\ref{MS_extended}) disk configurations. All simulations adopt coplanar, prograde geometries ($i=0^\circ$), which maximize the gravitational interaction between the perturber and the disk (Sect.~\ref{sec3}). For consistency with our earlier analysis, we use pericenter distances of 90~au for the \como\ model (Sect.~\ref{1enc_nice}) and 300~au for the \exto\ model (Sect.~\ref{sec3.1.1}).

\subsection{The effect of A- and B-type stars in compact disks}\label{MS_compact}

\begin{figure}
  \includegraphics[width=\columnwidth]{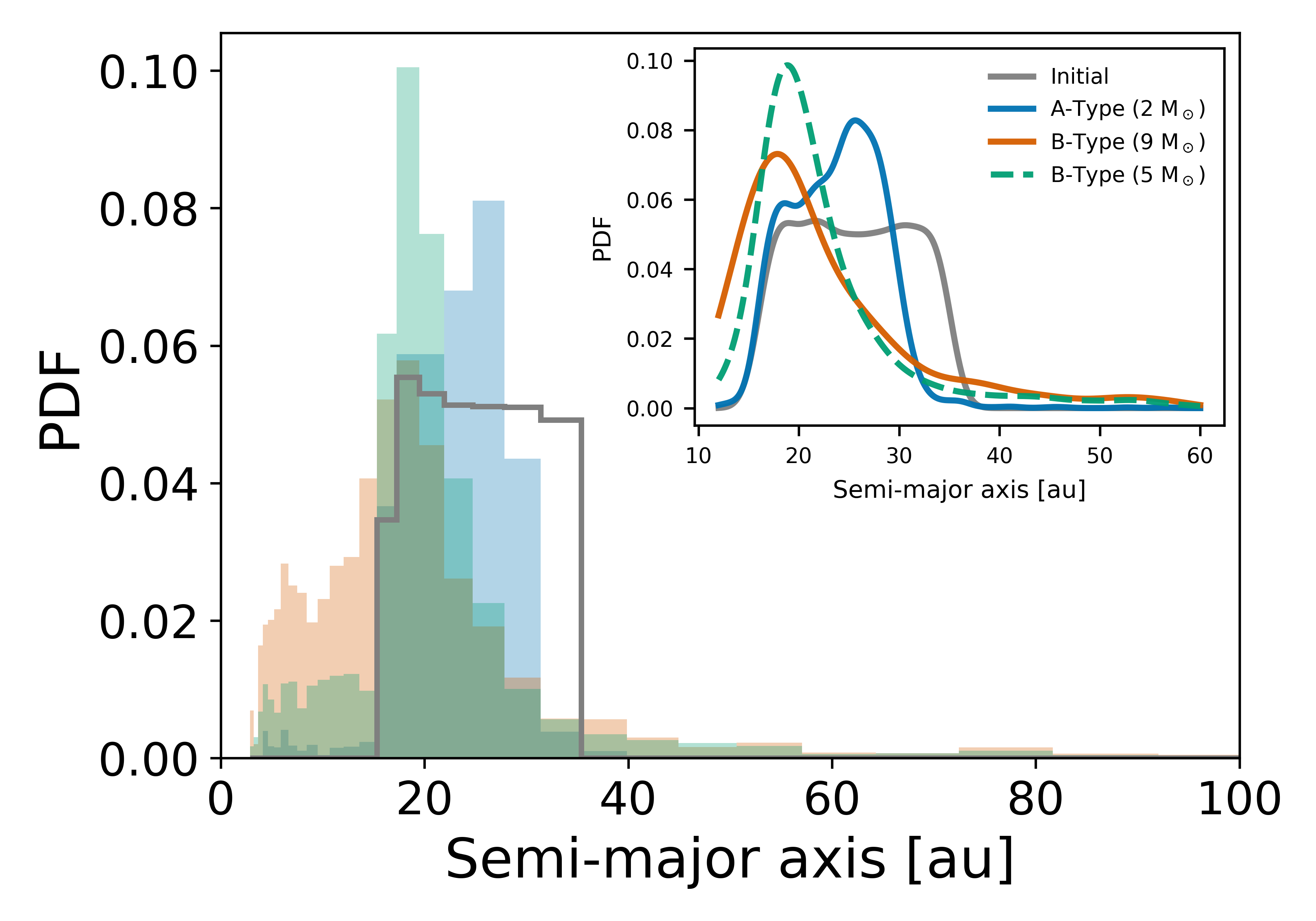}\\
  \includegraphics[width=\columnwidth]{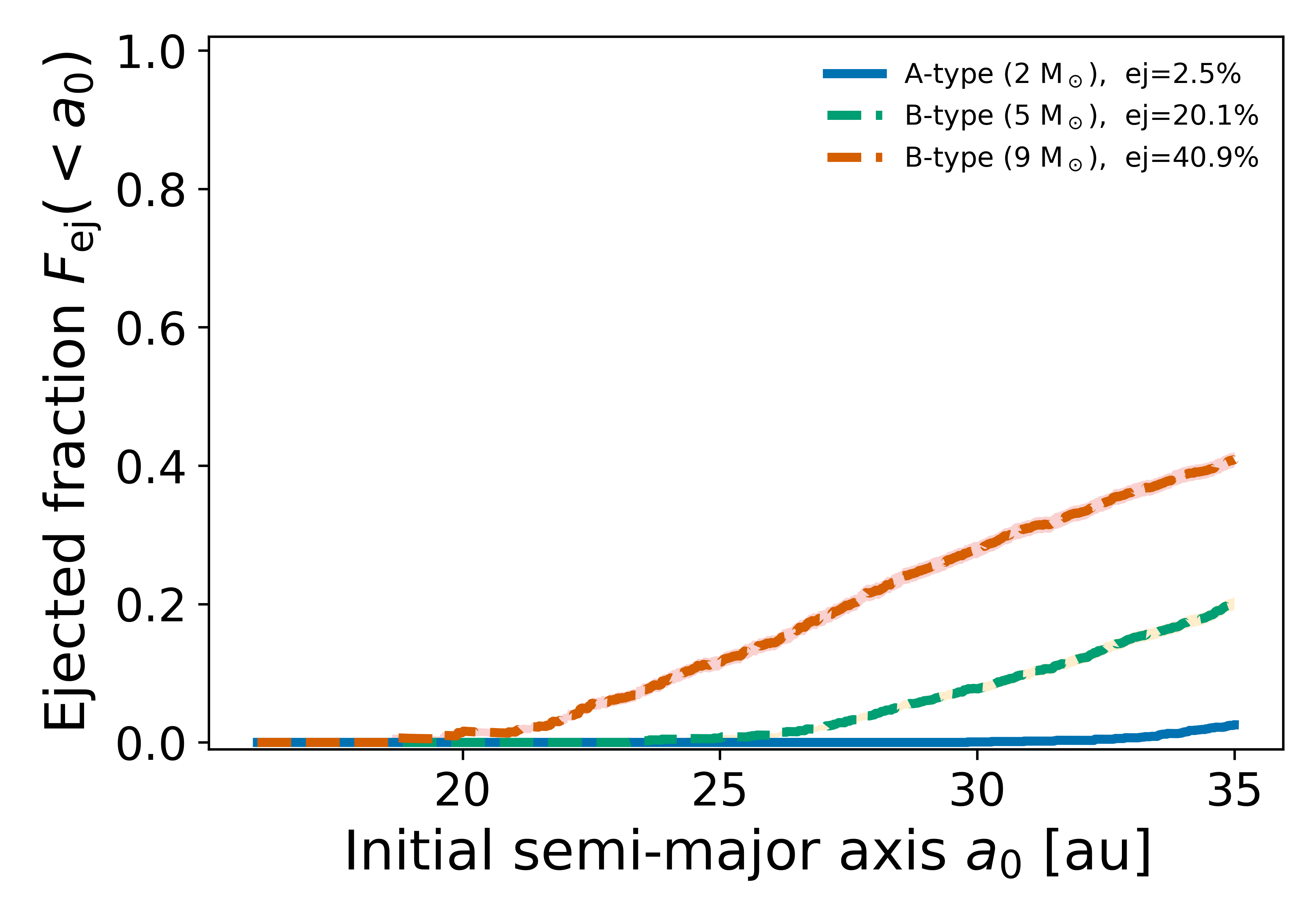}
  \caption{{Top:} Probability density function vs. semimajor axis for bound particles after a single stellar flyby. The histograms are shown for encounters with an A-type star ($2\,M_\odot$, blue), a B-type star ($5\,M_\odot$, green dashed and $9\,M_\odot$, orange), compared with the initial disk (gray). The inset shows kernel density estimates (KDEs). {Bottom:} Cumulative ejection fraction as a function of the initial semimajor axis $a_0$ for the same stellar flybys at a pericenter distance of $q_\star=90$~au. }
  \label{fig:hist_final}
\end{figure}

In Fig. \ref{fig:hist_final} we show the outcome of single encounters with A- and B-type stars for the \como model. This marks the threshold where the $2\,M_\odot$ A-type star begins to noticeably perturb the disk, ejecting $\simeq2.4\%$ of particles. At larger pericenters, the A-type has little effect, whereas B-type stars at the same distance remove up to $\sim40\%$ of the population. For comparison, at $q_\star=100$--140~au show negligible excitation, while a $1\,M_\odot$ perturber at 60~au ejects only $\sim90$ particles. Thus, the 90~au case provides a natural reference point for the onset of significant disruption for A- and B-type stars.

As expected, the A-type case ($2\,M_\odot$) is the least disruptive. Around $97.5\%$ of the $2000$ particles remain bound, with only $2.5\%$ ejected. The particles in the disk cluster around $a_{\rm med}=24.1$~au (18.4--28.3~au), with small eccentricities ($e_{\rm med}=0.15$) and nearly circular inclinations. About 10\% reach $e>0.5$, $<0.3\%$ exceed $i>30^\circ$, and only a few are ejected beyond the primordial disk ($\sim2\%$ to the scattered disk, $0.3\%$ to the inner Oort Cloud). The most distant particles lie at $a\simeq2300$~au, while ejected bodies acquire velocities of $v_\infty\simeq5$~km\,s$^{-1}$.

The $5\,M_\odot$ encounter is significantly more erosive. $20\%$ of the disk is ejected, with survivors concentrated inside $\sim25$~au. The median orbital elements are $a_{\rm med}=19.9$~au (16--84th range 16.4--28.3~au) and $e_{\rm med}=0.31$, with one third of particles reaching $e>0.5$. Inclination excitation remains low ($i_{\rm med}\simeq0^\circ$, $\sim1\%$ with $i>30^\circ$). About 9.5\% of survivors reach the scattered disk and 1.4\% the inner Oort Cloud, while the most distant particle extends to $\sim4500$~au. The ejected population leaves with $v_\infty\simeq7$~km\,s$^{-1}$.

The $9\,M_\odot$ B-type perturber is the most destructive as expected, removing $41\%$ of the disk and producing a sharp truncation at $a\simeq24$~au. Bound particles are tightly concentrated around 15--20~au, with $a_{\rm med}=18.5$~au (10.7--33.0~au) and $e_{\rm med}=0.45$. Nearly half of them have $e>0.5$, $2.7\%$ exceed $i>30^\circ$, and $\sim1\%$ flip to retrograde orbits. The scattered disk and inner Oort Cloud contain 12.5\% and 2.2\% of the particles, respectively, and the most extreme object reaches $a\simeq1.1\times10^5$~au. The ejected bodies escape with $v_\infty\simeq9.3$~km\,s$^{-1}$.

The cumulative ejection fractions $F_{\mathrm{ej}}(a_0)$ shown in the bottom panel of Fig. \ref{fig:hist_final} confirm that B-type encounters produce systematically stronger stripping at all $a_0$. The A-type flyby predominantly heats the outer disk ($a_0\gtrsim25$~au), while the B-type already perturbs material near the inner edge. Despite this, all four giant planets remain bound and nearly circular, underscoring the resilience of the planetary region in encounters with massive stars.

Overall, these results demonstrate the steep dependence of encounter strength on stellar mass. While the A-type flyby leaves the disk largely intact for a very close encounter (90 au), B-type encounters both erode and dynamically excite the population, truncating the disk and seeding long-period orbits. Although only a small fraction of particles reach Oort-like distances in single encounters, such events may nonetheless provide an efficient pathway for populating the early scattered disk and inner Oort Cloud in dense birth environments.

\subsection{The effect of A- and B-type stars in
extended disks}
\label{MS_extended}

\begin{figure}
  \includegraphics[width=\columnwidth]{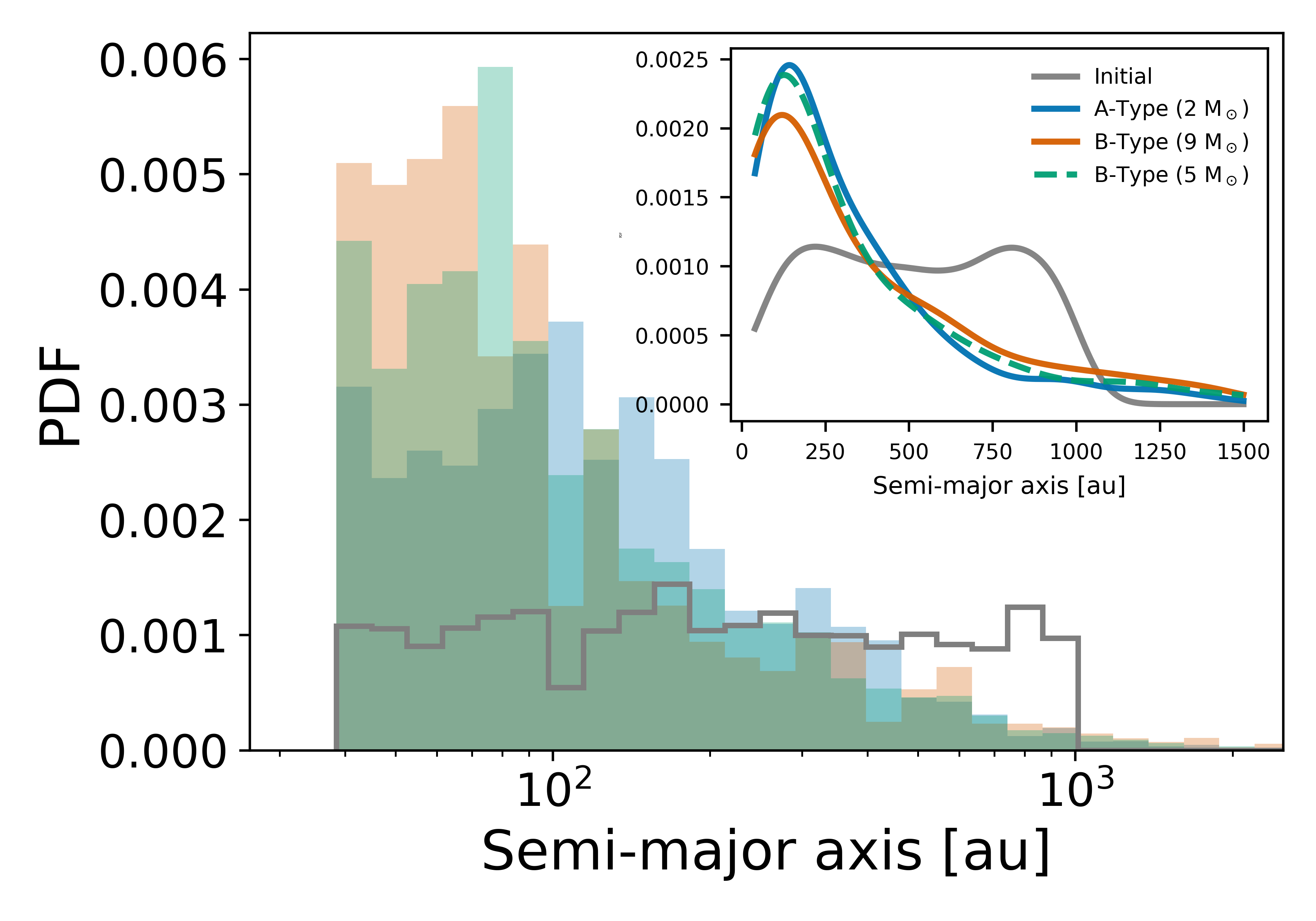}\\
  \includegraphics[width=\columnwidth]{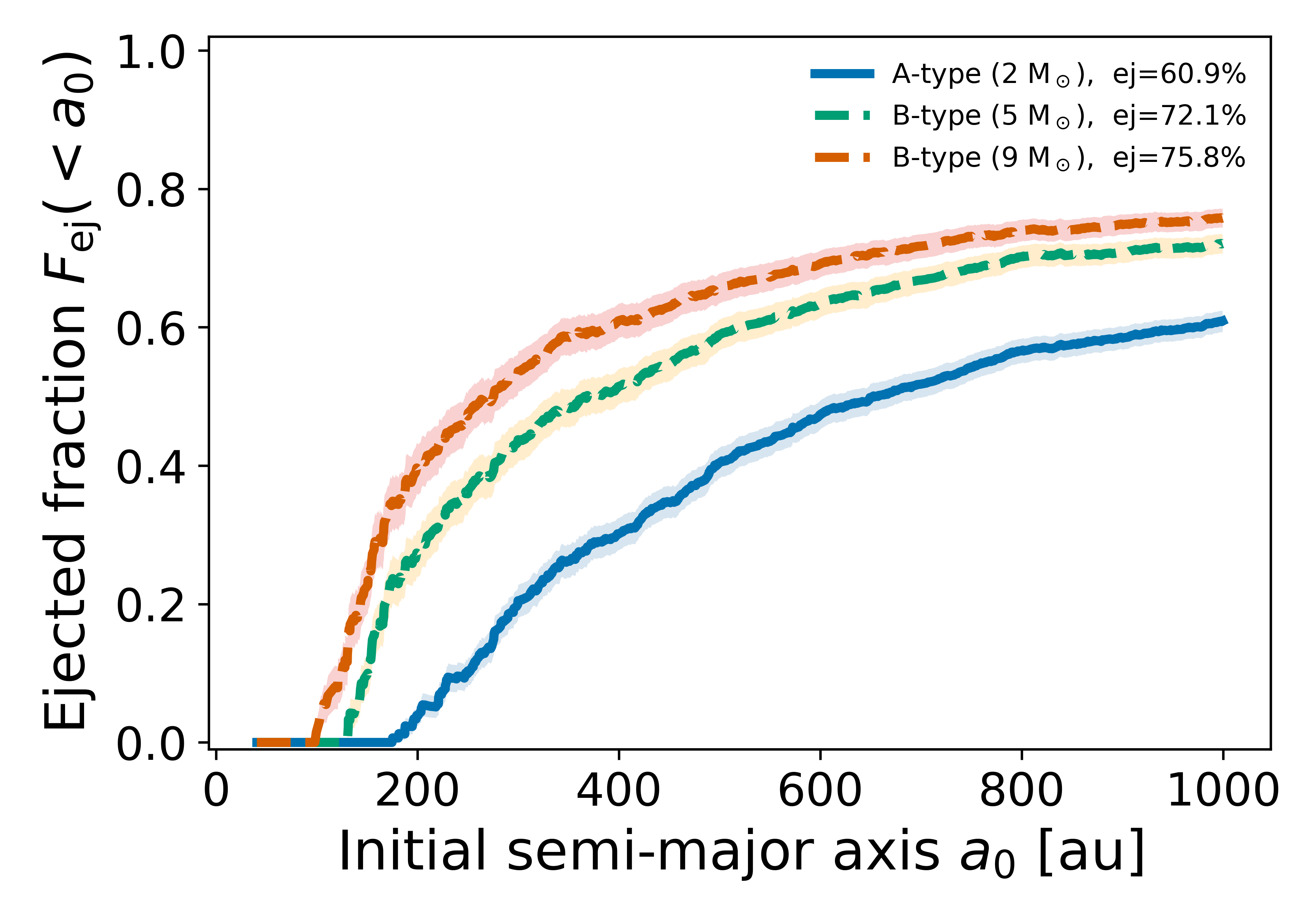}
  \caption{{Top:} Probability density function vs. semimajor axis for bound particles after a single stellar flyby for the \exto model. The histograms are shown for encounters with an A-type star ($2\,M_\odot$, blue), a B-type star ($5\,M_\odot$, green dashed and $9\,M_\odot$, orange), compared with the initial disk (gray). The inset shows kernel density estimates (KDEs). {Bottom:} Cumulative ejection fraction as a function of the initial semimajor axis $a_0$ for the same stellar flybys at a pericenter distance of $q_\star=300$~au. }
  \label{fig:hist_kde_AB5_ext}
\end{figure}

In Fig. \ref{fig:hist_kde_AB5_ext} we show that encounters with A- and B-type stars are highly disruptive in the \exto model. Between 60\% and 76\% of the initial particles are ejected into interstellar space. The efficiency of erosion increases systematically with stellar mass, with $f_{\rm ej}=0.61$ for a $2\,M_\odot$ perturber, $f_{\rm ej}=0.72$ for $5\,M_\odot$, and $f_{\rm ej}=0.76$ for $9\,M_\odot$.

Bound particles are confined to a progressively smaller radius with increasing stellar mass. The truncation radius, $a_{90}$, where 90\% of the survivors lie inward, decreases from $\sim 250$~au (A-type) to $\sim 150$~au (B5) and $\sim 125$~au (B9). Median semimajor axes are clustered around $a_{\rm med}=230$--290~au, with long tails extending to tens of thousands of au. Around $\sim1\%$ of the total population, reach $a\simeq2\times10^4$~au (A-type) and up to $\sim7\times10^4$~au (B5) or even $3.5\times10^4$~au (B9). These represent the first seeds of scattered-disk and Oort cloud-like populations.

The eccentricities of the reaming particles in the disk are strongly pumped, with median values $e_{\rm med}\simeq0.6$ in all cases and more than half exceeding $e>0.5$. Inclinations are widely dispersed, the A-type flyby produces $i_{\rm med}\simeq41^\circ$, with $\sim 58\%$ above $30^\circ$ and $\sim22\%$ retrograde. The $5\,M_\odot$ encounter drives $i_{\rm med}\simeq71^\circ$ with $\sim 66\%$ above $30^\circ$ and nearly 40\% retrograde, while the $9\,M_\odot$ case pushes the distribution to $i_{\rm med}\simeq94^\circ$ and $\sim52\%$ retrograde. For the $2\,M_\odot$ perturber, $90\%$ of survivors remain inside the initial disk, with only $\sim10\%$ scattered outward (3.2\% to $1000<a<5000$~au, 1.8\% to the inner Oort Cloud). In contrast, the $5$--$9\,M_\odot$ cases leave only $\sim80$\% of bound particles in the disk, while $\sim10$--15\% are placed in the scattered-disk region and up to 5\% in the inner Oort Cloud. No significant population reaches the outer Oort Cloud in these single encounters.

The ejected population leaves with characteristic asymptotic speeds of $v_\infty\simeq2$--3~km\,s$^{-1}$, somewhat lower than in the compact-disk case (Appendix~\ref{MS_compact}). This difference reflects the weaker binding of the extended disk: particles at large radii require only modest perturbations to escape, yielding high ejection fractions but relatively low excess velocities. These results show that single close encounters with massive stars (A- and B-type) are a highly efficient channel for producing interstellar comets from extended disks at very close passages ($\sim300$~au). While A-type stars already eject more than half the population, B-type encounters both erode the disk more strongly and drive more extreme excitation, seeding scattered-disk and Oort-like reservoirs and yielding $10^{11}$--$10^{12}$ ISO-scale bodies for a $20\,M_\oplus$ disk (Sect.~\ref{sec4}).

\end{appendix}

\end{document}